\documentclass{elsarticle}
\usepackage{needspace}
\usepackage{float}
\usepackage{graphicx} 
\usepackage{placeins}
\usepackage{hyperref} 
\usepackage{subcaption} 
\usepackage{tikz}
\usepackage{wrapfig} 
\usepackage{booktabs}
\usepackage{tabularx}
\usepackage{makecell} 
\usepackage{pifont}
\usepackage{amsmath} 
\usepackage{amssymb}
\usepackage{threeparttable}
\usepackage{bm}
\usepackage{multirow}
\usepackage{longtable}
\usepackage{pdflscape}

\newcommand{\cmark}{\ding{51}} 
\newcommand{\xmark}{\ding{55}} 

\usetikzlibrary{arrows.meta, positioning, shapes.geometric, fit}

\begin{document}

\begin{frontmatter}

\title{Rethinking Clinical Relevance in Chest X-ray Machine Learning: How Evaluation References Define Performance}

\author[1]{Panagiotis Fytas\corref{cor1}}
\ead{pf376@cam.ac.uk}
\author[4]{Ian Selby}
\author[3]{Clemens Karner}
\author[4]{Judith Babar}
\author[1]{Simon Baker}
\author[4]{Jake Beckford}
\author[4,6]{Timothy J Sadler}
\author[4]{Shahab Shahipasand}
\author[4]{Arthikkaa Thavakumar}
\author[4,6]{John Li Chen}
\author[4]{Alex Sawer}
\author[2]{Michael Roberts}
\author[6,7]{Jonathan Weir-McCall}
\author[5]{J.H.F. Rudd}
\author[2]{Carola-Bibiane Schönlieb}
\author[1]{Anna Korhonen}
\author[2,3]{Anna Breger}

\affiliation[1]{organization={Language Technology Lab (LTL), University of Cambridge},
country={United Kingdom}}
\affiliation[2]{organization={Department of Applied Mathematics and Theoretical Physics, University of Cambridge},
country={United Kingdom}}
\affiliation[3]{organization={Center for Medical Physics and Biomedical Engineering, Medical University of Vienna},
country={Austria}}
\affiliation[4]{organization={Cambridge University Hospitals NHS Foundation Trust},
country={United Kingdom}}
\affiliation[5]{organization={Department of Medicine, University of Cambridge},
country={United Kingdom}}
\affiliation[6]{organization={Department of Radiology, University of Cambridge},
country={United Kingdom}}
\affiliation[7]{organization={School of Biomedical Engineering and Imaging Sciences, King's College London},
country={United Kingdom}}

\cortext[cor1]{Corresponding author}

\begin{abstract}

Chest X-ray (CXR) machine learning relies heavily on automated evaluation using reference standards that aim to approximate clinical judgment. However, commonly used report-derived labels for pathology classification or generic image quality metrics for reconstruction may not reliably reflect clinical judgment. We systematically investigate how evaluation-reference choices affect model performance and ranking in both pathology classification and image quality assessment (IQA). To enable controlled comparison across evaluation references, we collected paired expert image- and report-derived labels for thoracic findings from a clinical cohort at Cambridge University Hospitals (CUH) and curated a subset of the public MIMIC-CXR dataset, along with expert ratings of diagnostic image quality. We show that for supervised image classifiers (ResNet, DenseNet), several zero-shot and fine-tuned vision-language models (e.g., MedKLIP, GLoRIA, and ConVIRT), changing the label source leads to substantial differences not only in performance estimates but also in model rankings. In parallel, alignment of IQA measures with expert judgment depends heavily on the choice of measure, and commonly used IQA metrics such as SSIM and PSNR often fail to align with expert assessments of diagnostic usability. Our results demonstrate that evaluation choices are crucial: they can determine which models and methods appear best and are therefore selected for further development or deployment. The selection of evaluation references should therefore be treated as a central component of clinical validity in CXR machine learning, and justified with respect to the pathology, imaging task, and intended downstream clinical use. The current code and information on access to the novel dataset are available at \url{https://github.com/PanagiotisFytas/XQA-Chest-X-ray}; the permanent ReShare link and access instructions will be added to the repository after the review of the repository is completed.

\end{abstract}

\end{frontmatter}
\section{Introduction}

The application of machine learning to chest X-ray (CXR) interpretation is often proposed as a clinical decision support tool to assist radiologists. Machine learning models are predominantly trained \citep{chexpert} and often evaluated  \citep{mimic-cxr, BUSTOS2020101797, reis2022brax} using labels derived from free-text radiology reports via automatic label extraction \citep{cohen2020limits}, including vision-language models (VLMs) \citep{zhang2022convirt, huang2021gloria, wang2022mgca, zhou2023mrm, zhou2022refers, wu2023medklip}. Within this heterogeneous training and evaluation landscape, \citet{rajpurkar2018deep} and \citet{chexpert} report that image classifiers trained on report-derived supervision labels occasionally exceed the performance of groups of radiologists when assessed against an evaluation reference derived directly from chest X-rays. At the same time, some works acknowledge disagreement between report-derived and image-derived labels, yet still treat expert image-derived labels as implicit ground truth \citep{gundel2021robust, visualchexbert}. However, radiology report-derived and image-derived labels encode different forms of clinical information: reports are shaped by clinical context, whereas expert image annotation is typically performed without access to patient history or indication. Apparent performance gains may therefore reflect alignment with a particular labeling convention rather than genuine clinical validity. This distinction is important because report-derived and image-derived labels are commonly used as references for the same CXR classification tasks, despite evidence that they can disagree substantially when applied to corresponding reports and images \citep{visualchexbert, mimic-cxr, chexpert}.
More broadly, systematic analyses have highlighted recurring methodological pitfalls in CXR machine learning, including dataset bias, confounding, unstable metrics, and limited external validation, that can undermine clinical suitability despite strong benchmark performance \citep{roberts2021common, cabitza2026almost}.

In this work, we ask whether conclusions about CXR machine learning are stable to the choice of evaluation reference for automated assessment. We study this question in two settings: pathology classification, where models are evaluated against image-derived or report-derived labels, and image quality assessment, where algorithms are typically judged using generic fidelity metrics rather than expert assessments of diagnostic usability \cite{breger2024study}.

To the best of our knowledge, publicly available CXR datasets typically provide only a single source of pathology labels. MIMIC-CXR \citep{mimic-cxr}, PadChest \citep{BUSTOS2020101797}, CheXpert \citep{chexpert}, and BRAX \citep{reis2022brax} provide labels automatically extracted from reports, whereas VinDr-CXR \citep{nguyen2022vindr} provides expert image-derived labels for thoracic findings. This limits the controlled comparison of report-derived and image-derived references. To address this gap, we curate XQA-Chest, which comprises paired expert image-derived and report-derived labels for pathology classification (XQA-RR) and expert ratings of diagnostic image quality under controlled degradations (XQA-IQA), thereby allowing us to study the impact of report-derived and image-derived labels on evaluation results. To test whether the same effects appear beyond a single clinical cohort, we additionally annotate a subset of MIMIC-CXR\footnote{A small subset of MIMIC-CXR reports has existing expert-annotated labels \citep{mimic-cxr}. However, these labels were produced for earlier MIMIC-CXR versions, and the associated report text does not always match that released in MIMIC-CXR 2.0 and later. Because Findings and Impression content may be combined or structured differently across versions, the mapping between the original expert labels and the current reports is not always clear. We therefore re-annotate the selected subset using a consistent protocol.} under the same image-only and report-only protocols. Methodological design and annotations were conducted in a clinical setting by a team of radiologists and radiographers from Cambridge University Hospitals NHS Foundation Trust, with varying levels of experience.

For pathology classification, we study how the choice of reference labels affects the ranking of a diverse set of deep learning models. Our experiments span two broad model families: supervised CXR classifiers and VLMs. The supervised classifiers are drawn from the TorchXRayVision framework \citep{cohen2022torchxrayvision} and include DenseNet-121 \citep{huang2017densely} and ResNet-50 \citep{he2016deep} models pretrained across a range of popular CXR datasets, as well as the JF-Healthcare probabilistic-CAM pooling model \citep{ye2020weakly}. In our experiments, we evaluate these supervised classifiers both with their original pretrained weights, without target-dataset adaptation, and after fine-tuning on the target dataset, enabling comparison between transferred and in-distribution supervised settings. The VLMs are drawn from the BenchX framework \citep{zhou2024benchx} and include  CONVIRT \citep{zhang2022convirt}, GLoRIA \citep{huang2021gloria}, MGCA \citep{wang2022mgca}, MRM \citep{zhou2023mrm}, REFERS \citep{zhou2022refers}, MedKLIP \citep{wu2023medklip}, MedCLIP \citep{wang2022medclip}, MFLAG \citep{liu2023mflag}, and PTUnifier \citep{chen2023ptunifier} models. For these models, we consider both zero-shot inference, using the pretrained image-text representations directly, and supervised variants, which we fine-tune on the target dataset.  We study whether model rankings are preserved when evaluated against varying label references, and whether ranking stability varies across pathologies. For image quality assessment, we analyze whether commonly used full-reference (FR) and no-reference (NR) IQA metrics preserve radiologists’ rankings of diagnostic image quality or favor distortions that are clinically irrelevant. Explored metrics include SSIM \citep{wang2004image}, PSNR \citep{gonzalez2008digital}, FSIM \citep{zhang2011fsim}, HaarPSI \citep{reisernhofer2018haarpsi}, NIQE \citep{mittal2012making}, SNR \citep{gonzalez2008digital} and HaarPSImed \citep{karner2025haarpsimed}.

\begin{figure}[ht]
    \centering
    \includegraphics[width=\textwidth]{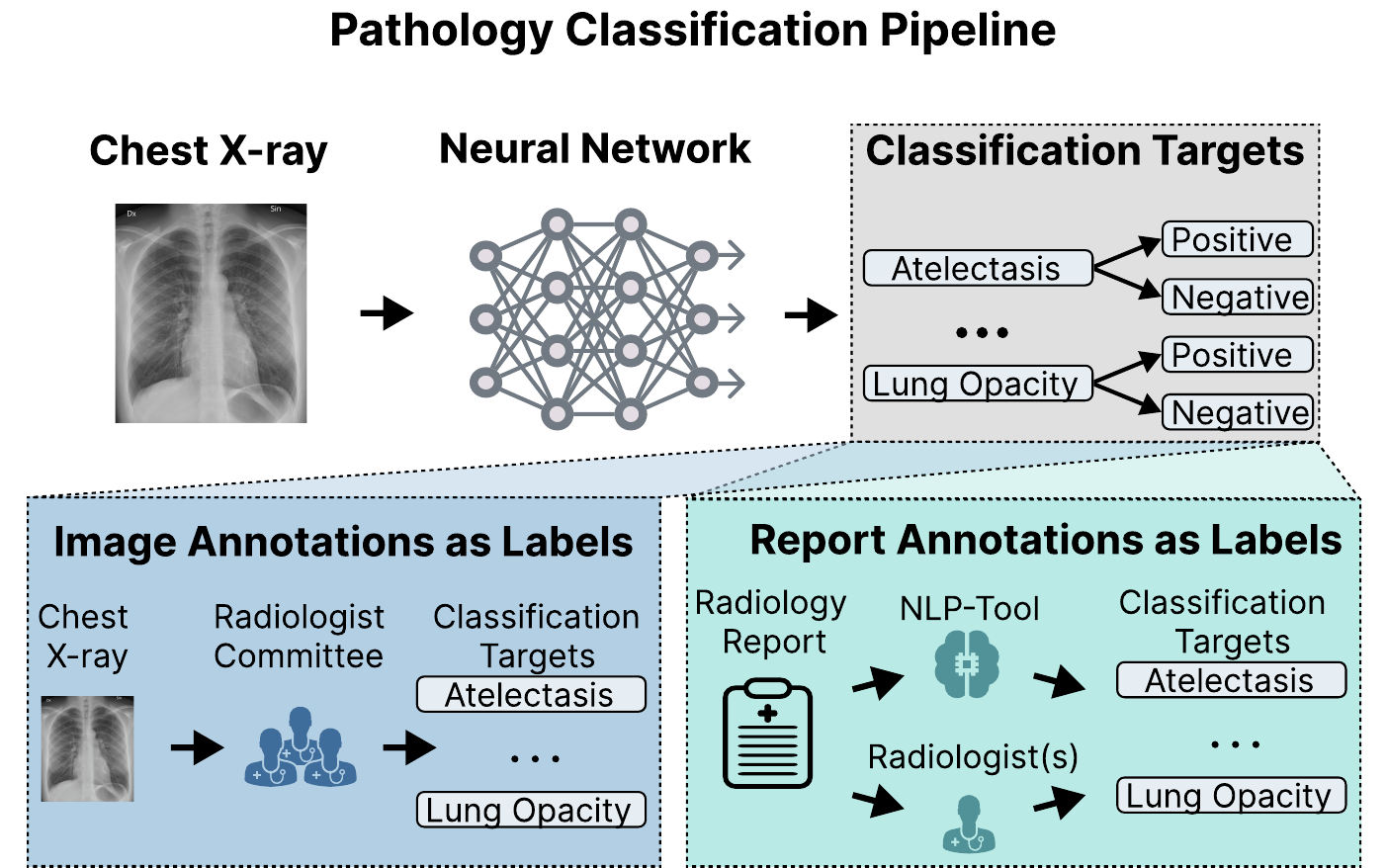}
    \caption{Overview of pathology classification with deep learning models. There are two sources of classification targets: labels derived either from expert image annotations or from (expert or automatic) radiology report annotations.}
    \label{fig:classification}
\end{figure}

\begin{figure}[ht]
    \centering
    \includegraphics[width=\textwidth]{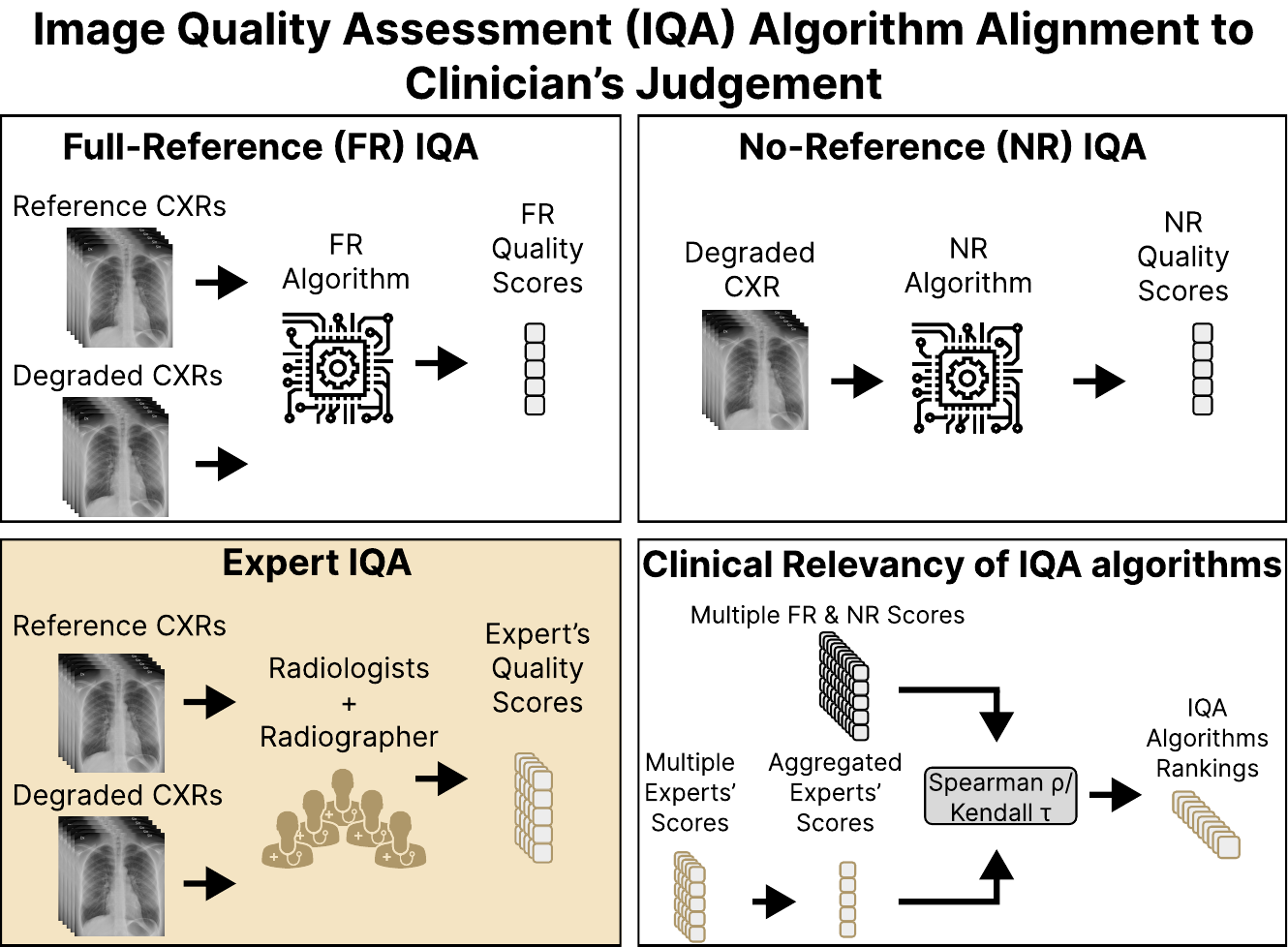}
    \caption{Overview of full-reference, no-reference, and clinician-preference-based IQA. The clinical relevance of IQA methods can be measured through ranking coefficient correlation measures between algorithmic and expert labels.}
    \label{fig:iqa}
\end{figure}

With the described data and machine learning models, we show that agreement between image-derived and report-derived references is strongly pathology-dependent and that these differences alter not only absolute performance estimates but also the relative ranking of models. For some pathologies, the model identified as best under one reference fails under another. We further show that the correlation of IQA metrics with expert judgments of diagnostic usability varies highly, and, for many commonly used metrics, it suggests that image outputs evaluated with those metrics do not guarantee clinical relevance. Together, these results demonstrate that common evaluation choices are consequential: the selected reference can determine which models and methods are judged best for clinically relevant downstream tasks. The design and choices of evaluation schemes, therefore, should be treated as a central component of clinical validity in CXR and, generally, medical machine learning.

\section{Methods}
We aim to determine how strongly the judgment of machine learning results depends on common choices in the underlying evaluation schemes. In particular, choices that are seen as standard practice and therefore, when employed, are usually not further challenged regarding their suitability. We analyze two tasks, pathology classification and image quality assessment, through ranking stability, i.e., whether changing the evaluation reference changes which models or methods are judged best.

\subsection{Chest X-ray Pathology Classification}

We study multi-label CXR pathology classification using a CheXpert/MIMIC-CXR-style label schema \citep{chexpert, mimic-cxr}. The full annotation schema comprises 14 finding categories: Atelectasis, Cardiomegaly, Edema, Enlarged cardiomediastinum, Consolidation, Fracture, Lung lesion, Lung opacity, Pleural effusion, Pleural other, Pneumonia, Pneumothorax, Support devices, and No finding. By focusing on pathologies, we exclude No finding and Support devices from our analysis, yielding 12 pathologies. We consider both supervised and zero-shot classification settings.

Each input image is a chest radiograph $\mathbf{x} \in \mathcal{X} \subset \mathbb{R}^{H \times W}$, where $\mathcal{X}$ denotes the space of grayscale images and $H$ and $W$ describe the size of the image. An image encoder $v_\theta : \mathcal{X} \rightarrow \mathbb{R}^d$ maps the input to a $d$-dimensional latent representation $\mathbf{v} = v_\theta(\mathbf{x})$, implemented by architectures such as DenseNet \citep{huang2017densely}, ResNet \citep{he2016deep}, or Vision Transformer \citep{dosovitskiy2021image} with some parameter space $\theta$.

In the supervised setting, a classifier head $f_\theta : \mathbb{R}^d \rightarrow \mathbb{R}^K$ produces pathology-specific logits, yielding predictions
\[
\mathbf{p} = (p_1,\dots, p_K) = \sigma\bigl(f_\theta(\mathbf{v})\bigr), \quad p_k \in (0,1), \text{ for } k \in \{1,\dots, K\}
\]
where $\sigma(\cdot)$ denotes the sigmoid activation applied independently to each pathology logit, and $K$ denotes the number of pathologies.

The image encoder $v_\theta$ can arise from different training paradigms: (i) classification models trained end-to-end in a fully supervised manner (e.g.\ TorchXRayVision DenseNet/ResNet encoders), and (ii) vision-language models (VLMs) that are first pretrained using image--text contrastive learning \citep{Radford2021LearningTV, zhang2022convirt} and subsequently fine-tuned for image classification. In both cases, training relies on task-specific labels, typically derived automatically from radiology reports.

Specifically, VLM pretraining learns a shared embedding space via contrastive learning \citep{oord2018representation, chen2020simple}. Given an image $\mathbf{x}$, an associated report text $\mathbf{t}$, and image/text encoders $v_\theta$ and $l_\theta$:
\[
\mathbf{v} = v_\theta(\mathbf{x}), \quad \mathbf{l} = l_\theta(\mathbf{t}),
\]
training maximizes similarity between $(\mathbf{v},\,\mathbf{l})$ matched pairs while minimizing similarity to other samples. A common objective is
\[
\mathcal{L} = - \log \frac{\exp\bigl(S(\mathbf{v}, \mathbf{l}) / \tau\bigr)}{\sum_{\mathbf{t'}} \exp\bigl(S(\mathbf{v}, l_\theta(\mathbf{t'})) / \tau\bigr)},
\]
with cosine similarity $S(\cdot,\cdot)$ and temperature $\tau \in  \mathbb{R}$.

In the zero-shot setting, predictions are obtained directly from this shared space. For each pathology $k$, a set of prompts $\mathcal{T}_k$ is constructed (e.g.\ \textit{``mild subsegmental atelectasis'', ``persistent atelectasis''}). Each prompt $\mathbf{t_k} \in \mathcal{T}_k$ is encoded as $\mathbf{l}_{\mathbf{t_k}} = l_\theta(\mathbf{{t_k}})$, and the score is computed as
\begin{equation*}
s_k = \max_{\mathbf{{t_k}} \in \mathcal{T}_k} S(\mathbf{v}, \mathbf{l}_{\mathbf{t_k}}),
\end{equation*}
which serves as a logit for pathology $k$.

For evaluation, each image is associated with a reference label vector:
\[
\mathbf{y}^{(r)} = (y^{(r)}_1,\dots, y^{(r)}_K), \quad y^{(r)}_k \in \{0,1\},
\]
where $r$ denotes the label source (e.g.\ expert image-derived or report-derived labels). Uncertain labels are excluded, following standard practice in benchmarks such as CheXpert \citep{chexpert}. Different choices of $r$ define the evaluation references studied in this work.

\subsubsection{Deep Learning Models and Evaluation Details}

Table~\ref{tab:model_overview} summarizes the evaluated models.  We consider two model families: (i) supervised CNN-based classification models and (ii) vision-language models (VLMs), evaluated in both supervised and zero-shot settings. Supervised models are trained on datasets with automatically extracted radiology report-derived labels, as shown in Table~\ref{tab:related_dataset_overview}, reflecting common practice in CXR machine learning.

We include publicly available supervised classification models from the Torch-XRayVision framework \citep{cohen2022torchxrayvision}, based on DenseNet-121 \citep{huang2017densely} and ResNet-50 \citep{he2016deep} architectures trained on individual and pooled popular CXR datasets. We also include the JF-Healthcare DenseNet-121 model with Probabilistic Class Activation Map (PCAM) pooling \citep{ye2020weakly}. These models are evaluated on XQA-RR both with their pretrained weights (without adaptation) and after fine-tuning (in-distribution). We fine-tune the models on a non-public CXR dataset from Cambridge University Hospitals (CUH) comprising approximately 13{,}000 images with labels automatically extracted by the RadPert labeler \citep{fytas2024can}, using the default TorchXRayVision hyperparameters. For MIMIC-CXR, we do not perform additional fine-tuning because TorchXRayVision already provides models pretrained on MIMIC-CXR as well as models trained on other CXR datasets, allowing comparison between in-domain and transferred settings without further adaptation.

We evaluate VLMs from the BenchX framework \citep{zhou2024benchx}, which provides a unified setup for pretraining a diverse set of CXR-specific VLMs under controlled conditions. Included models span a range of architectures and pretraining strategies: CONVIRT \citep{zhang2022convirt}, REFERS \citep{zhou2022refers}, PTUnifier \citep{chen2023ptunifier}, MGCA \citep{wang2022mgca}, GLoRIA \citep{huang2021gloria}, MedCLIP \citep{wang2022medclip}, MedKLIP \citep{wu2023medklip}, MFLAG \citep{liu2023mflag}, and MRM \citep{zhou2023mrm}.

VLMs are evaluated in both zero-shot and supervised settings. Zero-shot classification requires no further training. Conversely, supervised VLMs are fine-tuned for XQA-RR on the same CUH dataset described above and are denoted with a \texttt{-CUH} suffix (e.g., CONVIRT-CUH). For MIMIC-CXR, models fine-tuned on MIMIC-derived labels are denoted with a \texttt{-MIMIC} suffix. Training follows the hyperparameter settings provided by BenchX \citep{zhou2024benchx}.

Performance is quantified using the area under the receiver operating characteristic curve (ROC-AUC) \citep{hanley1982meaning}. We adopt ROC-AUC rather than threshold-dependent metrics because positive-label prevalence varies substantially across label sources. ROC-AUC measures the probability that a randomly chosen positive sample is ranked higher than a randomly chosen negative sample, and is invariant to class prevalence under fixed positive and negative score distributions \citep{brabec2020model}.
Thus, it removes changes in metric value arising purely from different positive-label rates and provides a suitable strategy for comparing model discrimination across reference label sets. In this setting, differences in ROC-AUC therefore reflect changes in how each reference source defines the positive and negative cases, rather than prevalence alone. In contrast, metrics such as the F1-score \citep{powers2020evaluation} depend on a chosen decision threshold and on the precision--recall trade-off. Because precision is directly affected by the ratio of positive to negative samples, F1-scores can change purely as a consequence of differing label prevalence, even when the model's underlying discrimination is unchanged \citep{williams2021effect,davis2006relationship}. We estimated ROC-AUC uncertainty using 1000-sample non-parametric bootstrapping over patient studies and report 95\% percentile confidence intervals \citep{efron1986bootstrap}.

\begin{table}[htbp]
\centering
\footnotesize
\setlength{\tabcolsep}{4pt}
\begin{tabularx}{\textwidth}{@{} l l >{\raggedright\arraybackslash}X l @{}}
\toprule
\textbf{Model} & \textbf{Backbone} & \makecell[l]{\textbf{Pretraining}\\\textbf{Datasets}} & \makecell[l]{\textbf{Training}\\\textbf{Objective}} \\
\midrule
D-RSNA \citep{cohen2022torchxrayvision}      & DenseNet-121          & RSNA Pneumonia & Supervised \\
D-NIH \citep{cohen2022torchxrayvision}       & DenseNet-121          & ChestX-ray14 & Supervised \\
D-PC \citep{cohen2022torchxrayvision}        & DenseNet-121          & PadChest & Supervised \\
D-CheX \citep{cohen2022torchxrayvision}      & DenseNet-121          & CheXpert & Supervised \\
D-MC$_{\text{ch}}$ \citep{cohen2022torchxrayvision} & DenseNet-121   & MIMIC-CXR (CheXpert labels) & Supervised \\
D-MC$_{\text{nb}}$ \citep{cohen2022torchxrayvision} & DenseNet-121   & MIMIC-CXR (NegBio labels) & Supervised \\
D-POOL \citep{cohen2022torchxrayvision}      & DenseNet-121          & ChestX-ray14, CheXpert, PadChest, MIMIC-CXR (CheXpert labels), OpenI, RSNA Pneumonia & Supervised \\
R-POOL \citep{cohen2022torchxrayvision}      & ResNet-50             & ChestX-ray14, CheXpert, PadChest, MIMIC-CXR (CheXpert labels), OpenI, RSNA Pneumonia & Supervised \\
JFH \citep{ye2020weakly}                     & DenseNet-121 + PCAM   & CheXpert & Supervised \\
CONVIRT \citep{zhang2022convirt}             & ResNet-50             & MIMIC-CXR & VL \\
REFERS \citep{zhou2022refers}                & ViT                   & MIMIC-CXR & VL \\
PTUNIFIER \citep{chen2023ptunifier}          & ViT                   & MIMIC-CXR & VL \\
V-MGCA \citep{wang2022mgca}                  & ViT                   & MIMIC-CXR & VL \\
R-MGCA \citep{wang2022mgca}                  & ResNet-50             & MIMIC-CXR & VL \\
GLORIA-G \citep{huang2021gloria}             & ResNet-50             & MIMIC-CXR & VL \\
GLORIA-B \citep{huang2021gloria}             & ResNet-50             & MIMIC-CXR & VL \\
V-MedCLIP \citep{wang2022medclip}            & ViT                   & MIMIC-CXR & VL \\
R-MedCLIP \citep{wang2022medclip}            & ResNet-50             & MIMIC-CXR & VL \\
MedKLIP \citep{wu2023medklip}                & ResNet-50             & MIMIC-CXR & VL \\
MFLAG \citep{liu2023mflag}                   & ResNet-50             & MIMIC-CXR & VL \\
MRM \citep{zhou2023mrm}                      & ViT                   & MIMIC-CXR & VL \\
\bottomrule
\end{tabularx}
\caption{Overview of evaluated chest X-ray models, their backbone architectures, training datasets, and training objectives.}
\label{tab:model_overview}
\end{table}

\subsubsection{Data and Annotation of Common Thoracic Findings}

\begin{figure}[ht]
    \centering
    \begin{subfigure}[t]{0.6443\textwidth}
        \vspace{0pt}
        \centering
        \includegraphics[width=\textwidth]{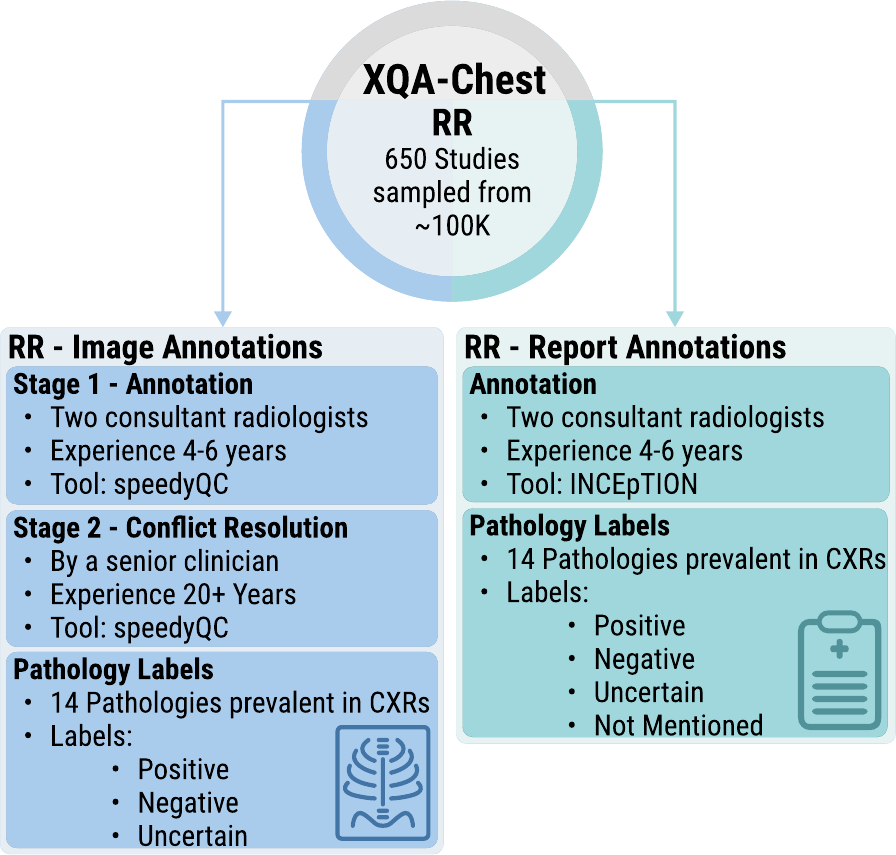}
        \caption{\textbf{XQA-RR:} enables comparison of chest X-ray pathology labeling sources.}
        \label{fig:xqa_rr}
    \end{subfigure}
    \hfill
    \begin{subfigure}[t]{0.3057\textwidth}
        \vspace{0pt}
        \centering
        \includegraphics[width=\textwidth]{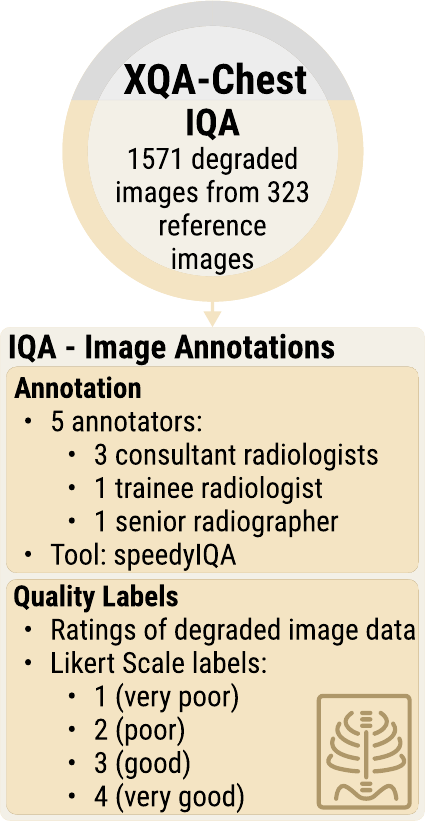}
        \caption{\textbf{XQA-IQA:} allows comparison of IQA metrics to expert judgments of clinical relevance of CXR images.}
        \label{fig:xqa_iqa}
    \end{subfigure}
    \caption{Overview XQA-Chest.}
    \label{fig:XQA}
\end{figure}

\begin{table}[htbp]
    \centering
    \begin{threeparttable}
    \begin{tabular}{lcccc}
        \toprule
        \multirow{2}{*}{\textbf{Dataset}} &
        \multicolumn{2}{c}{\textbf{Counts}} &
        \multicolumn{2}{c}{\textbf{Annotations}} \\
        \cmidrule(lr){2-3}
        \cmidrule(lr){4-5}
        & \textbf{Images} & \textbf{Reports} & \textbf{Images} & \textbf{Reports} \\
        \midrule
        
        ChestX-ray14 \citep{chestxray14}
        & 112k & \xmark & $\bm{\thicksim}$\tnote{1} & Automatic \\
        
        CheXpert \citep{chexpert}
        & 224k & 224k & $\bm{\thicksim}$\tnote{2} & Automatic \\
        
        MIMIC-CXR \citep{mimic-cxr}
        & 377k & 228k & \xmark & Automatic \\
        
        BRAX \citep{reis2022brax}
        & 40k & \xmark & \xmark & Automatic \\
        
        PadChest \citep{BUSTOS2020101797}
        & 169k & 110k & \xmark & Hybrid \\
        
        OpenI \citep{demner2015preparing}
        & 8k & 4k & \xmark & Manual \\
        
        RSNA Pneumonia \citep{shih2019augmenting}
        & 26k & \xmark & \cmark & \xmark \\
        
        VinDr-CXR \citep{nguyen2022vindr}
        & 18k & \xmark & \cmark & \xmark \\
        \midrule
        
        XQA-RR
        & 650 & 650 & \cmark & \cmark \\
        \bottomrule
    \end{tabular}
    \begin{tablenotes}
    \footnotesize
    \item[1] Image annotations have been annotated previously, but for a different label set than the report annotations, and are no longer publicly available \citep{majkowska2020chest}.
    \item[2] Image annotations exist for 500 images, for which reports and report annotations are nonexistent.
    \end{tablenotes}
    \end{threeparttable}
    \caption{Comparison of prevalent chest X-ray datasets with respect to availability of reports and annotation sources.}
    \label{tab:related_dataset_overview}
\end{table}

As observed in Table~\ref{tab:related_dataset_overview}, publicly available CXR datasets typically provide pathology labels from a single source, usually either automatic extraction from radiology reports or manual image annotation. This limits the ability to study how evaluation conclusions change across label sources. While some prior works examine disagreement between image-derived and report-derived labels on paired studies, they do not release such paired expertly-annotated labels as a public benchmark or systematically analyze how evaluation reference choice affects downstream conclusions; instead, expert image-derived labels are typically treated as the evaluation reference, despite the underlying assumption of their ground-truth status remaining largely untested \citep{gundel2021robust, visualchexbert}. To enable this analysis, we collected XQA-RR (Figure~\ref{fig:xqa_rr}) from a cohort of 650 patient studies acquired at Cambridge University Hospitals (CUH) between 2019 and 2021, each comprising a frontal chest radiograph (AP/PA) and a paired radiology report. Patient studies were sampled to ensure representation of both normal and abnormal examinations across 14 finding categories aligned with CheXpert/MIMIC-CXR conventions. Cases with missing reports, non-diagnostic image quality, or duplicate examinations from the same encounter were excluded, and sampling was stratified to reduce extreme class imbalance and fair demographic representation.

Images and reports were independently annotated by two consultant radiologists to avoid influence between image and report annotation sources. For image annotation, each radiograph was reviewed in an image-only setting (no report or clinical history) using SpeedyQC \citep{selbs_speedy_qc_2026}, and each finding was labeled as \emph{negative}, \emph{uncertain}, or \emph{positive}, with optional bounding box localization. Disagreements between the two annotators were resolved by a senior radiologist, producing adjudicated image-derived labels, which we refer to as conflict-resolved (CR) image-derived labels. Reports were annotated separately in INCEpTION \citep{klie-etal-2018-inception}, where the same two radiologists labeled the paired reports without viewing the images. For each finding category, reports were assigned one of four values: \emph{negative}, \emph{uncertain}, \emph{positive}, or \emph{not mentioned}.

To complement XQA-RR with a publicly available dataset, we additionally annotated a subset of MIMIC-CXR \citep{mimic-cxr}. We selected 735 frontal (AP\slash PA) chest radiographs from the MIMIC-CXR expert-annotated test set, augmenting the selection to better reflect the demographic distribution of the broader MIMIC-CXR population. All selected patient studies include paired radiology reports. Images were annotated by two radiologists using SpeedyQC under the same image-only protocol as CUH: each finding was labeled as \emph{negative}, \emph{uncertain}, or \emph{positive}, without access to the paired reports. Unlike CUH, no adjudication step was performed; instead, we derive a consensus label using automatic fusion of image annotator labels (Section~\ref{sec:label_agg}), which we validate against the conflict-resolved image-derived labels for XQA-RR. Reports were annotated manually by the same annotators using INCEpTION, independently for the Findings and Impressions sections of each report, without viewing the images. This yields two distinct report-derived label sets per study, enabling direct analysis of within-report label agreement as an additional source of annotation variability.

Because several pathologies are rare in these datasets, the primary comparative analysis is restricted to four selected pathologies with at least 50 positive cases in either the image-derived or report-derived reference: Atelectasis, Consolidation, Lung opacity, and Pleural effusion. Support devices and No finding are excluded from pathology-focused analyses, although the full annotation schema is retained for dataset-level summaries where stated explicitly. Positive prevalence for XQA-RR and MIMIC-CXR is summarized in Tables~\ref{tab:positive-prevalence} and~\ref{tab:positive-prevalence-mimic}, with full label distributions for the full annotation schema provided in Appendix Tables~\ref{tab:label-distribution-long} and~\ref{tab:label-distribution-long-mimic}.

We analyze differences between label sources using both omission analysis and agreement metrics. Omission analysis identifies cases where a positive finding in one label source is absent or negative in the other, providing insight into systematic differences in what each label source captures.

We quantify inter-annotator agreement and agreement between label sources using quadratically weighted Cohen's $\kappa$ \citep{cohen1960coefficient, cohen1968weighted}. Cohen's $\kappa$ measures agreement beyond chance and is defined as:
\begin{equation*}
\kappa = \frac{p_o - p_e}{1 - p_e},
\end{equation*}
where $p_o$ is the observed agreement and $p_e$ is the expected agreement under chance, estimated from the empirical marginal label distributions. To account for the ordinal structure of labels (negative $<$ uncertain $<$ positive), quadratically weighted $\kappa$ penalizes larger disagreements more strongly.

Because uncertainty handling can substantially affect agreement estimates, we additionally compute unweighted Cohen's $\kappa$ under two binary mappings, following \citet{visualchexbert}. In the optimistic mapping (High--$\kappa$), uncertain labels are treated as agreement with the corresponding definitive label, whereas in the pessimistic mapping (Low--$\kappa$), uncertain labels are treated as disagreement. These Low--$\kappa$ and High--$\kappa$ values are reported alongside the central quadratically weighted $\kappa$ estimate. Weighted F1-scores are reported in the Appendix for completeness.

For XQA-RR, we also performed a qualitative review of image--report label disagreements. The review was conducted by the adjudicating radiologist.

\subsubsection{Label aggregation and supervised fusion methods}
\label{sec:label_agg}

We evaluate several automatic label fusion methods for combining multiple image-derived labels into a single consensus label. Their agreement with the adjudicated reference is assessed on XQA-RR, where conflict-resolved image-derived labels provide an evaluation reference, and the resulting methods are then applied to MIMIC-CXR, where no adjudicated labels are available. Specifically, we derive a single consensus label from two image annotator labels using methods defined on the label space $\mathcal{Y}=\{\text{negative},\text{uncertain},\text{positive}\}$, applied independently for each pathology. We compare the following fusion methods:

\paragraph{Majority vote (MV)}
MV assigns each study the label chosen by the majority of annotators; ties resolve to \emph{uncertain}. It serves as a non-parametric baseline, treating all annotators equally.

\paragraph{Deterministic pairwise fusion}
For two annotators with labels $a,b\in\mathcal{Y}$, three fixed rules handle disagreements differently. FUSE\_OR resolves any conflict involving a positive to positive (optimistic); FUSE\_AND requires both annotators to agree on positive (conservative).

\paragraph{Dawid--Skene (DS) and MACE}
Both are probabilistic crowd models that jointly infer a latent consensus label and per-annotator reliability. Dawid--Skene \citep{dawid1979maximum} models each annotator by a full confusion matrix estimated via Expectation-maximization (EM); MACE \citep{hovy2013learning} additionally estimates annotator competence, down-weighting unreliable annotators.

\paragraph{Neural fusion (NN Fusion) training}
We trained a multilayer perceptron to predict conflict-resolved image-derived labels from the one-hot encoded label pair of both annotators, concatenated across all $K$ pathologies. The network is trained with pathology-wise class-weighted cross-entropy, with model selection via 5-fold cross-validation using macro-F1 on four key pathologies (Consolidation, Pleural effusion, Lung opacity, Atelectasis) as the stopping criterion. A pathology-specific variant (NN Fusion PS) replaces global predictions with dedicated specialist models for those four pathologies, combining shared statistical strength with pathology-level specialization.

\subsection{Image Quality Assessment}

Let $x \in \mathcal{X}$ denote a reference chest radiograph and $\tilde{x} \in \mathcal{X}$ a degraded version of the same image. An IQA metric can be viewed as a function:
\[
m : \mathcal{X} \times \mathcal{X} \rightarrow \mathbb{R},
\]
which assigns a quality score $q = m(x, \tilde{x})$ reflecting the perceived similarity or fidelity between the two images.

Given a set of degraded images $\{\tilde{x}_i\}_{i=1}^{n}$ derived from multiple reference images, each IQA metric induces a ranking over images according to their predicted quality scores. Independently, expert annotators provide ordinal ratings of diagnostic usability for the same images, inducing a reference ranking based on human judgment.

The IQA evaluation problem can therefore be formulated as the following: we assess how well the ranking induced by an IQA metric $m$ aligns with the ranking induced by expert opinion. In Subsection \ref{subsec:rank}, we introduce the statistics utilized for measuring ranking agreement.

We evaluate a range of full-reference (FR) and no-reference (NR) IQA metrics commonly used in medical imaging, including PSNR \citep{gonzalez2008digital}, SSIM variants \citep{wang2004image}, HaarPSI \citep{reisernhofer2018haarpsi}, SNR \citep{gonzalez2008digital}, and NIQE \citep{mittal2012making}. Most IQA measures differ in their sensitivity to structural distortions, contrast changes, and noise. However, they have not been designed for the medical image domain, and are rarely validated against expert judgment in clinical imaging \citep{breger2024study}.

We enable comparison between metric-based and expert-based quality judgments by constructing XQA-IQA (Figure \ref{fig:xqa_iqa}). We sample 323 reference frontal CXRs from CUH clinical scanners, each subjected to a set of controlled degradations directly from the scanner's visualization settings, such as contrast or sharpness, yielding a total of 1,571 altered images. The reference images were selected according to the preferred settings of the clinicians in charge. These images were rated independently by radiologists and reporting radiographers using a four-point Likert scale reflecting diagnostic usability using the SpeedyIQA app \citep{selbs_speedy_iqa_2026}. This design enables direct comparison between metric-based rankings and expert rankings of image quality, as well as analysis of inter-rater variability.

To account for individual rating biases, raw Likert scores are normalized per annotator using z-score standardization. Specifically, for annotator $a$ and image $i$, the z-score is:
\begin{equation*}
z_{i,a} = \frac{r_{i,a} - \mu_a}{\sigma_a},
\label{eq:zscore}
\end{equation*}
where $r_{i,a}$ is the raw rating, and $\mu_a$ and $\sigma_a$ are the mean and standard deviation of annotator $a$'s ratings across all images. The z-score Mean Opinion Score (z-MOS) for image $i$ is then defined as the mean z-score across all $A$ annotators:
\begin{equation*}
\text{z-MOS}_i = \frac{1}{A} \sum_{a=1}^{A} z_{i,a}.
\label{eq:mos}
\end{equation*}

\subsection{Ranking Stability Metrics}
\label{subsec:rank}
We focus on rank-based analyses rather than absolute performance to determine how the pathology classification evaluation reference and IQA method affect conclusions about model performance and image quality, respectively. We employ Spearman’s rank correlation coefficient (SRCC) \citep{spearman1904} and Kendall’s rank correlation coefficient (KRCC) \citep{kendall1938} to quantify ranking stability across evaluation references.

\paragraph{Spearman’s Rank Correlation Coefficient (SRCC)}
Given two rankings \( \{R_i\}_{i=1}^n \) and \( \{S_i\}_{i=1}^n \) over the same set of \( n \) models or image outputs, SRCC is defined as:
\begin{equation*}
\rho_s
\;=\;
1 - \frac{6 \sum_{i=1}^{n} d_i^2}{n(n^2 - 1)},
\label{eq:srcc}
\end{equation*}
where \( d_i = R_i - S_i \) denotes the difference between the ranks assigned to item \( i \) under the two evaluation references. SRCC takes values in the interval \([-1,1]\), with higher values indicating stronger agreement between rankings.

\paragraph{Kendall’s Rank Correlation Coefficient (KRCC)}
KRCC measures the correspondence between two rankings by considering all unordered pairs of items. For a given pair \((i,j)\), the rankings are said to be \emph{concordant} if both rankings agree on their ordering, and \emph{discordant} otherwise. Let \( C \) and \( D \) denote the number of concordant and discordant pairs, respectively. KRCC is then defined as:
\begin{equation*}
\tau
\;=\;
\frac{C - D}{\frac{1}{2}n(n-1)}.
\label{eq:krcc}
\end{equation*}

For pathology classification ranking analyses on XQA-RR and MIMIC-CXR, SRCC and KRCC uncertainty was estimated using 1000 bootstrap replicates over patient studies \citep{efron1986bootstrap}. In each replicate, we computed ROC-AUCs for each model and evaluation reference, derived the corresponding model rankings, and calculated rank correlations between
references. We report 95\% percentile confidence intervals from the
resulting bootstrap distributions.

\section{Results}

We first present results for pathology classification, examining how the choice of evaluation reference affects labels and conclusions about model performance. We then report the image quality assessment results, comparing algorithmic metrics against expert judgments of diagnostic usability.

\subsection{Image Classification Models}

We begin by analyzing differences between image-derived and report-derived label sources, including omission patterns, agreement, and uncertainty sensitivity. We then examine whether these differences affect ROC-AUC estimates and model rankings.

\subsubsection{Quantitative Analysis of XQA-RR and MIMIC-CXR Labels.}

Agreement results are presented in the main text using quadratically weighted $\kappa$ as the central estimate, with Low--$\kappa$ and High--$\kappa$ indicating unweighted Cohen's $\kappa$ under pessimistic and optimistic uncertainty mappings. Weighted F1 results follow the same qualitative trends and are reported in the Appendix for completeness. Unless otherwise stated, reported $\kappa$ values refer to the central quadratically weighted $\kappa$ estimate.

\begin{figure}[!htbp]
    \centering
    \begin{subfigure}[b]{0.47\textwidth}
        \centering
        \includegraphics[width=\textwidth]{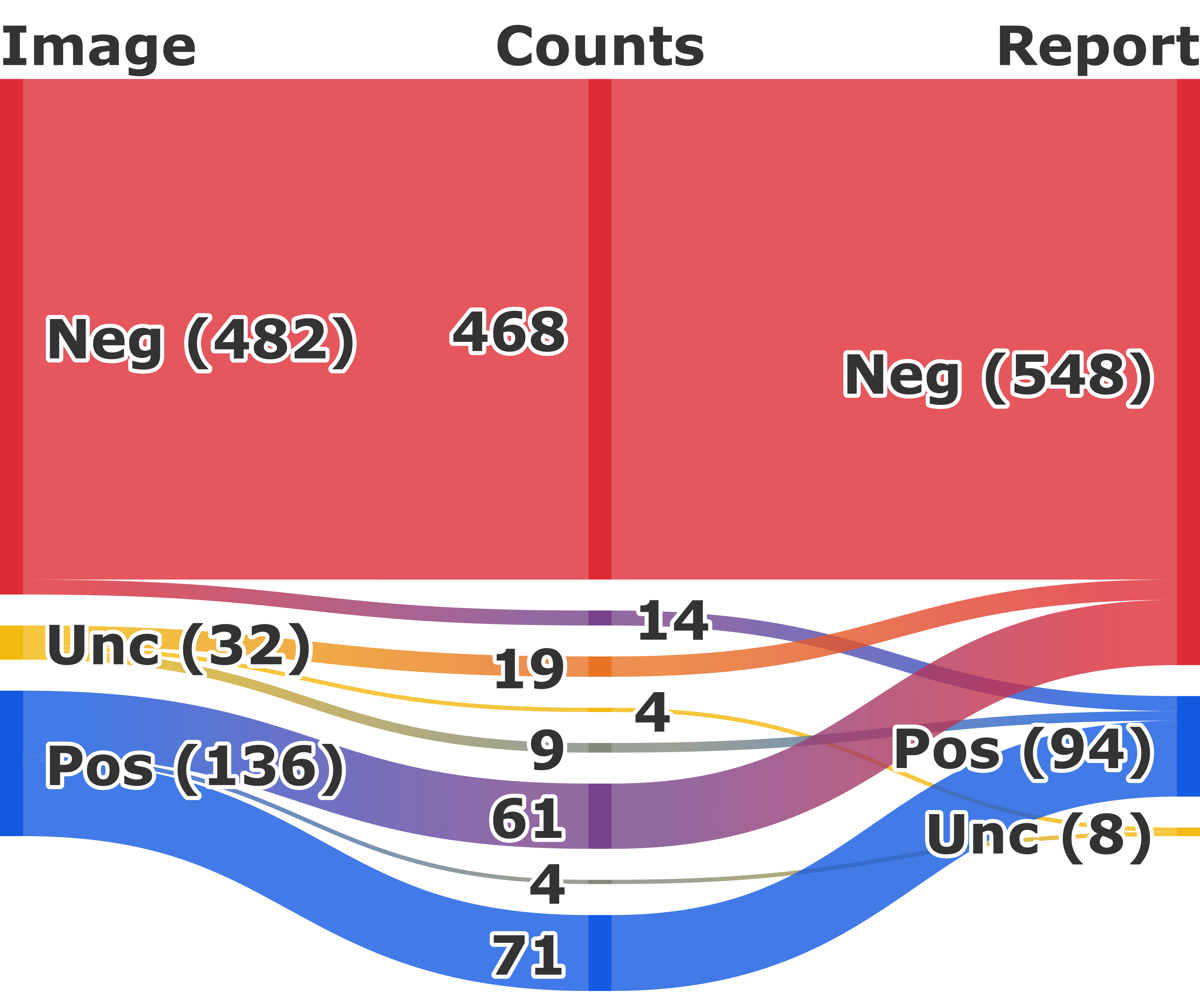}
        \subcaption{Pleural effusion}
        \label{fig:sankey_plef}
    \end{subfigure}
    \hfill
    \begin{subfigure}[b]{0.47\textwidth}
        \centering
        \includegraphics[width=\textwidth]{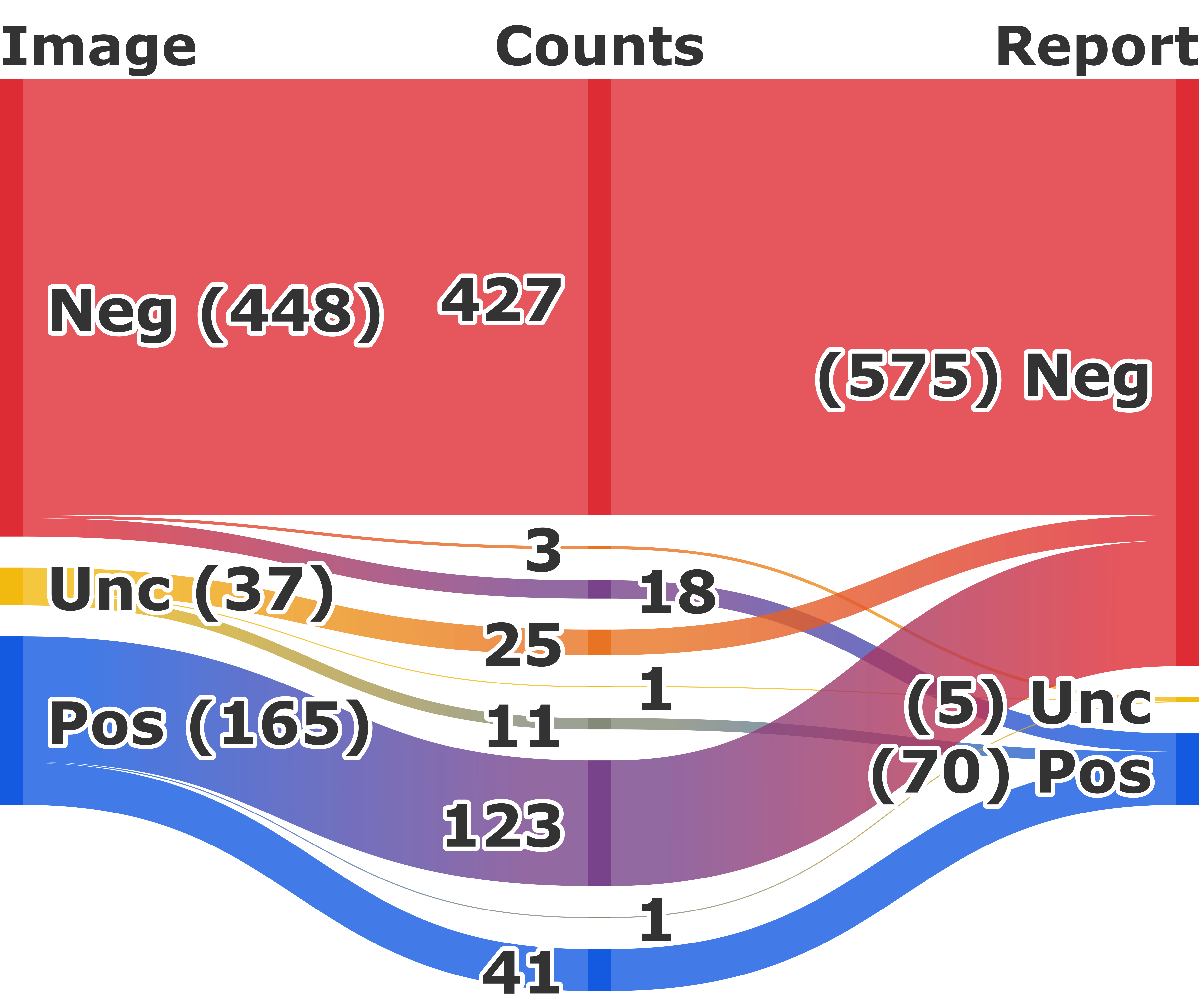}
        \subcaption{Atelectasis}
        \label{fig:sankey_atelectasis}
    \end{subfigure}

    \vspace{0.5em}

    \begin{subfigure}[b]{0.47\textwidth}
        \centering
        \includegraphics[width=\textwidth]{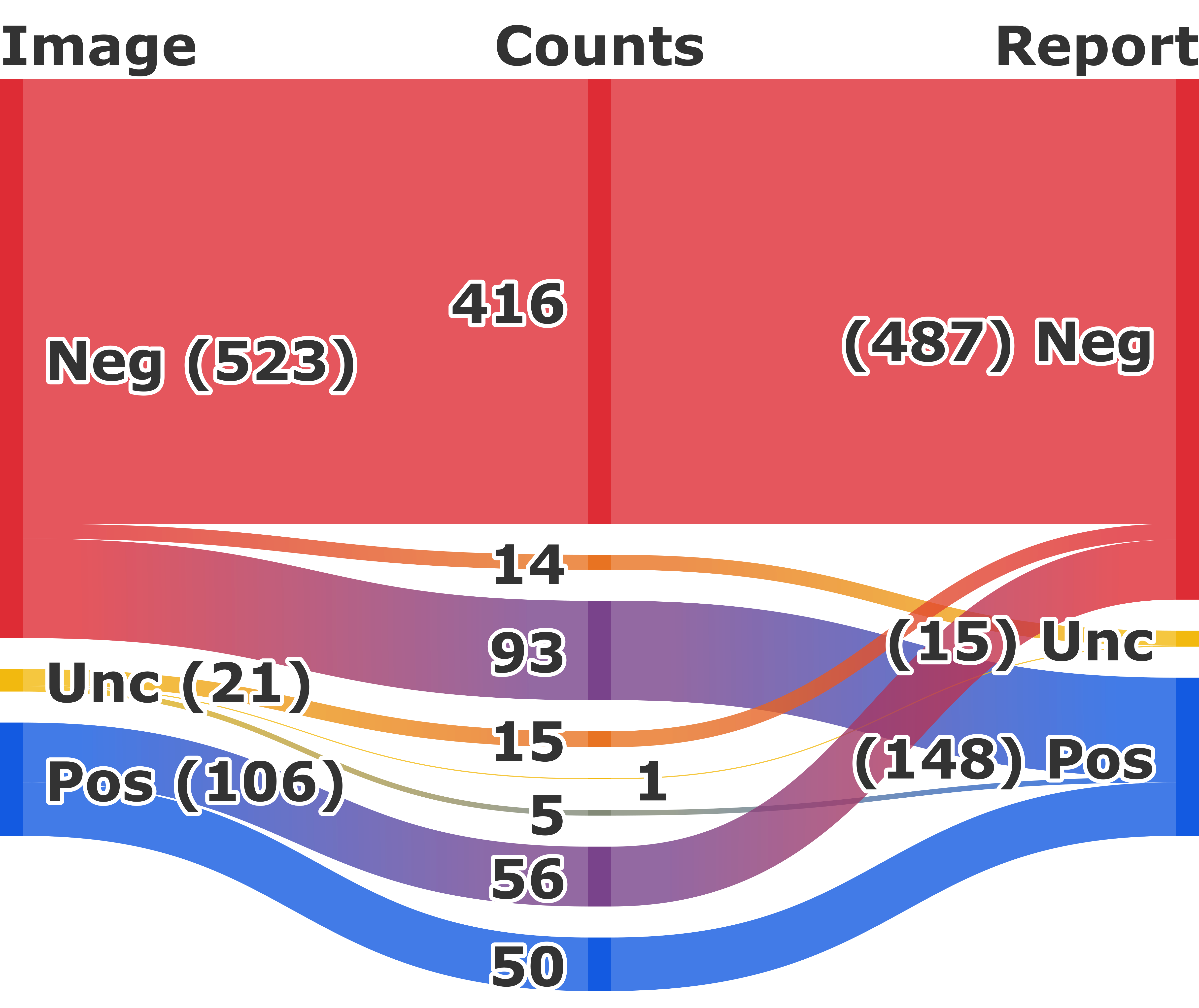}
        \subcaption{Consolidation}
        \label{fig:sankey_consolidation}
    \end{subfigure}
    \hfill
    \begin{subfigure}[b]{0.47\textwidth}
        \centering
        \includegraphics[width=\textwidth]{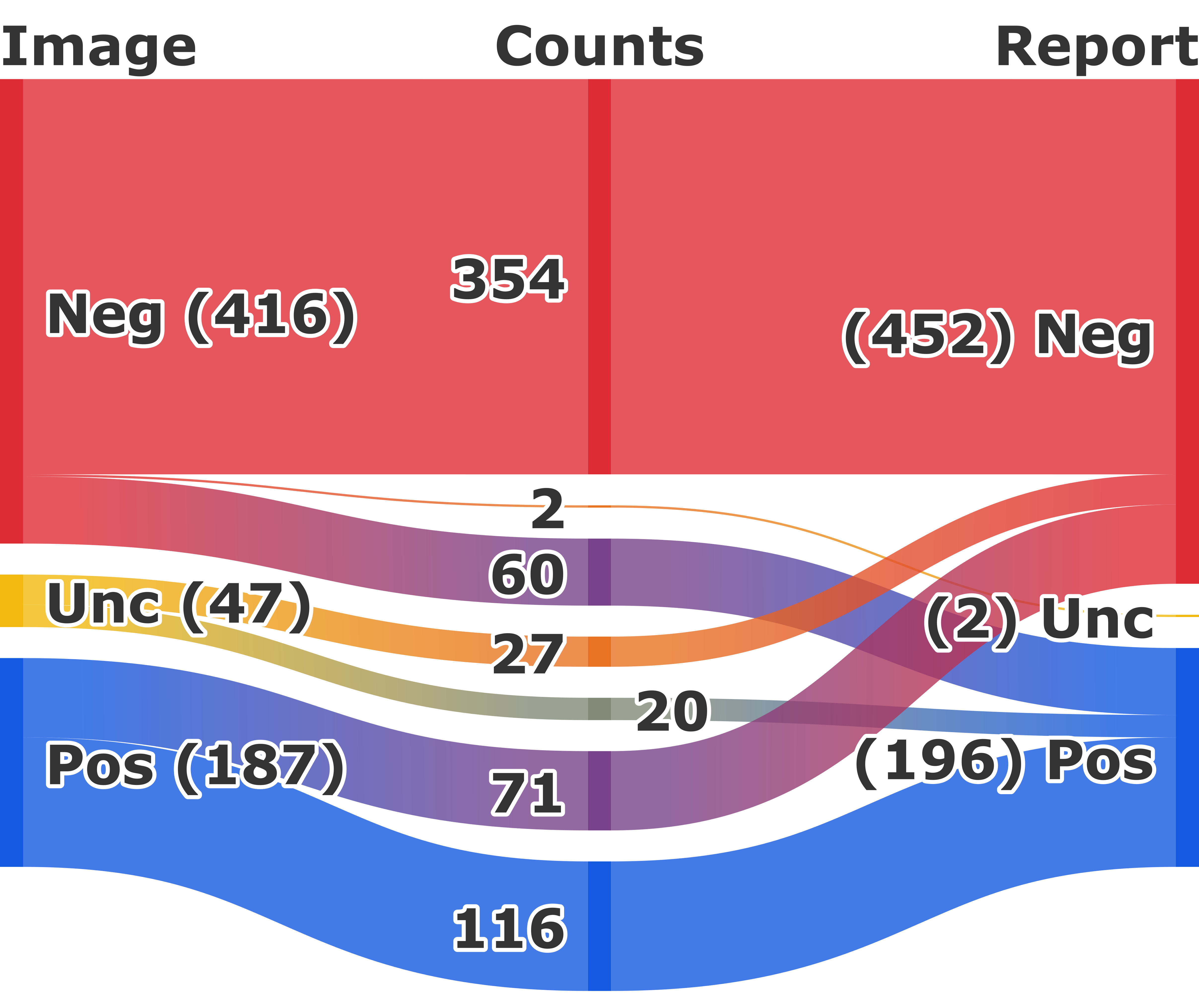}
        \subcaption{Lung opacity}
        \label{fig:sankey_luop}
    \end{subfigure}

    \vspace{0.5em}

    \begin{subfigure}[b]{0.47\textwidth}
        \centering
        \includegraphics[width=\textwidth]{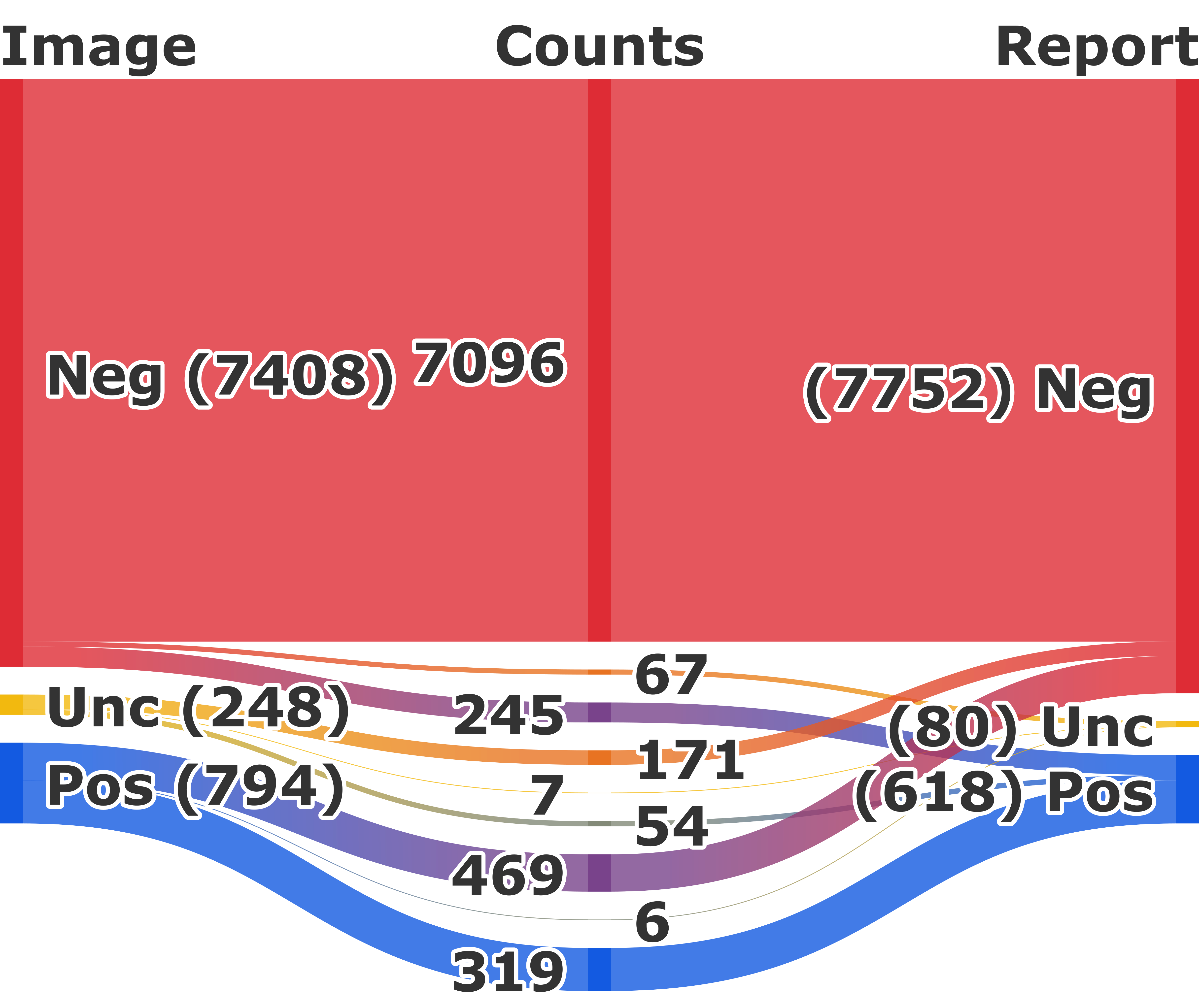}
        \subcaption{Overall across 12 pathologies.}
        \label{fig:sankey_overall}
    \end{subfigure}
    \caption{Sankey diagrams showing agreement and omission patterns between conflict-resolved (CR) image-derived labels and report-derived labels from annotator 1. Each flow represents the number of cases across the three label states (positive (Pos), uncertain (Unc), negative (Neg)), revealing the types and frequencies of agreement and disagreement between label sources. Results are shown for pathologies with more than 50 positive labels in either source (Figures~\ref{fig:sankey_plef}--\ref{fig:sankey_luop}), as well as across all 12 pathologies (Figure~\ref{fig:sankey_overall}).}
    \label{fig:sankey}
\end{figure}

Bidirectional omission patterns show substantial asymmetry between image-derived and report-derived references (Figure~\ref{fig:sankey}). These differences are strongly pathology-dependent. For Atelectasis (Figure~\ref{fig:sankey_atelectasis}), most positive conflict-resolved (CR) pathologies are not labeled as positive in the reports ($\frac{123}{165} = 75\%$). For Consolidation (Figure~\ref{fig:sankey_consolidation}), the opposite pattern is observed: most positive report-derived labels are not labeled as positive in the CR image-derived labels ($\frac{93}{148} = 63\%$). Across all 12 pathologies (Figure~\ref{fig:sankey_overall}), 469 of 794 positive CR image-derived labels (59\%) were labeled negative in the reports, whereas 245 of 618 positive report-derived labels (40\%) were labeled negative in the CR image-derived labels. CR image-derived labels also contain substantially more uncertain values than report-derived labels (248 vs.\ 80), corresponding to uncertain-to-positive ratios of 31\% and 13\%, respectively. These asymmetric patterns of omission and uncertainty indicate that the two label sources differ both in the frequency of positive labels and in the findings they capture.

Agreement is substantially lower for image annotators than for report annotators across the four selected pathologies (Figure~\ref{fig:cuh_agreement}; Tables~\ref{tab:cuh_images_ann1_vs_ann2} and~\ref{tab:cuh_reports_ann1_vs_ann2} in the Appendix). Among image annotators, Atelectasis has the lowest quadratically weighted $\kappa$ agreement ($\kappa=0.15$), indicating the greatest ambiguity, whereas Pleural effusion has the highest ($\kappa=0.67$). Image inter-annotator agreement is also sensitive to uncertainty handling, as shown by the wide low--high $\kappa$ intervals. By contrast, the report inter-annotator agreement is consistently high ($\kappa$: $0.80$--$0.94$) and shows narrower uncertainty intervals, indicating greater internal reproducibility.

\begin{figure}[!htbp]
    \centering
    \includegraphics[width=0.82\textwidth]{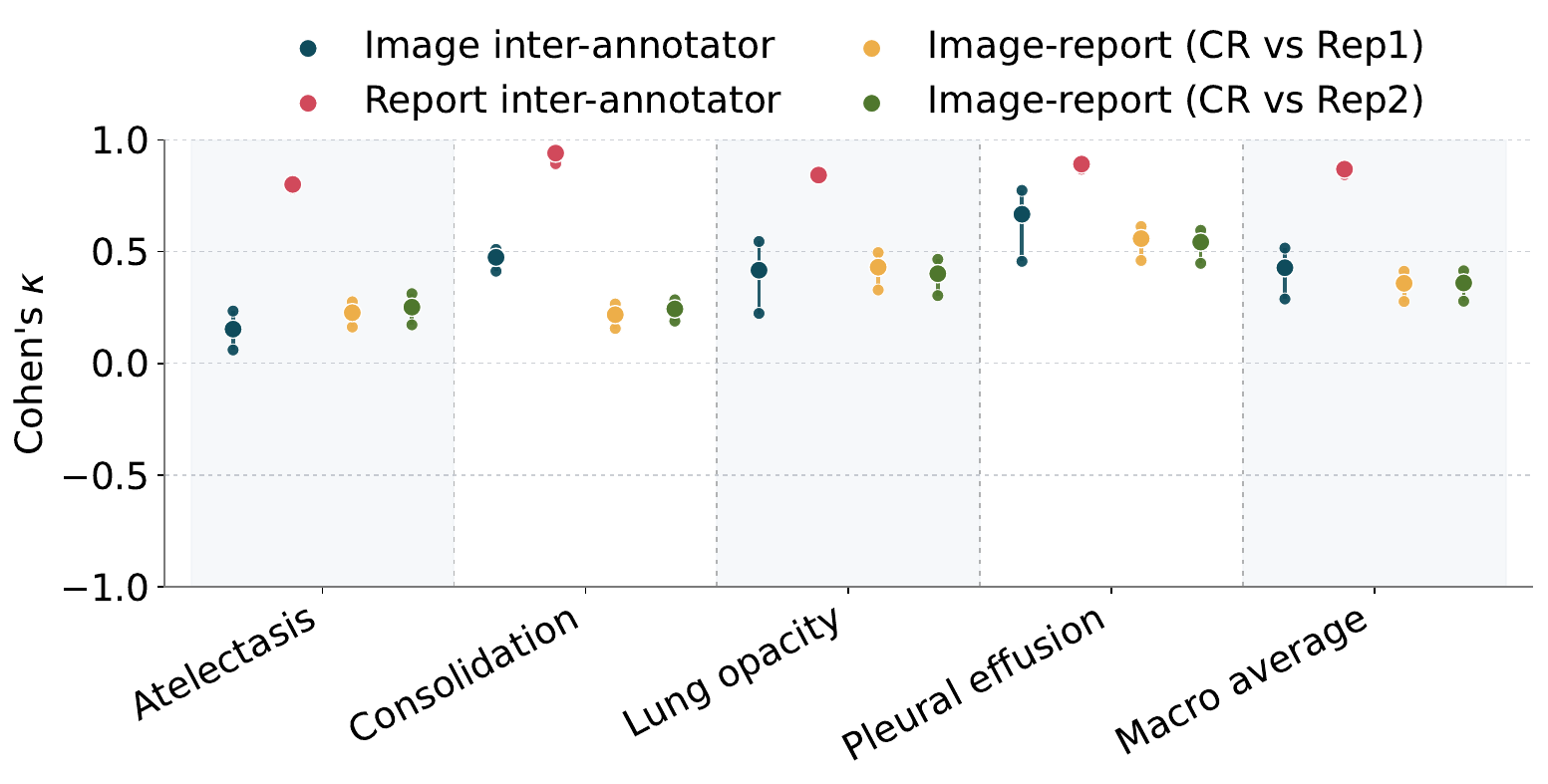}
    \caption{XQA-RR inter-annotator and cross-source label agreement summarized using $\kappa$-based agreement statistics: image inter-annotator, report inter-annotator, and cross-source conflict-resolved (CR) image-derived labels vs.\ report-derived labels from both annotators (Rep1/Rep2). Intervals indicate unweighted Cohen's $\kappa$ under pessimistic and optimistic uncertainty mappings, denoted Low--$\kappa$ and High--$\kappa$, respectively; central points indicate the quadratically weighted $\kappa$ computed using the original label values. Macro-average scores are calculated across the four selected pathologies. The corresponding weighted F1 results, which show the same qualitative pattern, are reported in Appendix Figure~\ref{fig:cuh_agreement_f1_app}.}
    \label{fig:cuh_agreement}
\end{figure}

Cross-source agreement between conflict-resolved (CR) image-derived labels and radiology report-derived labels is lower than report inter-annotator agreement and remains similar across the two report annotators (Table~\ref{tab:cuh_conflict_resolver_vs_report_ann1_ann2}). Consolidation shows the lowest cross-source agreement ($\kappa \approx 0.22$), despite having the highest report inter-annotator agreement ($\kappa=0.94$), whereas Pleural effusion shows the highest cross-source agreement ($\kappa \approx 0.55$).

\begin{figure}[!htbp]
    \centering
    \includegraphics[width=0.82\textwidth]{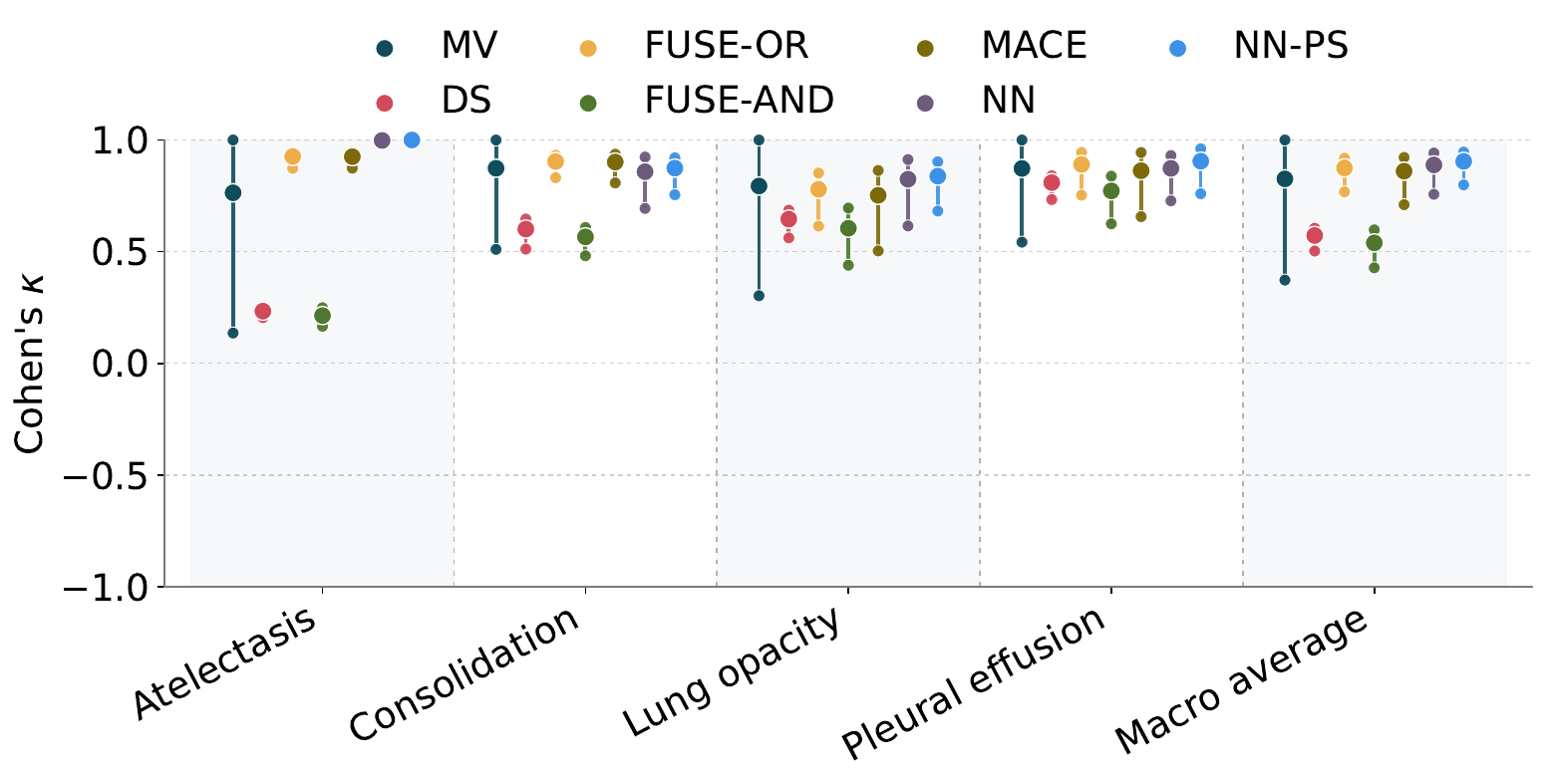}
    \caption{Agreement between each image-derived label fusion method and the XQA-RR conflict-resolved (CR) image-derived labels, summarized using $\kappa$-based agreement statistics. Intervals indicate unweighted Cohen's $\kappa$ under pessimistic and optimistic uncertainty mappings, denoted Low--$\kappa$ and High--$\kappa$, respectively; central points indicate the quadratically weighted $\kappa$ computed using the original label values. Macro-average scores are calculated across the four selected pathologies. The corresponding weighted F1 results, which follow the same qualitative trends, are reported in Appendix Figure~\ref{fig:cuh_fusion_agreement_f1_app}.}
    \label{fig:cuh_fusion_agreement_k}
\end{figure}

Figure~\ref{fig:cuh_fusion_agreement_k} reports agreement between each aggregation algorithm and the CR reference; the corresponding weighted F1 results and the detailed tables are shown in Appendix Figure~\ref{fig:cuh_fusion_agreement_f1_app} and Table~\ref{tab:cuh_image_algorithms_vs_conflict_resolver}. With two annotators, Majority Vote (MV), which maps disagreements to \emph{uncertain}, yields nominally high quadratically weighted $\kappa$ values (0.76--0.87) but extremely wide Low--$\kappa$ to High--$\kappa$ intervals, with $\kappa_{\text{low}}$ as low as 0.14 for Atelectasis. This reflects strong sensitivity to uncertainty handling, because all MV disagreements are assigned to the uncertain class. FUSE\_AND and DS-EM perform poorly on the harder pathologies (Atelectasis $\kappa=0.21$ and $0.23$, respectively; Consolidation $\kappa=0.57$ and $0.60$). FUSE\_OR, MACE, and both neural fusion variants achieve the highest quadratically weighted $\kappa$ agreement with CR alongside tighter uncertainty intervals: FUSE\_OR and MACE reach $\kappa\approx0.90$ for both Consolidation and Atelectasis. NN Fusion and NN Fusion (PS) further improve on Atelectasis (both with $\kappa=1.00$) and achieve a macro-average $\kappa$ of $0.89$--$0.90$ across the four selected pathologies---the highest of all methods.

\begin{figure}[!htbp]
    \centering
    \includegraphics[width=0.82\textwidth]{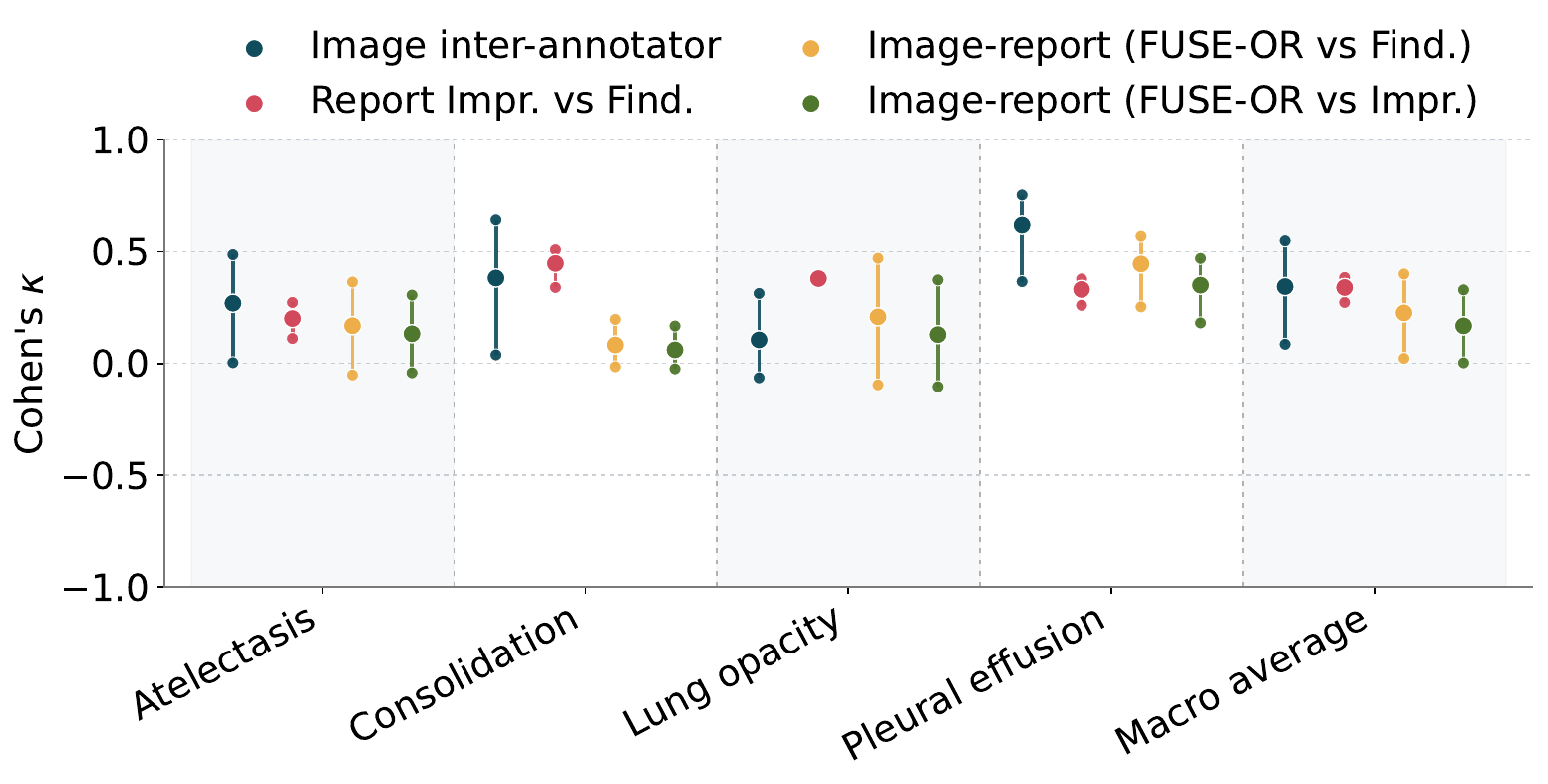}
\caption{MIMIC-CXR image inter-annotator agreement, agreement across labels derived from the Impression (Impr.) and Findings (Find.) sections, and cross-source agreement between fused image-derived labels (FUSE\_OR) and labels derived from the report Findings and Impressions sections, summarized using $\kappa$-based agreement statistics. Intervals indicate unweighted Cohen's $\kappa$ under pessimistic and optimistic uncertainty mappings, denoted Low--$\kappa$ and High--$\kappa$, respectively; central points indicate quadratically weighted $\kappa$ computed using the original three-class label values. Macro-average scores are calculated across the four selected pathologies. The corresponding weighted F1 results, which show the same qualitative pattern, are reported in Appendix Figure~\ref{fig:mimic_agreement_f1_app}.}
\label{fig:mimic_agreement}
\end{figure}

Figure~\ref{fig:mimic_agreement} summarizes agreement scores on MIMIC-CXR for image annotators, between labels derived from the Findings and Impressions sections of the same radiology report, and between FUSE\_OR image aggregation labels and the labels derived from each report section. We focus on FUSE\_OR because it showed strong agreement with the conflict-resolved XQA-RR reference while remaining a simple, non-black-box fusion rule. Tables~\ref{tab:mimic_images_ann1_vs_ann2}--\ref{tab:mimic_image_fusions_vs_impressions} in the Appendix provide detailed metrics for additional fusion methods. The image inter-annotator agreement pattern mirrors XQA-RR: quadratically weighted $\kappa$ agreement is low across the four selected pathologies, with Pleural effusion again the most consistently annotated finding ($\kappa=0.62$). However, the pathology-specific profile differs: Lung opacity shows substantially lower agreement on MIMIC-CXR ($\kappa=0.11$) than on XQA-RR ($\kappa=0.42$), while Atelectasis is slightly higher ($\kappa=0.27$ vs.\ $0.15$).

Labels derived from the Impressions and Findings sections, although originating from a single document written by the same radiologist, disagree substantially: macro-average $\kappa=0.31$, with Atelectasis as low as $\kappa=0.20$ and Pleural effusion at $\kappa=0.33$. This shows that report-derived labels can vary substantially depending on which section of the report is annotated.

For fused image-derived labels and each report section, agreement is low across both report sections. Against the Findings section labels, FUSE\_OR agrees with Consolidation with rates of $\kappa = 0.08$, Lung opacity $\kappa = 0.21$, and Atelectasis $\kappa = 0.17$; Pleural effusion is again the most stable ($\kappa = 0.45$). Agreement against labels derived from the Impressions section is even lower for all four pathologies (with corresponding $\kappa$ scores of $0.06$, $0.13$, $0.13$ and $0.35$), reflecting  additional divergence introduced by the Impressions section.

\subsubsection{Qualitative analysis of XQA-RR disagreements.}

The adjudicating radiologist of XQA-RR noted that many disagreements stemmed from the difficulty of interpreting imaging signs without access to clinical history. For example, Lung opacity is a broad descriptor of increased density, whereas Consolidation refers more specifically to airspace filling processes that may arise from infection, pulmonary edema, or pulmonary hemorrhage. In routine clinical reporting, radiologists often use contextual information to distinguish between these possibilities. In the absence of such context, annotators may differ in whether they assign a broad label (e.g.\ Lung opacity) or a more specific one (e.g.\ Consolidation or Edema). Other disagreements reflected a distinction between describing visible imaging signs and inferring higher-level diagnostic impressions.

\subsubsection{Model performance across evaluation references.}

ROC-AUC results for all models across the four selected pathologies and label sources are reported in Tables~\ref{tab:cuh_atelectasis_combined_auc}--\ref{tab:mimic_s_overall_1_combined_auc} in the Appendix. Some near-random performance is expected from the training label sets of the corresponding models: D-RSNA performs non-randomly only for Lung opacity, consistent with its training on the RSNA Pneumonia Detection Challenge, whereas D-NIH, D-PC, and JFH produce non-informative predictions for Lung opacity because this finding is not included in their training label sets (Table~\ref{tab:model_overview}). Several zero-shot vision-language models also yield AUC values below 0.500 for Atelectasis and Consolidation, indicating that their similarity scores are inversely associated with the positive label for these pathologies.

The choice of label source affects apparent absolute ROC-AUC values. On XQA-RR, models achieve a mean AUC of $0.77$ against CR image-derived labels for Lung opacity, compared with $0.73$--$0.74$ against report-derived labels (Table~\ref{tab:cuh_lung_opacity_combined_auc}). A similar directional pattern holds for Consolidation ($0.77$ vs.\ $0.71$--$0.71$; Table~\ref{tab:cuh_consolidation_combined_auc}). These differences indicate that evaluation-reference choice can change the apparent absolute performance of models, motivating the ranking-stability analysis below.

\subsubsection{Model ranking stability across evaluation references.}

\begin{figure}[tb]
    \centering
    \includegraphics[width=0.84\textwidth]{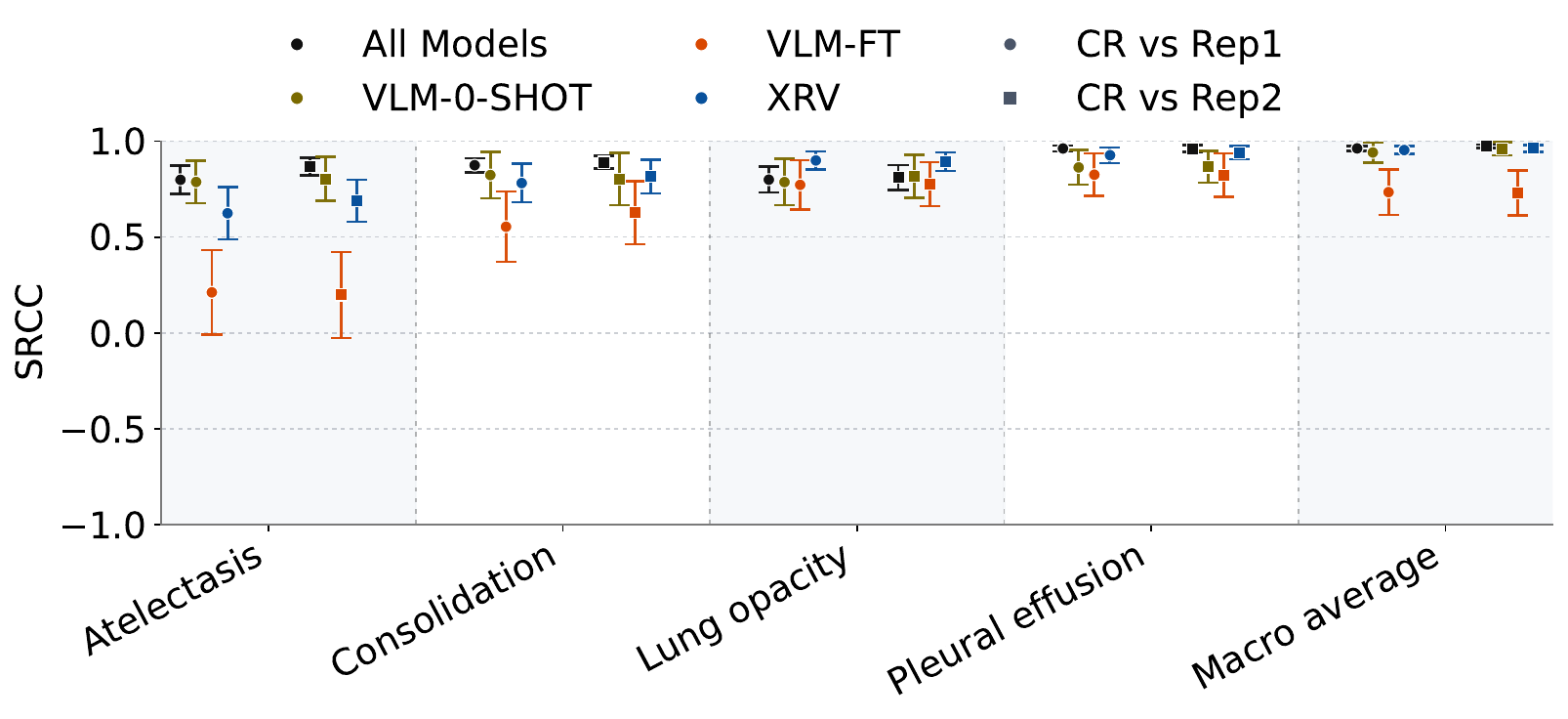}
    \caption{XQA-RR rank correlation coefficients between model rankings induced by each pair of label sources, measured by Spearman rank correlation coefficient (SRCC). CR denotes conflict-resolved image-derived labels; Rep1/Rep2 denote report-derived labels from the two report annotators. Results are broken down by pathology and model group. Macro average refers to rankings induced by ROC-AUC macro-averaged across the four selected pathologies. The corresponding Kendall rank correlation coefficients (KRCC), which show the same qualitative pattern, are reported in Appendix Figure~\ref{fig:cuh_rcc_krcc_app}.}
    \label{fig:cuh_rcc}
\end{figure}

\begin{figure}[tb]
    \centering
    \includegraphics[width=0.84\textwidth]{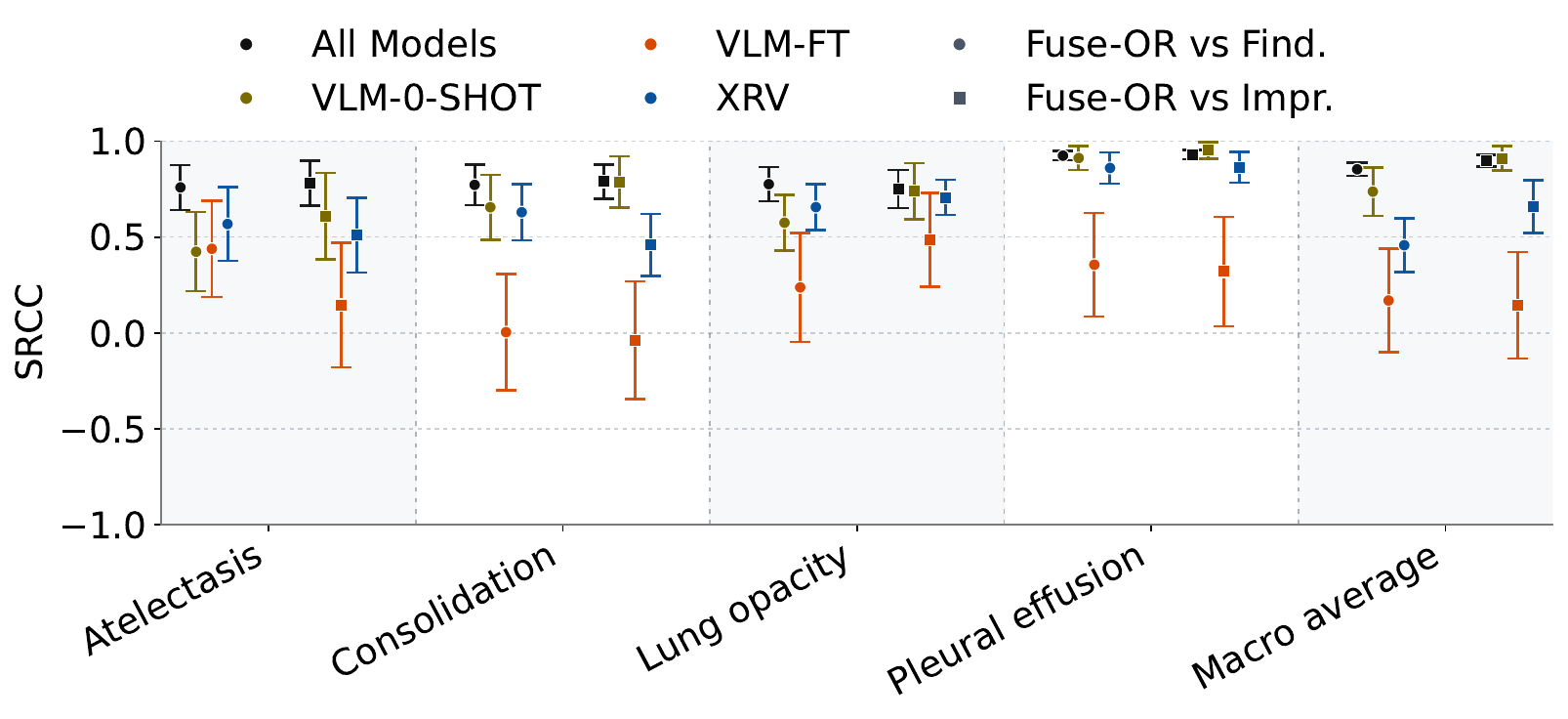}
    \caption{MIMIC-CXR rank correlation coefficients between model rankings induced by each pair of label sources, measured by Spearman rank correlation coefficient (SRCC). FUSE\_OR denotes automatically fused image-derived labels; Find.\ and Impr.\ denote labels derived from the Findings and Impressions sections of the radiology report. Results are broken down by pathology and model group. Macro average refers to rankings induced by ROC-AUC macro-averaged across the four selected pathologies. The corresponding Kendall rank correlation coefficients (KRCC), which show the same qualitative pattern, are reported in Appendix Figure~\ref{fig:mimic_rcc_krcc_app}.}
    \label{fig:mimic_rcc}
\end{figure}

To determine whether label source affects not only absolute ROC-AUC but also the relative ordering of models, Figures~\ref{fig:cuh_rcc} and~\ref{fig:mimic_rcc} report Spearman rank correlation coefficients (SRCC) between model rankings induced by each label source for XQA-RR and MIMIC-CXR, respectively. The corresponding Kendall rank correlation coefficients (KRCC), which show the same qualitative trends, are reported in the Appendix (Figures~\ref{fig:cuh_rcc_krcc_app} and~\ref{fig:mimic_rcc_krcc_app}). Full numerical results are in Tables~\ref{tab:cuh_all_pairs_spearman_all_groups}--\ref{tab:mimic_all_pairs_kendalltau_all_groups} in the Appendix. Values close to 1 indicate that two label sources induce similar model rankings, whereas lower values indicate greater sensitivity of model ordering to the evaluation reference.

Across the selected pathologies, SRCC values broadly track agreement between label sources, but the strength of this relationship varies by model family (Figures~\ref{fig:cuh_rcc} and~\ref{fig:mimic_rcc}; Tables~\ref{tab:cuh_all_pairs_spearman_all_groups}--\ref{tab:mimic_all_pairs_kendalltau_all_groups}). In the All-models comparison, Pleural effusion, which has the highest cross-source label agreement, also shows the highest rank stability (SRCC $\approx 0.96$). Atelectasis and Consolidation, which have lower cross-source label agreement, show lower overall SRCC values (Atelectasis $\approx 0.80$--$0.87$, Consolidation $\approx 0.88$--$0.89$). This instability is particularly important for fine-tuned VLMs (VLM-FT), which include several of the highest-performing models in absolute ROC-AUC terms: for Atelectasis, VLM-FT rankings induced by CR image-derived labels and report-derived labels are close to uncorrelated (Atelectasis SRCC $\approx 0.21 \pm 0.22$). Instability within this family, therefore, directly affects which models appear best under each evaluation reference. XRV models are generally more stable across label sources than VLM-FT models, especially for pathologies with higher cross-source agreement. The corresponding KRCC results follow the same qualitative pattern and are reported in the Appendix.

Lung opacity presents a notable departure from the agreement--rank-stability pattern. Despite having moderate quadratically weighted $\kappa$ cross-source label agreement on XQA-RR ($\kappa = 0.42$), its All-models cross-source SRCC ($0.80$--$0.81$) is comparable to that of Atelectasis ($0.80$--$0.87$), which has lower cross-source agreement. This indicates that cross-source label agreement alone does not fully explain ranking stability. One possible explanation is that models achieve similar AUCs for Lung opacity (Table~\ref{tab:cuh_lung_opacity_combined_auc}), so small absolute performance differences translate into larger rank changes.

\begin{figure}[t]
\centering

\begin{subfigure}[b]{0.4\textwidth}
    \centering
    \includegraphics[width=\textwidth]{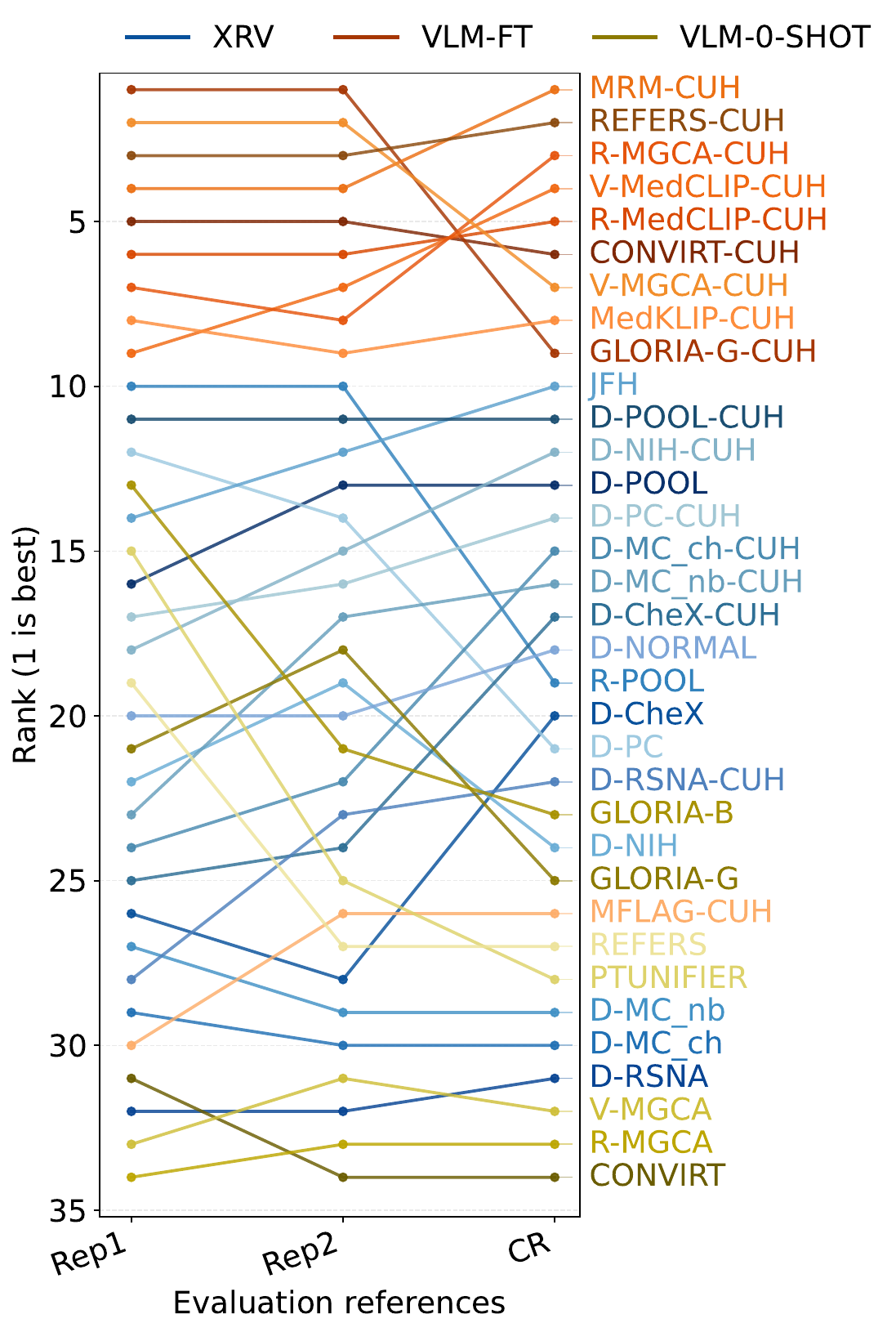}
    \caption{XQA-RR}
    \label{fig:cuh_rankings_atelectasis}
\end{subfigure}
\hfill
\begin{subfigure}[b]{0.4\textwidth}
    \centering
    \includegraphics[width=\textwidth]{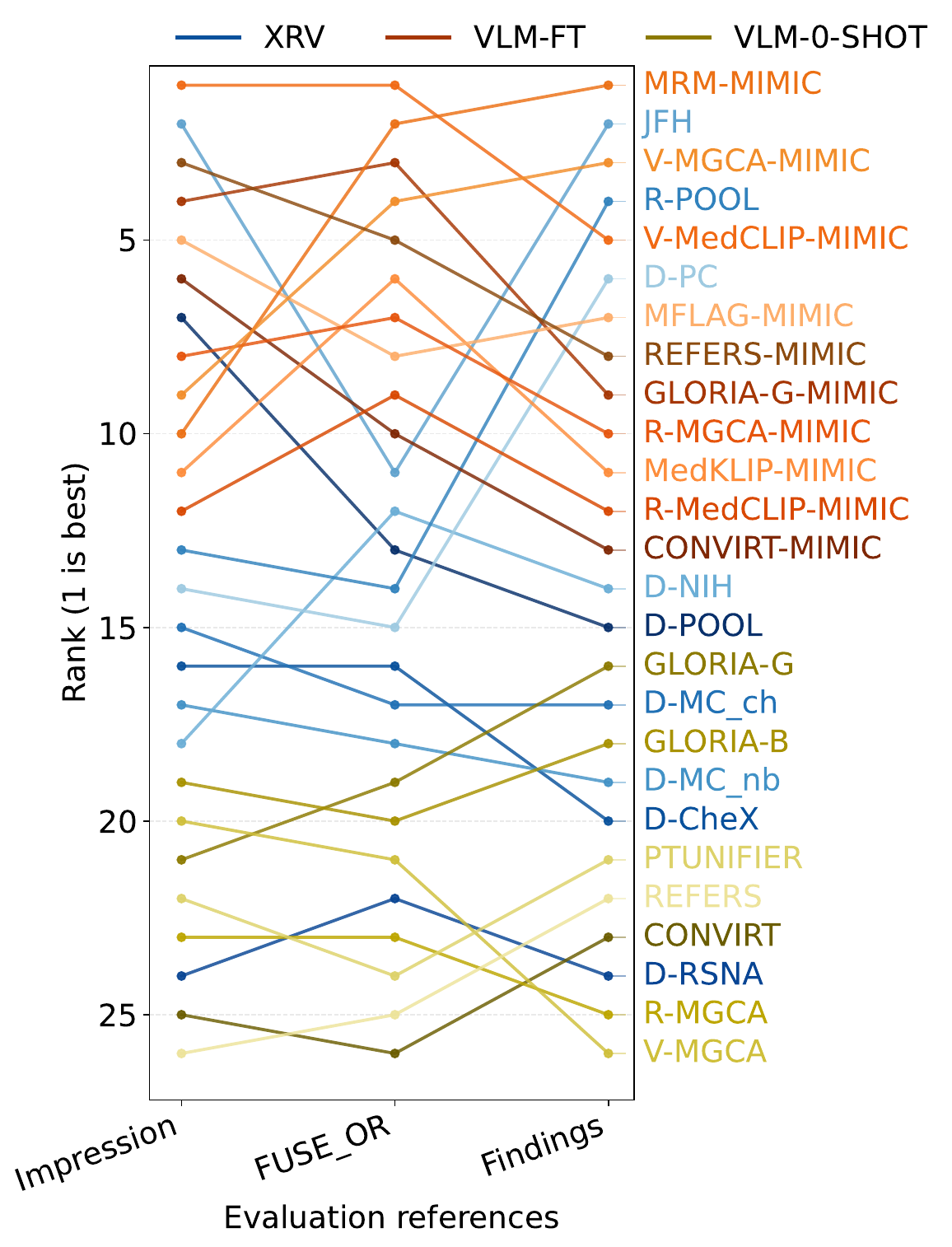}
    \caption{MIMIC-CXR}
    \label{fig:mimic_rankings_atelectasis}
\end{subfigure}
\caption{Model rankings across label sources for \textbf{Atelectasis}.}
\label{fig:rankings_atelectasis}

\end{figure}

\begin{figure}[t]
\centering

\begin{subfigure}[b]{0.4\textwidth}
    \centering
    \includegraphics[width=\textwidth]{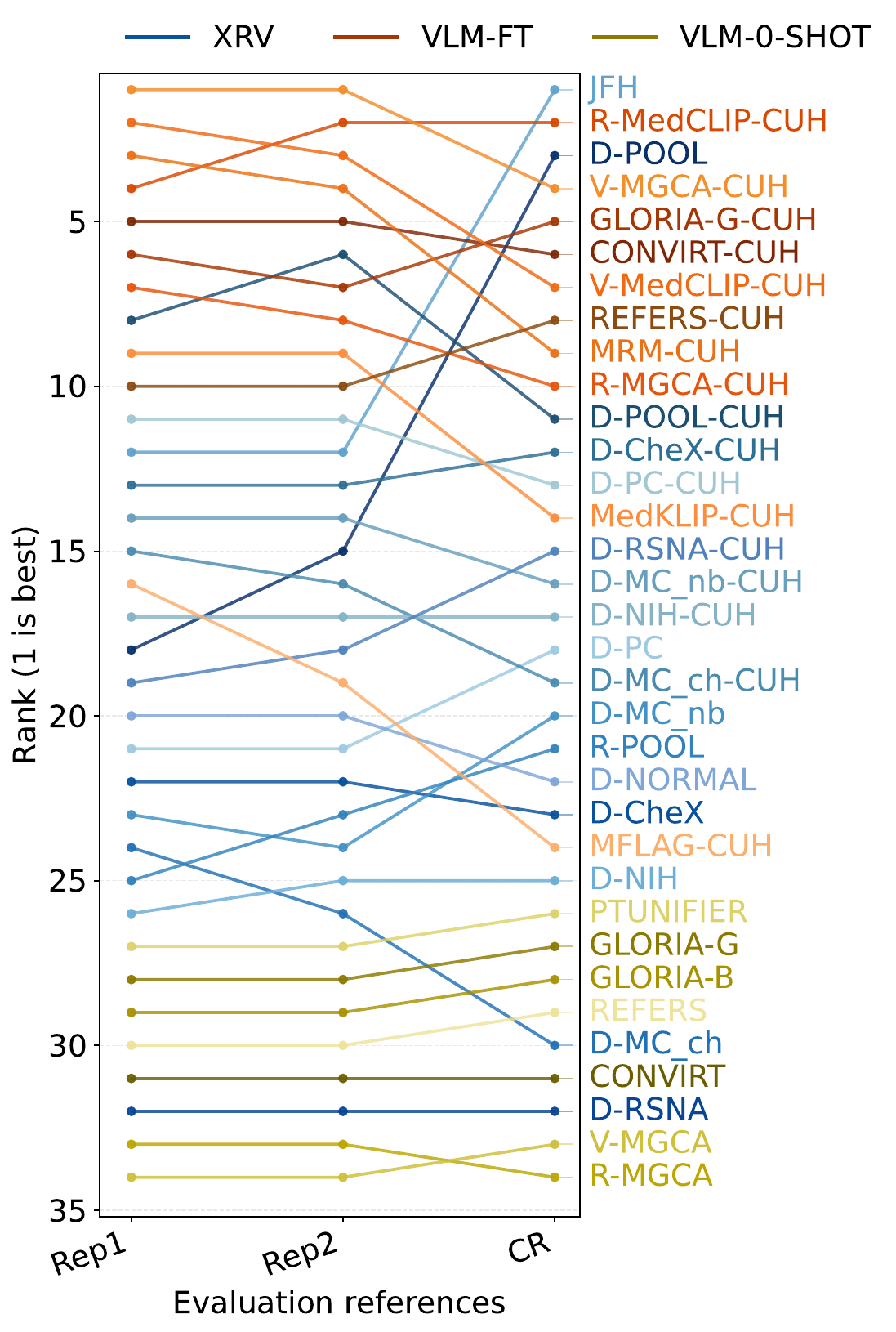}
    \caption{XQA-RR}
    \label{fig:cuh_rankings_consolidation}
\end{subfigure}
\hfill
\begin{subfigure}[b]{0.4\textwidth}
    \centering
    \includegraphics[width=\textwidth]{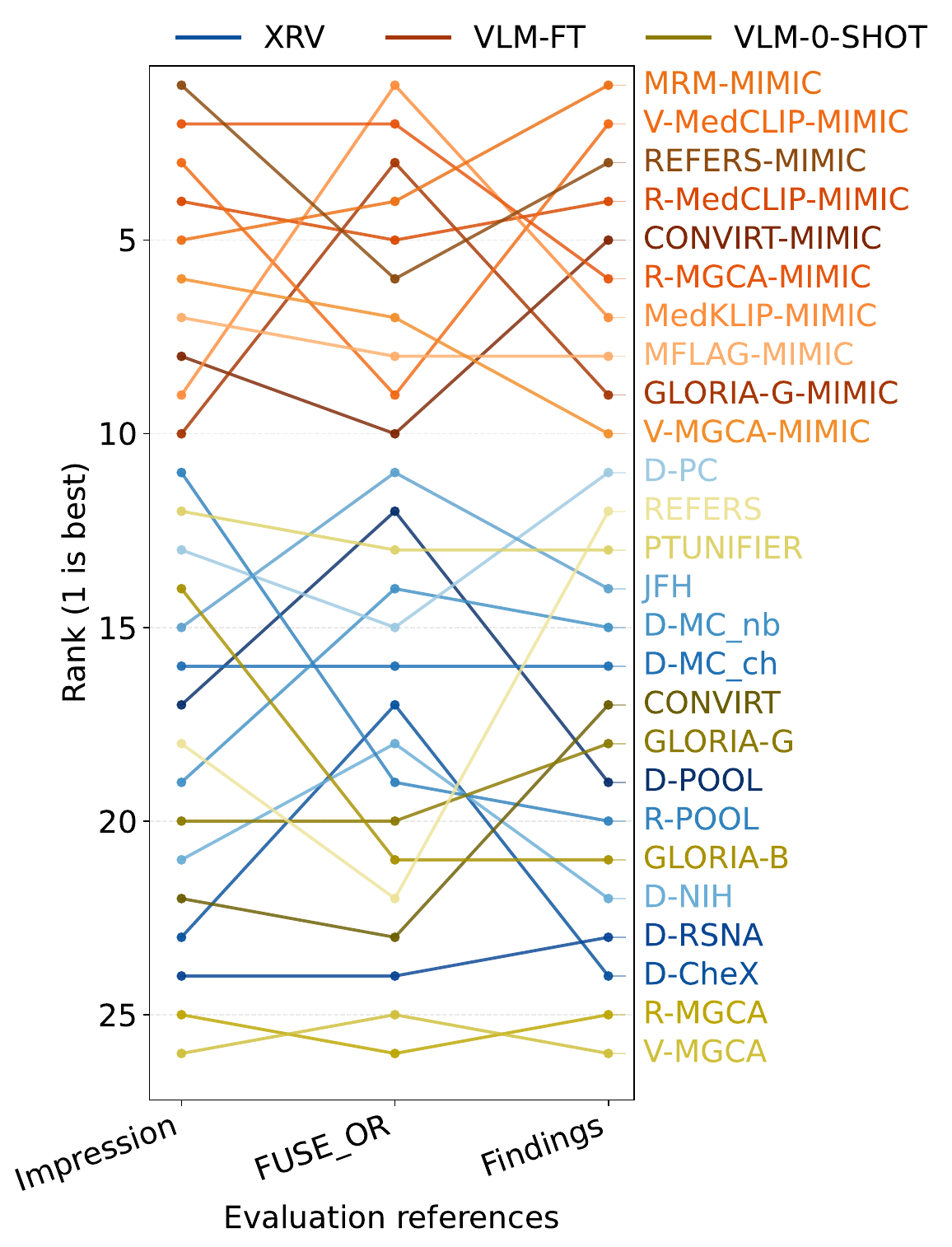}
    \caption{MIMIC-CXR}
    \label{fig:mimic_rankings_consolidation}
\end{subfigure}
\caption{Model rankings across label sources for \textbf{Consolidation}.}
\label{fig:rankings_consolidation}

\end{figure}

\begin{figure}[t]
\centering

\begin{subfigure}[b]{0.4\textwidth}
    \centering
    \includegraphics[width=\textwidth]{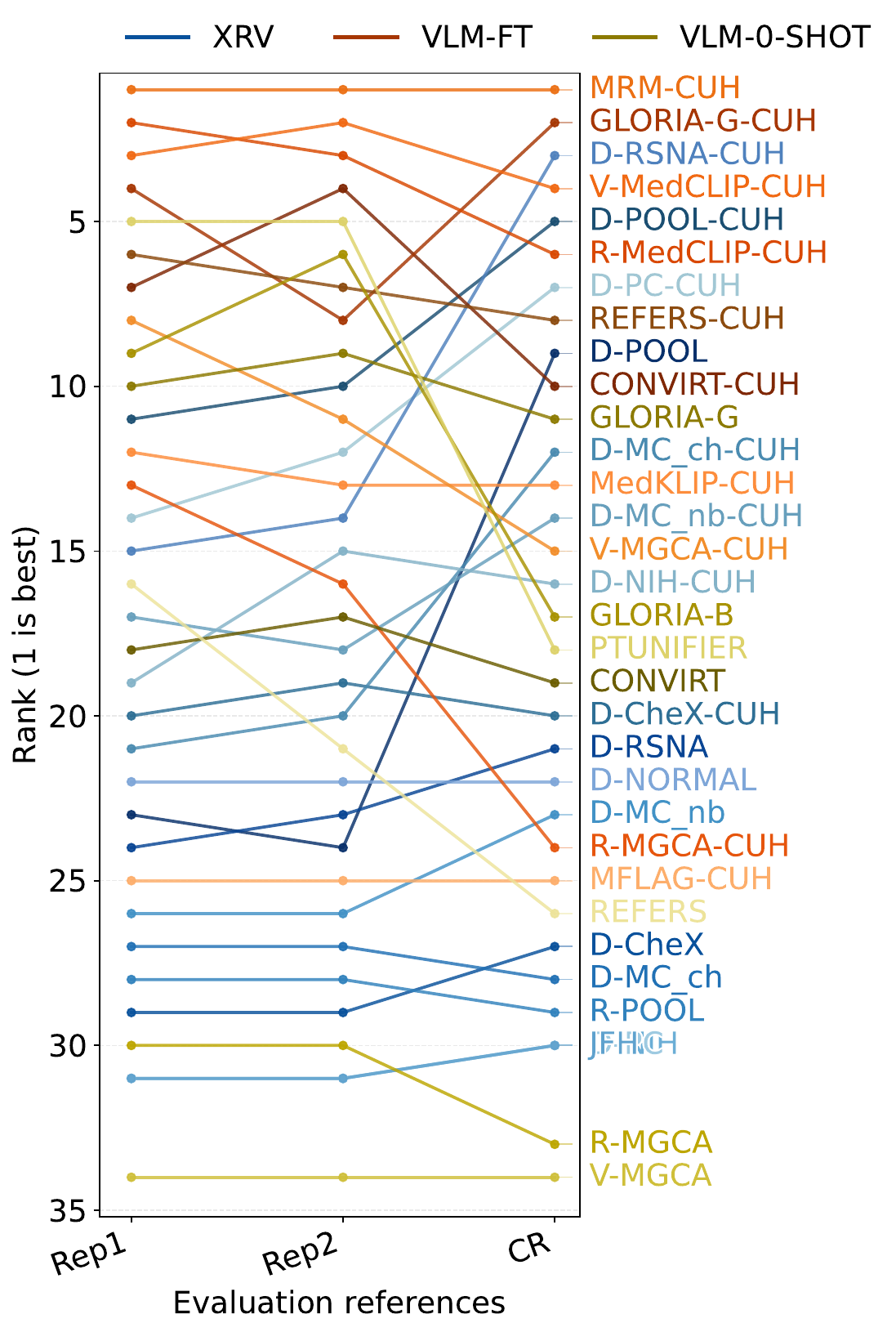}
    \caption{XQA-RR}
    \label{fig:cuh_rankings_lung_opacity}
\end{subfigure}
\hfill
\begin{subfigure}[b]{0.4\textwidth}
    \centering
    \includegraphics[width=\textwidth]{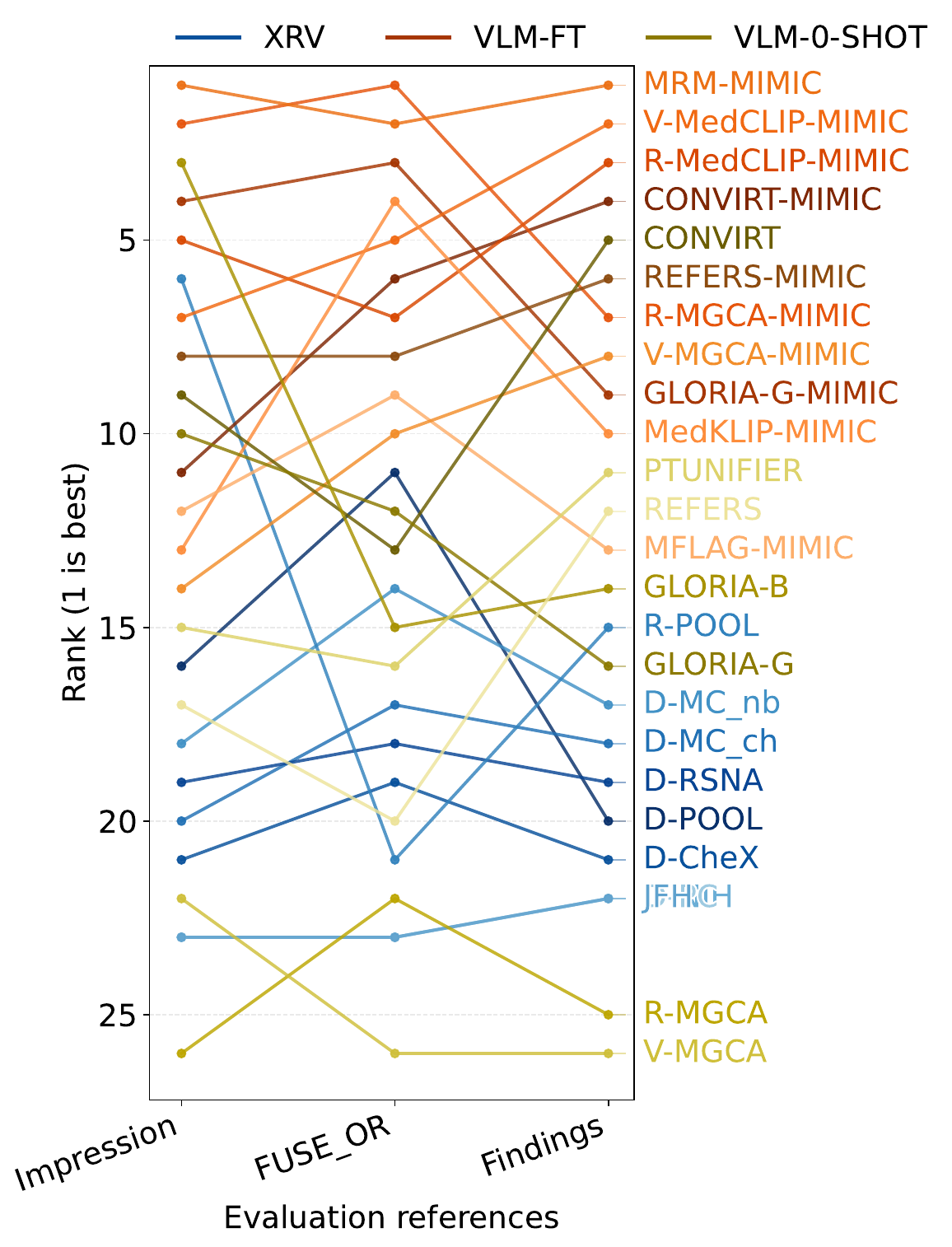}
    \caption{MIMIC-CXR}
    \label{fig:mimic_rankings_lung_opacity}
\end{subfigure}
\caption{Model rankings across label sources for \textbf{Lung opacity}.}
\label{fig:rankings_lung_opacity}

\end{figure}

\begin{figure}[t]
\centering

\begin{subfigure}[b]{0.4\textwidth}
    \centering
    \includegraphics[width=\textwidth]{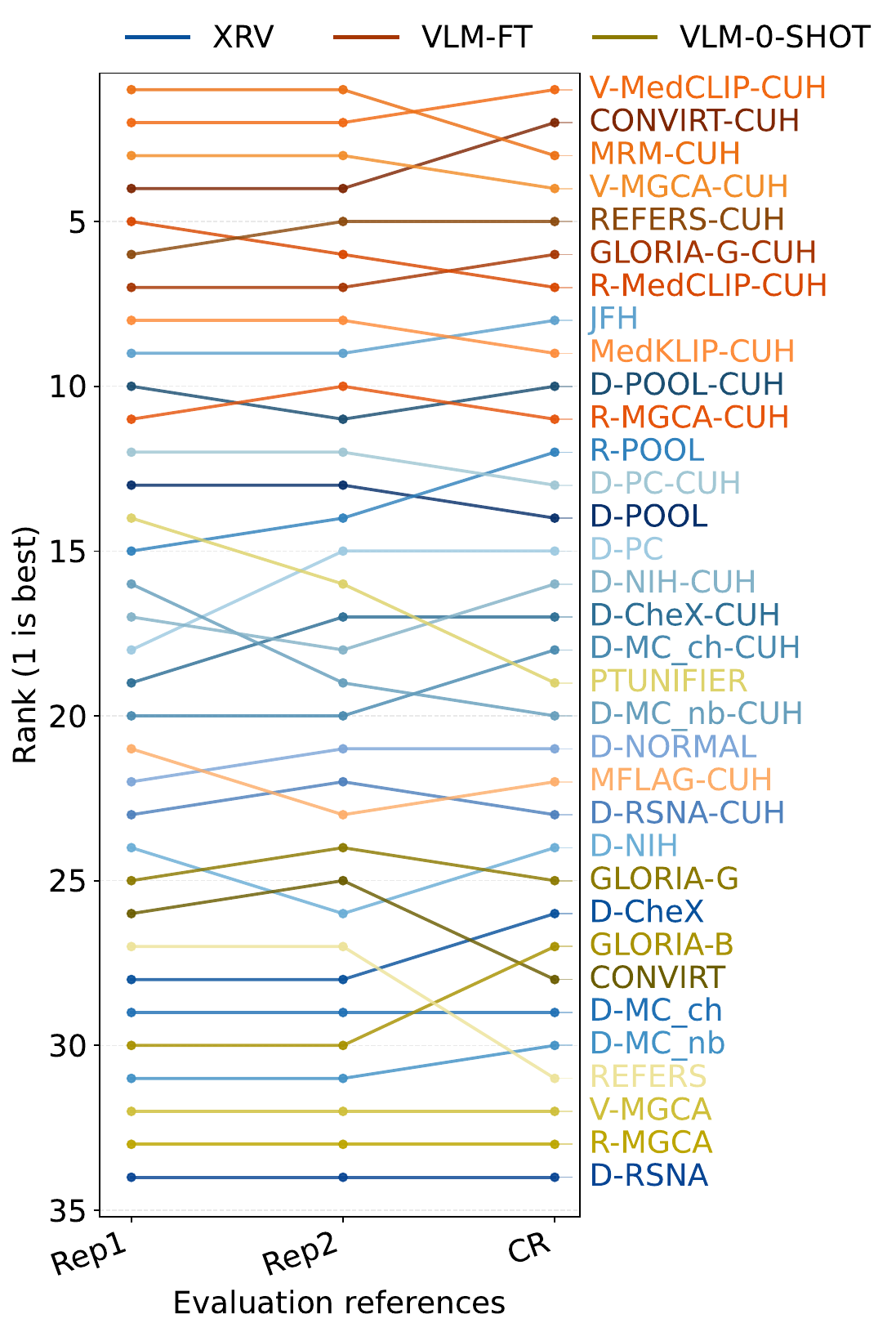}
    \caption{XQA-RR}
    \label{fig:cuh_rankings_pleural_effusion}
\end{subfigure}
\hfill
\begin{subfigure}[b]{0.4\textwidth}
    \centering
    \includegraphics[width=\textwidth]{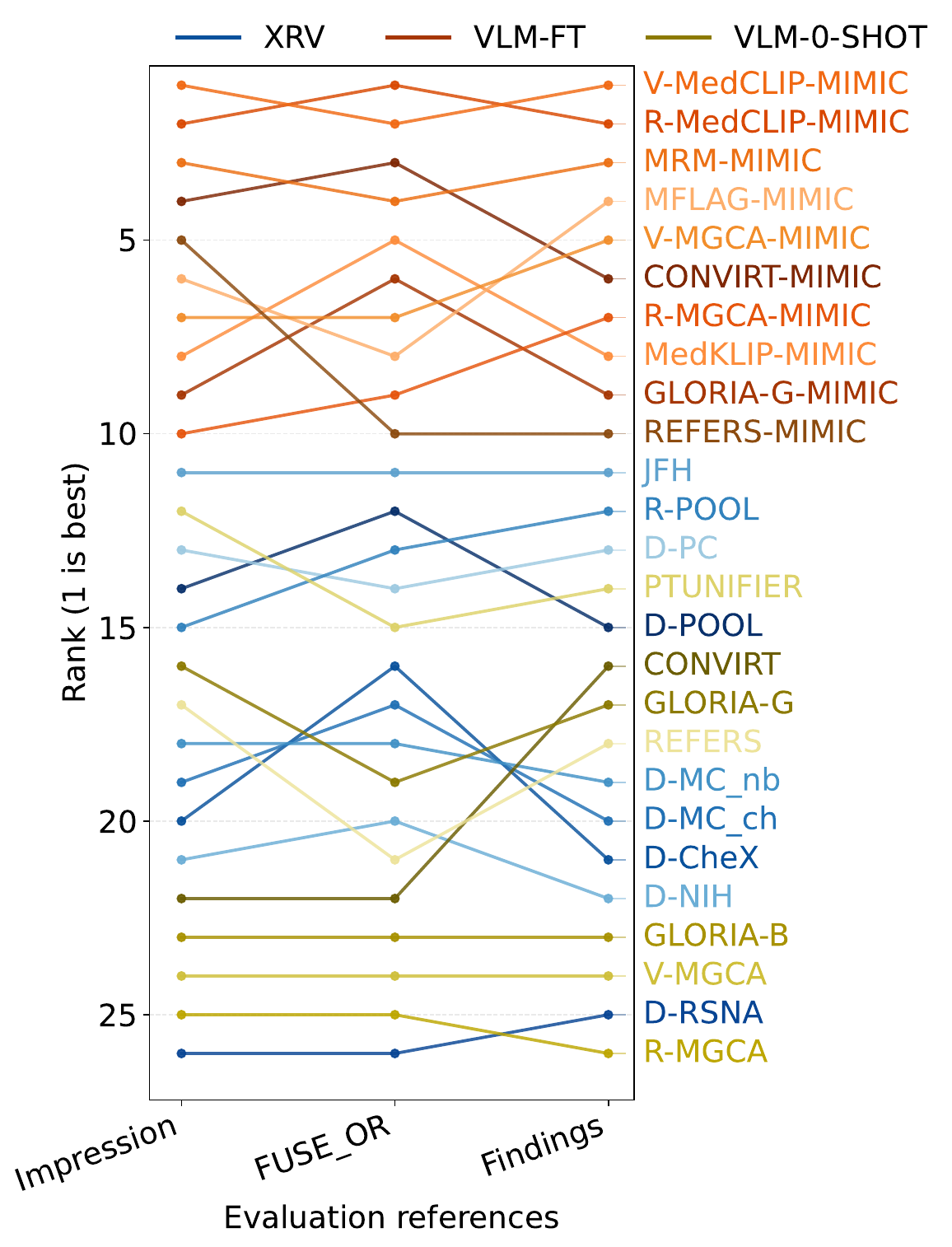}
    \caption{MIMIC-CXR}
    \label{fig:mimic_rankings_pleural_effusion}
\end{subfigure}
\caption{Model rankings across label sources for \textbf{Pleural effusion}.}
\label{fig:rankings_pleural_effusion}

\end{figure}

\begin{figure}[t]
\centering

\begin{subfigure}[b]{0.4\textwidth}
    \centering
    \includegraphics[width=\textwidth]{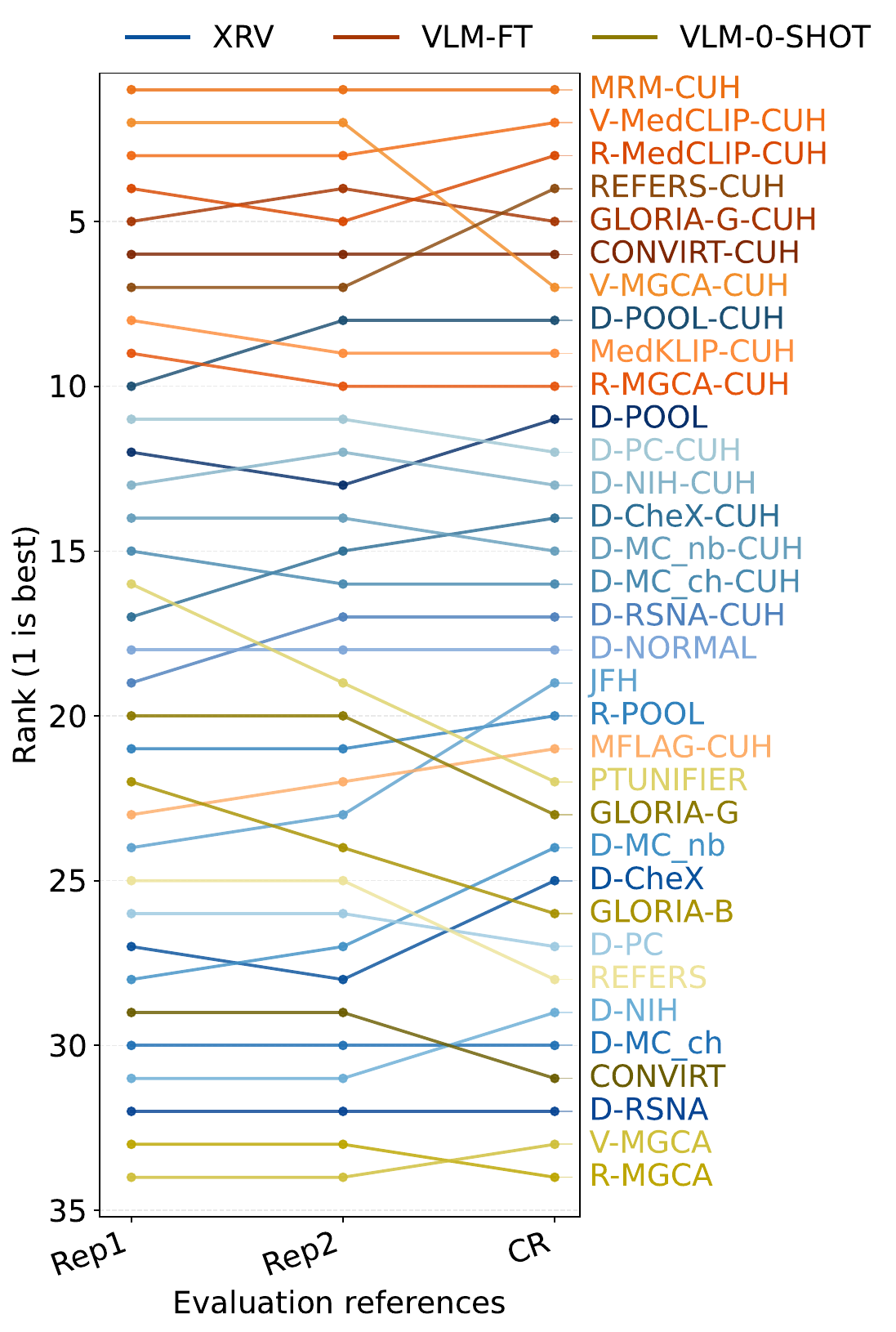}
    \caption{XQA-RR}
    \label{fig:cuh_rankings_global}
\end{subfigure}
\hfill
\begin{subfigure}[b]{0.4\textwidth}
    \centering
    \includegraphics[width=\textwidth]{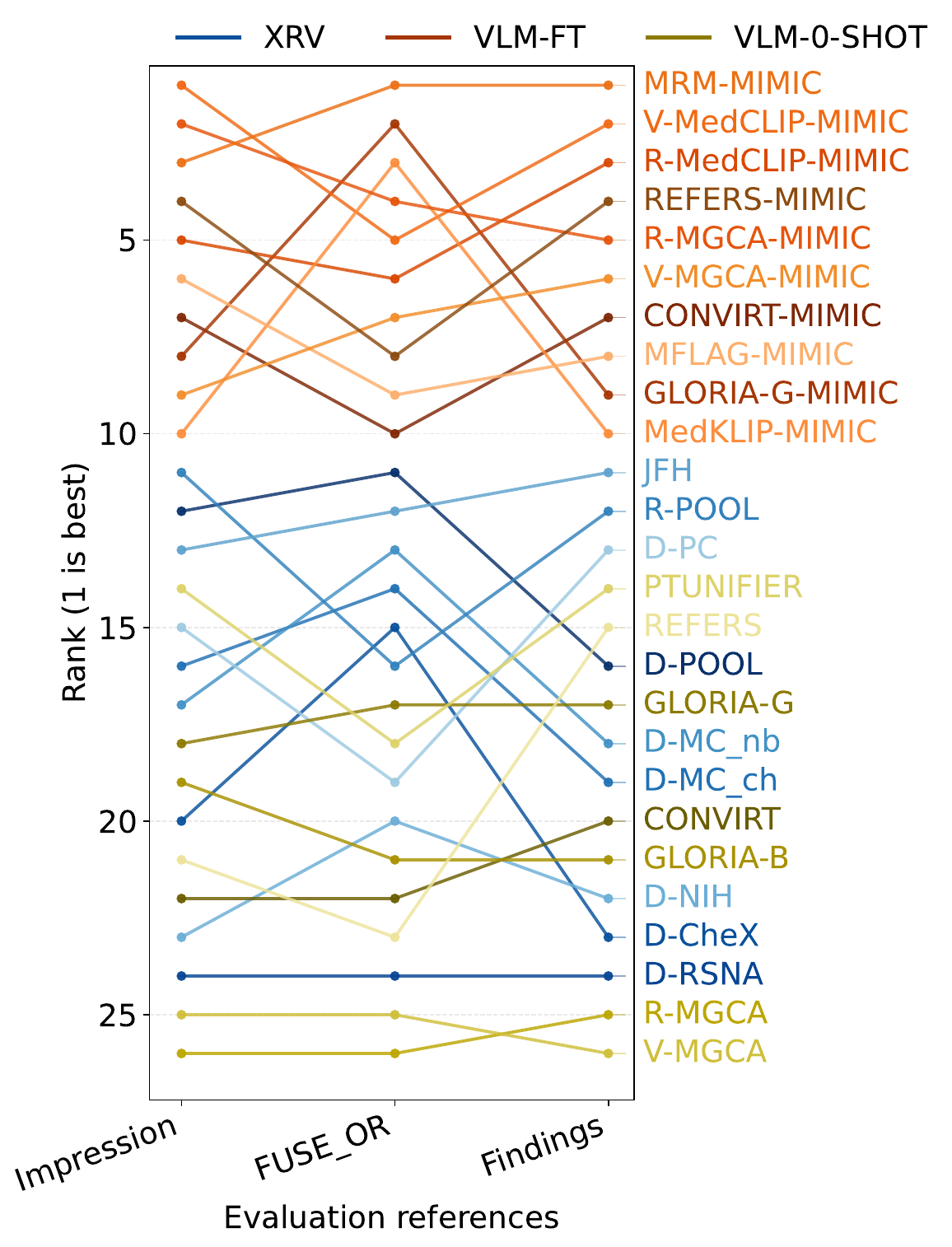}
    \caption{MIMIC-CXR}
    \label{fig:mimic_rankings_global}
\end{subfigure}
\caption{Model rankings across label sources, macro-averaged across the four selected pathologies.}
\label{fig:rankings_global}

\end{figure}

Aggregate rank correlations can obscure practically important instability at the top of the ranking, where model selection decisions are made. Figures~\ref{fig:cuh_rankings_atelectasis}--\ref{fig:cuh_rankings_global} visualize XQA-RR model rankings across label sources for each pathology and for the macro-average across the four selected pathologies. Figures~\ref{fig:mimic_rankings_atelectasis}--\ref{fig:mimic_rankings_global} show the equivalent rankings for MIMIC-CXR.

On XQA-RR, the choice of evaluation reference often changes which model appears best. For Atelectasis (Figure~\ref{fig:cuh_rankings_atelectasis}), GLORIA-G-CUH ranks first under both report-derived labels, yet falls to 9\textsuperscript{th} when evaluated against CR image-derived labels. Conversely, MRM-CUH ranks first under CR image-derived labels but falls to 4\textsuperscript{th} when evaluated against both report-derived references. For Consolidation (Figure~\ref{fig:cuh_rankings_consolidation}), the model ranked first under CR image-derived labels is ranked 13\textsuperscript{th} under the report-derived labels. For Lung opacity (Figure~\ref{fig:cuh_rankings_lung_opacity}), MRM-CUH remains the top-ranked model across all evaluation references, although substantial reordering is still observed among the remaining high-performing models. Pleural effusion shows the greatest top-rank stability, consistent with its higher cross-source agreement (Figure~\ref{fig:cuh_rankings_pleural_effusion}). In the macro-average across the four selected pathologies, MRM-CUH remains consistently top-ranked, but instability persists among near-best alternatives (Figure~\ref{fig:cuh_rankings_global}). Thus, reference choice affects not only the overall ordering of models, but also which models would be selected as best-in-class or near-best alternatives.

A similar pattern is observed on MIMIC-CXR (Figures~\ref{fig:mimic_rankings_atelectasis}--\ref{fig:mimic_rankings_global}). Changing the label source (FUSE\_OR vs.\ Findings vs.\ Impressions) produces larger changes for Atelectasis and Consolidation than for Pleural effusion.

On XQA-RR, the rankings induced by the two report annotators are highly consistent across the four selected pathologies, indicating that evaluation against report-derived labels is internally reproducible across annotators even when it diverges from evaluation against CR image-derived labels. On MIMIC-CXR, however, rankings derived from the Findings and Impressions section labels of the same report, as annotated by the same radiologist, are substantially less consistent with each other (Figures~\ref{fig:mimic_rankings_atelectasis}--\ref{fig:mimic_rankings_global}; Figures~\ref{fig:mimic_impr_atelectasis}--\ref{fig:mimic_impr_global} in the Appendix). Thus, the report section used for label derivation is itself a source of evaluation variability and can affect conclusions about which models perform best.

\subsection{IQA}

Table~\ref{interrater} reports inter-rater agreement between each annotator's ranking of degraded images and the z-MOS, measured by SRCC and KRCC. Most annotators show high agreement with the z-MOS ranking, with SRCC values ranging from $0.83$ to $0.88$. Ann.~4 is the sole exception, achieving a notably lower SRCC of $0.65$.

\begin{table}[!htbp]
\centering
\resizebox{\linewidth}{!}{%
\begin{tabular}{lllllll}
\toprule
 & Ann.~1 & Ann.~2 & Ann.~3 & Ann.~4 & Ann.~5 & z-MOS \\
\midrule
Ann.~1          & ---    & 0.69 / 0.60 & 0.65 / 0.57 & 0.48 / 0.43 & 0.74 / 0.67 & 0.88 / 0.76 \\
Ann.~2          & ---    & ---         & 0.72 / 0.65 & 0.34 / 0.28 & 0.70 / 0.61 & 0.83 / 0.69 \\
Ann.~3          & ---    & ---         & ---         & 0.46 / 0.40 & 0.69 / 0.60 & 0.86 / 0.73 \\
Ann.~4          & ---    & ---         & ---         & ---         & 0.51 / 0.46 & 0.65 / 0.54 \\
Ann.~5          & ---    & ---         & ---         & ---         & ---         & 0.87 / 0.74 \\
z-MOS           & ---    & ---         & ---         & ---         & ---         & ---         \\
\bottomrule
\end{tabular}}
\caption{SRCC/KRCC between each annotator’s ranking and the z-score MOS ranking for the CXR images in the XQA-IQA dataset.}
\label{interrater}
\end{table}

\FloatBarrier

\begin{wraptable}{r}{0.46\textwidth}
\vspace{0em}
\centering
\resizebox{0.44\textwidth}{!}{%
\begin{tabular}{ll}
\toprule
FR-IQA & \\
\midrule
SSIM-matlab & 0.78 / 0.57 \\
MS-SSIM-matlab & 0.84 / 0.63 \\
IW-SSIM-matlab & 0.80 / 0.59 \\
CW-SSIM-matlab & 0.49 / 0.35 \\
PSNR-matlab & 0.52 / 0.37 \\
RMSE-matlab & 0.66 / 0.48 \\
DISTS-matlab & 0.82 / 0.60 \\
DISTS-python-pytorch & 0.82 / 0.60 \\
DSS-matlab & 0.64 / 0.47 \\
FSIM-matlab & 0.83 / 0.61 \\
GMSD-matlab & 0.86 / 0.65 \\
HaarPSI-matlab & 0.87 / 0.66 \\
HaarPSI-python-pytorch & 0.84 / 0.63 \\
HaarPSImed-python-pytorch & 0.85 / 0.65 \\
LPIPS-python & 0.84 / 0.63 \\
MDSI-matlab & 0.83 / 0.62 \\
VIF-python-pytorch & 0.59 / 0.40 \\
VSI-matlab & 0.83 / 0.62 \\
UQI-python & 0.50 / 0.35 \\
AMBE-matlab & 0.44 / 0.32 \\
\midrule
NR-IQA & \\
\midrule
niqe-matlab & 0.41 / 0.28 \\
SNR-python & 0.21 / 0.14 \\
PAQ2PIQ-python & 0.61 / 0.43 \\
piqe-matlab & 0.42 / 0.20 \\
\bottomrule
\end{tabular}}
\caption{SRCC/KRCC between the z-MOS ranking and rankings induced by IQA measures for CXR images in XQA-IQA.}
\label{iqacorr}
\vspace{-3.0em}
\end{wraptable}

Table~\ref{iqacorr} presents the SRCC and KRCC between the z-MOS ranking and the rankings induced by each IQA metric. Among FR-IQA metrics, HaarPSI achieves the highest correlation with expert judgment ($\text{SRCC} = 0.87$, $\text{KRCC} = 0.66$), followed closely by GMSD ($0.86$ /$0.65$) and the medically adapted HaarPSImed variant ($0.85 / 0.65$). MS-SSIM, HaarPSI (Python), and LPIPS also perform competitively, all reaching an SRCC of $0.84$. In contrast, several FR-IQA metrics show substantially weaker correlation: CW-SSIM ($0.49$), UQI ($0.50$), PSNR ($0.52$), and AMBE ($0.44$) fall well below the top performers. NR-IQA metrics perform poorly across the board, with PAQ2PIQ being the only one to achieve moderate correlation ($0.61 / 0.43$), while SNR ($0.21 / 0.14$), NIQE ($0.41 / 0.28$), and PIQE ($0.42 / 0.20$) show negligible agreement with expert rankings.

\section{Discussion}

The central finding of this study is that evaluation reference choice is not a technical detail: it can change the scientific conclusion about which CXR models and image processing methods perform best. We show this in two settings that are usually studied separately---pathology classification and image quality assessment---but that share the same underlying problem: reliance on proxy evaluation targets whose alignment with clinical judgment is rarely tested. Across both tasks, changing the evaluation reference changes benchmark rankings in ways that are systematic, clinically interpretable, and consequential for model selection.

In pathology classification, image-derived and report-derived labels are not interchangeable. Image inter-annotator agreement is substantially lower than report inter-annotator agreement, reflecting the different nature of the two annotation tasks. Image annotators must interpret visual signs without access to clinical context, whereas report annotators label text that already encodes clinical history. The qualitative review of XQA-RR disagreements supports this interpretation. Annotators varied in whether they apply broad pathologies (e.g.\ Lung opacity) or more specific ones (e.g.\ Consolidation).

The resulting disagreement rates are pathology dependent. Atelectasis is the most ambiguous pathology for image annotators ($\kappa = 0.15$): its radiographic appearance depends heavily on the clinical context. Pleural effusion, by contrast, is the most consistently labeled pathology across all label sources, reflecting its comparatively visually distinct radiological appearance that can be recognized without patient history. In Consolidation, report annotators agree almost perfectly ($\kappa = 0.94$), yet cross-source agreement is the lowest among all selected pathologies ($\kappa \approx 0.22$). Distinguishing consolidation from lung opacity often requires inferring an underlying etiology (infection, edema, hemorrhage) that might be available in the radiology report but not from the image. Image annotators, therefore, frequently label the same finding as opacity, producing systematic divergence even when report annotators are fully consistent with one another.

Cross-source disagreements are consistent across both report annotators, suggesting that they reflect a structural difference between label sources rather than an artifact of individual raters (Table~\ref{tab:cuh_conflict_resolver_vs_report_ann1_ann2} in the Appendix). The Sankey \citep{schmidt2008sankey} visualizations (Figure~\ref{fig:sankey}) further reveal that omission patterns are bidirectional and pathology-specific: neither label source consistently captures all positives found by the other. Together, these patterns suggest that image-derived and report-derived labels draw on partially distinct sources of clinical information even when they are used to evaluate the same nominal pathology.

Comparing CUH and MIMIC-CXR further shows that pathology-specific annotation difficulty varies across datasets and institutions. Lung opacity, for example, shows
substantially lower image inter-annotator agreement on MIMIC-CXR than on CUH, while other pathologies show the opposite pattern.

The MIMIC-CXR results also show that report-derived labels are not a single, uniform reference source. Labels derived from the Findings and Impressions sections of the same report can differ substantially, and these differences can propagate to model rankings. This should not be interpreted only as annotation noise. Findings sections usually describe radiographic observations, whereas Impressions correspond to a clinically grounded interpretation of the examination and thus may incorporate context that is inaccessible from the image alone. Evaluation studies using report-derived labels should, therefore, specify which report section is used and justify whether that section is appropriate for the intended task.

High aggregate rank correlations do not necessarily imply stable model selection. The most consequential instability occurs at the top of the ranking, where practical decisions are made. For Atelectasis and Consolidation, the top-ranked model under one reference is ranked substantially lower under another, whereas Pleural effusion shows much greater stability, consistent with its higher cross-source agreement. The pattern of rank instability across pathologies broadly mirrors the pattern of cross-source disagreement, supporting the interpretation that reference choice affects rankings most strongly where image-derived and report-derived labels encode different clinical information. In practice, a researcher or clinician may select a different best-in-class system depending on whether the evaluation is based on image-derived or report-derived labels.

At the model-family level, rank stability reflects both cross-source label agreement and the distribution of model performance within each family. XRV models maintain stable rankings across label sources for well-agreed pathologies, whereas VLM-FT models show near-random rank agreement for Atelectasis ($\text{SRCC} = 0.21 \pm 0.22$), consistent with its low cross-source agreement. This is important because fine-tuned VLMs include several of the strongest models in absolute ROC-AUC terms: improved absolute performance under one reference does not guarantee stable ranking under another. One possible explanation for the greater stability of XRV rankings is stronger performance separation within this family: pooled or target-adapted XRV models perform substantially better than models not fine-tuned on the target datasets, producing a more stable ordering across references. By contrast, VLM-FT models are all fine-tuned in a similar setting and cluster more closely in performance, making their rankings more sensitive to small reference-dependent AUC differences. Zero-shot VLMs constitute a notable exception to the broader agreement--rank-stability pattern: despite Lung opacity having moderate cross-source agreement, zero-shot models show lower rank stability for this pathology than agreement levels alone would predict. This exception also affects the All-models group. One plausible explanation is score compression: if models achieve similar Lung opacity AUCs, small absolute differences can translate into larger rank fluctuations.

Evaluation against report-derived labels is internally reproducible but not interchangeable with evaluation against image-derived labels. On XQA-RR, the two report annotators induce highly consistent model rankings, showing that report-derived evaluation is stable across annotators even when it diverges from image-derived evaluation. We do not argue that image-derived or report-derived labels are universally superior. Rather, these references answer different questions: image-derived labels reflect what is visually recoverable from the radiograph alone, whereas report-derived labels reflect a clinically contextualized interpretation intended to guide care. Treating these references as interchangeable risks drawing misleading conclusions about clinical relevance. This distinction also affects absolute performance estimates: for pathologies with larger image--report disagreement, models tend to achieve higher AUC against image-derived references than against report-derived references, suggesting that image-derived evaluation may give a more optimistic estimate for models trained to detect visually recoverable findings. Whereas recent stress-testing work shows that multimodal health-AI benchmarks can conceal fragility under input perturbations \citep{gu2026evaluating}, our results identify a distinct source of fragility: the clinical target encoded by the evaluation reference itself.

Our IQA results reinforce the same conclusion from a complementary angle: widely used automated evaluation metrics do not necessarily reflect clinically meaningful judgments of image quality. This is particularly important for image outputs produced by black-box methods, such as neural-network-based reconstruction or generative models, which are often evaluated primarily through automated IQA measures. Poor alignment between metric scores and expert judgment may therefore lead to misleading conclusions about clinical usability and to rankings that fail to reflect even straightforward quality differences \citep{breger2025study}. This concern is consistent with prior work showing that standard IQA metrics often fail to capture expert assessments of medical image quality \citep{breger2024study}.

Among the full-reference metrics, HaarPSI, GMSD, and HaarPSImed align most closely with expert diagnostic quality rankings, whereas commonly used metrics such as PSNR and CW-SSIM show substantially weaker agreement. Metric selection in IQA studies is therefore not a neutral choice: algorithms that rank highly under PSNR or SNR may differ from those preferred by expert radiologists, and the choice of metric shapes which post-processing methods appear clinically effective.

\section{Limitations}

The employed data set XQA-RR contains 650 CUH and 735 MIMIC-CXR patient studies, each comprising a frontal chest radiograph and associated report. This sample size supports the agreement and ranking analyses presented, but remains limited for rare pathologies, several of which were excluded from the primary analysis due to low positive counts. Image-derived labels relied on two annotators per dataset; for MIMIC-CXR, no adjudication step was performed, and FUSE\_OR fusion served as a proxy for conflict-resolved image-derived labels, a choice validated against CUH but not equivalent to third-annotator adjudication. In addition, annotations for both datasets were produced by radiologists from the same institution, so annotation style and labeling conventions may reflect a particular clinical culture rather than a universal standard. Finally, the XQA-IQA evaluation uses controlled synthetic degradations rather than the complex real-world image quality issues encountered in clinical practice, which may limit the generalization of the metric rankings to operational settings.

Future work should examine how automatically extracted report labels, including those produced by large language models, affect these conclusions. However, improved report extraction alone does not eliminate the need to specify whether the intended reference is an image-visible abnormality, a report-level clinical interpretation, or diagnostic usability.

\section{Summary}

We investigated whether conclusions about chest X-ray machine learning are stable to the choice of evaluation reference. This is important because CXR image classifiers are often trained and evaluated using labels derived from different sources, while image quality assessment (IQA) methods are commonly judged using generic metrics rather than expert assessments of diagnostic usability.

In pathology classification, evaluation against image-derived versus report-derived labels leads to systematic differences in performance estimates and model rankings. These effects are pathology-dependent: image-derived and report-derived labels appear more interchangeable for visually well-defined findings such as Pleural effusion, but diverge more strongly for findings such as Atelectasis, Consolidation, and Lung opacity. These results do not support a single general recommendation for the use of either image-derived or report-derived labels. Instead, the evaluation reference should be justified for each pathology and intended application. Image-derived labels are more appropriate for assessing visual detectability from the radiograph alone, whereas report-derived labels are more appropriate for assessing alignment with the clinically contextualized interpretation recorded in the radiology report.

In image quality assessment, automated metrics correlate inconsistently with expert judgments of diagnostic usability. Some full-reference metrics align relatively well with expert rankings, whereas commonly used metrics such as PSNR and several no-reference measures align weakly with clinical preference. Thus, IQA metrics should not be treated as clinical proxies without validation against expert judgment.

Taken together, these results demonstrate that the choice of evaluation reference is central to assessing CXR machine learning. Especially when black box models are judged mainly by benchmark performance, conclusions about utility depend critically on whether the chosen reference is clinically meaningful. Convenient proxy targets, including generic IQA metrics and interchangeable image- or report-derived labels, should therefore be used only with explicit justification.

To support further research on clinically grounded evaluation, code and current data-access information are available at \url{https://github.com/PanagiotisFytas/XQA-Chest-X-ray}. XQA-Chest comprises 650 patient studies with paired expert image-derived and report-derived labels for pathology classification, together with expert ratings of diagnostic image quality from five clinicians for 1{,}571 degraded images generated from 323 reference images. The dataset will be released through ReShare following completion of repository review, and the permanent ReShare link and access instructions will be added to the GitHub repository. The resource is intended to support reproducible work on clinically grounded evaluation references for CXR machine learning.

\section{Ethics}

This study was conducted under ethical approval for the study ``AI-assisted diagnosis and prognostication in COVID-19'' (REC: 20/HRA/2504; IRAS: 282705). The study used retrospective anonymised or pseudo-anonymised data, and the research team did not have access to direct personal identifiers. Informed consent was not obtained because the study was retrospective and used anonymized data. Data handling was conducted in accordance with the NHS Code of Practice on Confidentiality, UK GDPR, and the Data Protection Act 2018.

\section{Code availability}

Code used for the evaluation experiments is available at \url{https://github.com/PanagiotisFytas/XQA-Chest-X-ray}.

\section{Data availability}

The XQA-Chest annotations and image quality assessment ratings are currently undergoing repository review for release through ReShare. The permanent ReShare link, access conditions, and instructions will be added to \url{https://github.com/PanagiotisFytas/XQA-Chest-X-ray} following completion of repository review.

The MIMIC-CXR component of the study is based on the publicly available MIMIC-CXR dataset, subject to the access conditions of PhysioNet. The annotation labels produced for the study will be made available through the project GitHub repository. The non-public CUH data used for model fine-tuning cannot be shared because they derive from clinical data governed by institutional, ethical, and data-protection restrictions.

\section{Funding}

P.F. was supported by an award from the Claudia Harding Foundation through the University of Cambridge Harding Distinguished Postgraduate Scholars Programme. A.B. acknowledges support from the EU/EFPIA Innovative Medicines Initiative 2 Joint Undertaking through the DRAGON project (Grant No. 101005122). A.B. and C.K. acknowledge support from the Austrian Science Fund (FWF) through project T1307. This work was supported by UK Research and Innovation through the EQUATE grant (EP/Y031350/1) awarded to A.K. J.H.F.R. is part-supported by the NIHR Cambridge Biomedical Research Centre (NIHR203312) and the British Heart Foundation Centre of Research Excellence (RE/24/130011).

The funders had no role in study design, data collection, analysis, interpretation, manuscript preparation, or the decision to submit the article for publication.

\section{Author contributions}

Panagiotis Fytas: Conceptualization, Methodology, Software, Formal analysis, Data curation, Visualization, Writing -- original draft, Writing -- review and editing. Ian Selby: Conceptualization, Investigation, Data curation, Software. Clemens Karner: Methodology, Software, Formal analysis, Writing -- review and editing. Judith Babar: Investigation, Data curation, Validation, Writing -- review and editing. Simon Baker: Supervision, Writing -- review and editing. Jake Beckford: Data curation, Writing -- review and editing. Timothy J. Sadler: Data curation, Writing -- review and editing. Shahab Shahipasand: Data curation, Writing -- review and editing. Arthikkaa Thavakumar: Data curation, Writing -- review and editing. John Li Chen: Data curation, Writing -- review and editing. Alex Sawer: Data curation, Writing -- review and editing. Michael Roberts: Supervision, Writing -- review and editing. Jonathan Weir-McCall: Investigation, Writing -- review and editing. J.H.F. Rudd: Investigation, Writing -- review and editing. Carola-Bibiane Sch\"onlieb: Supervision, Funding acquisition, Writing -- review and editing. Anna Korhonen: Supervision, Funding acquisition, Writing -- review and editing. Anna Breger: Conceptualization, Methodology, Investigation, Data curation, Software, Formal analysis, Supervision, Writing -- original draft, Writing -- review and editing.

\section{Declaration of Generative AI and AI-Assisted Technologies in the Manuscript Preparation Process}

During the preparation of this work, the authors used ChatGPT (OpenAI) to assist in improving the clarity, readability, and organization of the manuscript text. No AI tool was used to generate experimental results, perform analyses, or make scientific conclusions. After using this tool/service, the authors reviewed and edited the content as needed and take full responsibility for the content of the published article.

\bibliographystyle{elsarticle-harv}
\bibliography{references/references}

\clearpage

\appendix

\setlength{\LTpre}{6pt}
\setlength{\LTpost}{12pt}

\section{Dataset Label Prevalence}

\begin{table}[H]
\centering
\small
\setlength{\tabcolsep}{5pt}
\begin{tabular}{lccccc}
\toprule
Pathology & CR & Img1 & Img2 & Rep1 & Rep2 \\
\midrule
No Finding & 121 (18.6\%) & 91 (14.0\%) & 129 (19.8\%) & 16 (2.5\%) & 20 (3.1\%) \\
Enlarg. Card. & 1 (0.2\%) & 0 (0.0\%) & 1 (0.2\%) & 2 (0.3\%) & 1 (0.2\%) \\
Cardiomegaly & 21 (3.2\%) & 153 (23.5\%) & 21 (3.2\%) & 19 (2.9\%) & 17 (2.6\%) \\
Lung Opacity & 187 (28.8\%) & 197 (30.3\%) & 131 (20.2\%) & 200 (30.8\%) & 171 (26.3\%) \\
Lung Lesion & 4 (0.6\%) & 4 (0.6\%) & 3 (0.5\%) & 6 (0.9\%) & 7 (1.1\%) \\
Edema & 4 (0.6\%) & 1 (0.2\%) & 1 (0.2\%) & 11 (1.7\%) & 11 (1.7\%) \\
Consolidation & 106 (16.3\%) & 110 (16.9\%) & 66 (10.2\%) & 154 (23.7\%) & 151 (23.2\%) \\
Pneumonia & 9 (1.4\%) & 0 (0.0\%) & 10 (1.5\%) & 1 (0.2\%) & 11 (1.7\%) \\
Atelectasis & 165 (25.4\%) & 44 (6.8\%) & 165 (25.4\%) & 58 (8.9\%) & 68 (10.5\%) \\
Pneumothorax & 4 (0.6\%) & 5 (0.8\%) & 7 (1.1\%) & 7 (1.1\%) & 8 (1.2\%) \\
Pleural Eff. & 136 (20.9\%) & 122 (18.8\%) & 114 (17.5\%) & 92 (14.2\%) & 93 (14.3\%) \\
Pleural Other & 11 (1.7\%) & 9 (1.4\%) & 5 (0.8\%) & 17 (2.6\%) & 14 (2.2\%) \\
Fracture & 25 (3.8\%) & 14 (2.2\%) & 23 (3.5\%) & 24 (3.7\%) & 19 (2.9\%) \\
Support Dev. & 274 (42.2\%) & 309 (47.5\%) & 274 (42.2\%) & 284 (43.7\%) & 279 (42.9\%) \\

\bottomrule
\end{tabular}
\caption{XQA-RR positive label prevalence per CheXpert pathology and label source, presented as counts (percentage). Label sources included conflict-resolved image labels (CR), labels from individual image annotators 1 and 2 (Img1, Img2), and labels from individual report annotators (Rep1, Rep2).}
\label{tab:positive-prevalence}
\end{table}

\begin{table}[H]
\centering
\small
\setlength{\tabcolsep}{5pt}
\begin{tabular}{lcccc}
\toprule
Pathology & Img1 & Img2 & Find & Impr \\
\midrule
No Finding & 43 (5.9\%) & 19 (2.6\%) & 0 (0.0\%) & 58 (7.9\%)  \\
Enlarged Card. & 21 (2.9\%) & 5 (0.7\%) & 22 (3.0\%) & 16 (2.2\%)  \\
Cardiomegaly & 27 (3.7\%) & 9 (1.2\%) & 188 (25.6\%) & 101 (13.7\%)  \\
Lung Opacity & 23 (3.1\%) & 150 (20.4\%) & 213 (29.0\%) & 142 (19.3\%)  \\
Lung Lesion & 25 (3.4\%) & 2 (0.3\%) & 41 (5.6\%) & 37 (5.0\%)  \\
Edema & 7 (1.0\%) & 3 (0.4\%) & 106 (14.4\%) & 102 (13.9\%)  \\
Consolidation & 181 (24.6\%) & 40 (5.4\%) & 37 (5.0\%) & 44 (6.0\%)  \\
Pneumonia & 5 (0.7\%) & 0 (0.0\%) & 36 (4.9\%) & 54 (7.3\%)  \\
Atelectasis & 186 (25.3\%) & 59 (8.0\%) & 179 (24.4\%) & 98 (13.3\%)  \\
Pneumothorax & 15 (2.0\%) & 13 (1.8\%) & 23 (3.1\%) & 13 (1.8\%)  \\
Pleural Eff. & 176 (23.9\%) & 209 (28.4\%) & 182 (24.8\%) & 161 (21.9\%)  \\
Pleural Other & 19 (2.6\%) & 5 (0.7\%) & 33 (4.5\%) & 15 (2.0\%) \\
Fracture & 24 (3.3\%) & 23 (3.1\%) & 41 (5.6\%) & 22 (3.0\%) \\
Support Dev. & 176 (23.9\%) & 361 (49.1\%) & 235 (32.0\%) & 132 (18.0\%)  \\

\bottomrule
\end{tabular}
\caption{MIMIC-CXR positive label prevalence per CheXpert pathology and label source, presented as counts (percentage). Label sources include labels from individual image annotators (Img1 and Img2) and labels from annotations of findings (Find) and impression (Impr) radiology reports sections.}
\label{tab:positive-prevalence-mimic}
\end{table}

\clearpage
\begingroup
\small
\setlength{\tabcolsep}{4pt}
\begin{longtable}[c]{llccc}
\toprule
Pathology & Source & 0 & 1 & 2 \\
\midrule
\endfirsthead

\toprule
Pathology & Source & 0 & 1 & 2 \\
\midrule
\endhead

\midrule
\multicolumn{5}{r}{\textit{Continued on next page}} \\
\endfoot

\bottomrule
\\ \caption{XQA-RR label distribution per CheXpert pathology and label source. Label sources included conflict-resolved image labels (CR), labels from individual image annotators 1 and 2 (Img1, Img2), and labels from individual report annotators (Rep1, Rep2).
Values are shown as count (percentage).}
\label{tab:label-distribution-long}
\endlastfoot
\makecell[l]{No Finding}
& CR & 526 (80.9\%) & 3 (0.5\%) & 121 (18.6\%) \\
& Img1 & 558 (85.8\%) & 1 (0.2\%) & 91 (14.0\%) \\
& Img2 & 519 (79.8\%) & 2 (0.3\%) & 129 (19.8\%) \\
& Rep1 & 0 (0.0\%) & 0 (0.0\%) & 16 (2.5\%) \\
& Rep2 & 1 (0.2\%) & 0 (0.0\%) & 20 (3.1\%) \\
\midrule
\makecell[l]{Enlarged Cardiomediastinum}
& CR & 646 (99.4\%) & 3 (0.5\%) & 1 (0.2\%) \\
& Img1 & 647 (99.5\%) & 3 (0.5\%) & 0 (0.0\%) \\
& Img2 & 646 (99.4\%) & 3 (0.5\%) & 1 (0.2\%) \\
& Rep1 & 96 (14.8\%) & 24 (3.7\%) & 2 (0.3\%) \\
& Rep2 & 90 (13.8\%) & 23 (3.5\%) & 1 (0.2\%) \\
\midrule
\makecell[l]{Cardiomegaly}
& CR & 592 (91.1\%) & 37 (5.7\%) & 21 (3.2\%) \\
& Img1 & 457 (70.3\%) & 40 (6.2\%) & 153 (23.5\%) \\
& Img2 & 592 (91.1\%) & 37 (5.7\%) & 21 (3.2\%) \\
& Rep1 & 99 (15.2\%) & 10 (1.5\%) & 19 (2.9\%) \\
& Rep2 & 100 (15.4\%) & 10 (1.5\%) & 17 (2.6\%) \\
\midrule
\makecell[l]{Lung Opacity}
& CR & 416 (64.0\%) & 47 (7.2\%) & 187 (28.8\%) \\
& Img1 & 400 (61.5\%) & 53 (8.2\%) & 197 (30.3\%) \\
& Img2 & 461 (70.9\%) & 58 (8.9\%) & 131 (20.2\%) \\
& Rep1 & 8 (1.2\%) & 2 (0.3\%) & 200 (30.8\%) \\
& Rep2 & 7 (1.1\%) & 1 (0.2\%) & 171 (26.3\%) \\
\midrule
\makecell[l]{Lung Lesion}
& CR & 642 (98.8\%) & 4 (0.6\%) & 4 (0.6\%) \\
& Img1 & 625 (96.2\%) & 21 (3.2\%) & 4 (0.6\%) \\
& Img2 & 641 (98.6\%) & 6 (0.9\%) & 3 (0.5\%) \\
& Rep1 & 33 (5.1\%) & 2 (0.3\%) & 6 (0.9\%) \\
& Rep2 & 34 (5.2\%) & 2 (0.3\%) & 7 (1.1\%) \\
\midrule
\makecell[l]{Edema}
& CR & 622 (95.7\%) & 24 (3.7\%) & 4 (0.6\%) \\
& Img1 & 645 (99.2\%) & 4 (0.6\%) & 1 (0.2\%) \\
& Img2 & 614 (94.5\%) & 35 (5.4\%) & 1 (0.2\%) \\
& Rep1 & 24 (3.7\%) & 4 (0.6\%) & 11 (1.7\%) \\
& Rep2 & 23 (3.5\%) & 2 (0.3\%) & 11 (1.7\%) \\
\midrule
\makecell[l]{Consolidation}
& CR & 523 (80.5\%) & 21 (3.2\%) & 106 (16.3\%) \\
& Img1 & 533 (82.0\%) & 7 (1.1\%) & 110 (16.9\%) \\
& Img2 & 572 (88.0\%) & 12 (1.8\%) & 66 (10.2\%) \\
& Rep1 & 132 (20.3\%) & 13 (2.0\%) & 154 (23.7\%) \\
& Rep2 & 137 (21.1\%) & 6 (0.9\%) & 151 (23.2\%) \\
\midrule
\makecell[l]{Pneumonia}
& CR & 610 (93.8\%) & 31 (4.8\%) & 9 (1.4\%) \\
& Img1 & 650 (100.0\%) & 0 (0.0\%) & 0 (0.0\%) \\
& Img2 & 575 (88.5\%) & 65 (10.0\%) & 10 (1.5\%) \\
& Rep1 & 1 (0.2\%) & 1 (0.2\%) & 1 (0.2\%) \\
& Rep2 & 8 (1.2\%) & 8 (1.2\%) & 11 (1.7\%) \\
\midrule
\makecell[l]{Atelectasis}
& CR & 448 (68.9\%) & 37 (5.7\%) & 165 (25.4\%) \\
& Img1 & 586 (90.2\%) & 20 (3.1\%) & 44 (6.8\%) \\
& Img2 & 448 (68.9\%) & 37 (5.7\%) & 165 (25.4\%) \\
& Rep1 & 2 (0.3\%) & 0 (0.0\%) & 58 (8.9\%) \\
& Rep2 & 34 (5.2\%) & 3 (0.5\%) & 68 (10.5\%) \\
\midrule
\makecell[l]{Pneumothorax}
& CR & 640 (98.5\%) & 6 (0.9\%) & 4 (0.6\%) \\
& Img1 & 644 (99.1\%) & 1 (0.2\%) & 5 (0.8\%) \\
& Img2 & 638 (98.2\%) & 5 (0.8\%) & 7 (1.1\%) \\
& Rep1 & 121 (18.6\%) & 1 (0.2\%) & 7 (1.1\%) \\
& Rep2 & 131 (20.2\%) & 2 (0.3\%) & 8 (1.2\%) \\
\midrule
\makecell[l]{Pleural Effusion}
& CR & 482 (74.2\%) & 32 (4.9\%) & 136 (20.9\%) \\
& Img1 & 495 (76.2\%) & 33 (5.1\%) & 122 (18.8\%) \\
& Img2 & 480 (73.8\%) & 56 (8.6\%) & 114 (17.5\%) \\
& Rep1 & 100 (15.4\%) & 3 (0.5\%) & 92 (14.2\%) \\
& Rep2 & 120 (18.5\%) & 6 (0.9\%) & 93 (14.3\%) \\
\midrule
\makecell[l]{Pleural Other}
& CR & 638 (98.2\%) & 1 (0.2\%) & 11 (1.7\%) \\
& Img1 & 640 (98.5\%) & 1 (0.2\%) & 9 (1.4\%) \\
& Img2 & 640 (98.5\%) & 5 (0.8\%) & 5 (0.8\%) \\
& Rep1 & 2 (0.3\%) & 4 (0.6\%) & 17 (2.6\%) \\
& Rep2 & 23 (3.5\%) & 2 (0.3\%) & 14 (2.2\%) \\
\midrule
\makecell[l]{Fracture}
& CR & 623 (95.8\%) & 2 (0.3\%) & 25 (3.8\%) \\
& Img1 & 635 (97.7\%) & 1 (0.2\%) & 14 (2.2\%) \\
& Img2 & 625 (96.2\%) & 2 (0.3\%) & 23 (3.5\%) \\
& Rep1 & 6 (0.9\%) & 0 (0.0\%) & 24 (3.7\%) \\
& Rep2 & 8 (1.2\%) & 0 (0.0\%) & 19 (2.9\%) \\
\midrule
\makecell[l]{Support Devices}
& CR & 357 (54.9\%) & 19 (2.9\%) & 274 (42.2\%) \\
& Img1 & 339 (52.2\%) & 2 (0.3\%) & 309 (47.5\%) \\
& Img2 & 357 (54.9\%) & 19 (2.9\%) & 274 (42.2\%) \\
& Rep1 & 1 (0.2\%) & 1 (0.2\%) & 284 (43.7\%) \\
& Rep2 & 1 (0.2\%) & 1 (0.2\%) & 279 (42.9\%) \\
\end{longtable}
\endgroup

\begingroup
\small
\setlength{\tabcolsep}{4pt}
\begin{longtable}[c]{llccc}
\toprule
Pathology & Source & 0 & 1 & 2 \\
\midrule
\endfirsthead

\toprule
Pathology & Source & 0 & 1 & 2 \\
\midrule
\endhead

\midrule
\multicolumn{5}{r}{\textit{Continued on next page}} \\
\endfoot

\bottomrule
\\ \caption{MIMIC-CXR label distribution per CheXpert pathology and label source. Label sources include labels from individual image annotators (Img1 and Img2) and labels from annotations of findings (Find) and impression (Impr) radiology reports sections.
Values are shown as count (percentage).}
\label{tab:label-distribution-long-mimic}
\endlastfoot
\makecell[l]{No Finding}
& Img1 & 682 (92.8\%) & 10 (1.4\%) & 43 (5.9\%) \\
& Img2 & 659 (89.7\%) & 57 (7.8\%) & 19 (2.6\%) \\
& Find & 2 (0.3\%) & 0 (0.0\%) & 0 (0.0\%) \\
& Impr & 0 (0.0\%) & 0 (0.0\%) & 58 (7.9\%) \\
\midrule
\makecell[l]{Enlarged Cardiomediastinum}
& Img1 & 665 (90.5\%) & 49 (6.7\%) & 21 (2.9\%) \\
& Img2 & 719 (97.8\%) & 11 (1.5\%) & 5 (0.7\%) \\
& Find & 146 (19.9\%) & 103 (14.0\%) & 22 (3.0\%) \\
& Impr & 12 (1.6\%) & 28 (3.8\%) & 16 (2.2\%) \\
\midrule
\makecell[l]{Cardiomegaly}
& Img1 & 655 (89.1\%) & 53 (7.2\%) & 27 (3.7\%) \\
& Img2 & 701 (95.4\%) & 25 (3.4\%) & 9 (1.2\%) \\
& Find & 69 (9.4\%) & 83 (11.3\%) & 188 (25.6\%) \\
& Impr & 33 (4.5\%) & 10 (1.4\%) & 101 (13.7\%) \\
\midrule
\makecell[l]{Lung Opacity}
& Img1 & 628 (85.4\%) & 84 (11.4\%) & 23 (3.1\%) \\
& Img2 & 412 (56.1\%) & 173 (23.5\%) & 150 (20.4\%) \\
& Find & 42 (5.7\%) & 0 (0.0\%) & 213 (29.0\%) \\
& Impr & 4 (0.5\%) & 0 (0.0\%) & 142 (19.3\%) \\
\midrule
\makecell[l]{Lung Lesion}
& Img1 & 628 (85.4\%) & 82 (11.2\%) & 25 (3.4\%) \\
& Img2 & 666 (90.6\%) & 67 (9.1\%) & 2 (0.3\%) \\
& Find & 10 (1.4\%) & 9 (1.2\%) & 41 (5.6\%) \\
& Impr & 0 (0.0\%) & 14 (1.9\%) & 37 (5.0\%) \\
\midrule
\makecell[l]{Edema}
& Img1 & 705 (95.9\%) & 23 (3.1\%) & 7 (1.0\%) \\
& Img2 & 647 (88.0\%) & 85 (11.6\%) & 3 (0.4\%) \\
& Find & 79 (10.7\%) & 16 (2.2\%) & 106 (14.4\%) \\
& Impr & 62 (8.4\%) & 21 (2.9\%) & 102 (13.9\%) \\
\midrule
\makecell[l]{Consolidation}
& Img1 & 438 (59.6\%) & 116 (15.8\%) & 181 (24.6\%) \\
& Img2 & 544 (74.0\%) & 151 (20.5\%) & 40 (5.4\%) \\
& Find & 109 (14.8\%) & 23 (3.1\%) & 37 (5.0\%) \\
& Impr & 26 (3.5\%) & 13 (1.8\%) & 44 (6.0\%) \\
\midrule
\makecell[l]{Pneumonia}
& Img1 & 646 (87.9\%) & 84 (11.4\%) & 5 (0.7\%) \\
& Img2 & 702 (95.5\%) & 33 (4.5\%) & 0 (0.0\%) \\
& Find & 70 (9.5\%) & 48 (6.5\%) & 36 (4.9\%) \\
& Impr & 54 (7.3\%) & 81 (11.0\%) & 54 (7.3\%) \\
\midrule
\makecell[l]{Atelectasis}
& Img1 & 459 (62.4\%) & 90 (12.2\%) & 186 (25.3\%) \\
& Img2 & 549 (74.7\%) & 127 (17.3\%) & 59 (8.0\%) \\
& Find & 5 (0.7\%) & 30 (4.1\%) & 179 (24.4\%) \\
& Impr & 3 (0.4\%) & 38 (5.2\%) & 98 (13.3\%) \\
\midrule
\makecell[l]{Pneumothorax}
& Img1 & 715 (97.3\%) & 5 (0.7\%) & 15 (2.0\%) \\
& Img2 & 718 (97.7\%) & 4 (0.5\%) & 13 (1.8\%) \\
& Find & 335 (45.6\%) & 10 (1.4\%) & 23 (3.1\%) \\
& Impr & 115 (15.6\%) & 4 (0.5\%) & 13 (1.8\%) \\
\midrule
\makecell[l]{Pleural Effusion}
& Img1 & 469 (63.8\%) & 90 (12.2\%) & 176 (23.9\%) \\
& Img2 & 441 (60.0\%) & 85 (11.6\%) & 209 (28.4\%) \\
& Find & 243 (33.1\%) & 31 (4.2\%) & 182 (24.8\%) \\
& Impr & 37 (5.0\%) & 26 (3.5\%) & 161 (21.9\%) \\
\midrule
\makecell[l]{Pleural Other}
& Img1 & 697 (94.8\%) & 19 (2.6\%) & 19 (2.6\%) \\
& Img2 & 711 (96.7\%) & 19 (2.6\%) & 5 (0.7\%) \\
& Find & 3 (0.4\%) & 4 (0.5\%) & 33 (4.5\%) \\
& Impr & 6 (0.8\%) & 4 (0.5\%) & 15 (2.0\%) \\
\midrule
\makecell[l]{Fracture}
& Img1 & 708 (96.3\%) & 3 (0.4\%) & 24 (3.3\%) \\
& Img2 & 695 (94.6\%) & 17 (2.3\%) & 23 (3.1\%) \\
& Find & 48 (6.5\%) & 6 (0.8\%) & 41 (5.6\%) \\
& Impr & 5 (0.7\%) & 2 (0.3\%) & 22 (3.0\%) \\
\midrule
\makecell[l]{Support Devices}
& Img1 & 540 (73.5\%) & 19 (2.6\%) & 176 (23.9\%) \\
& Img2 & 360 (49.0\%) & 14 (1.9\%) & 361 (49.1\%) \\
& Find & 10 (1.4\%) & 0 (0.0\%) & 235 (32.0\%) \\
& Impr & 5 (0.7\%) & 0 (0.0\%) & 132 (18.0\%) \\
\end{longtable}
\endgroup

\clearpage
\section{Agreement Tables}

\begin{table}[H]
\centering
\begin{tabular}{lrrrrrr}
\toprule
Pathology & $\kappa_{\text{w}}$ & $\kappa_{\text{low}}$ & $\kappa_{\text{high}}$ & F1 & F1$_{\text{low}}$ & F1$_{\text{high}}$ \\
\midrule
Atelectasis & 0.15 & 0.06 & 0.23 & 0.62 & 0.62 & 0.72 \\
Consolidation & 0.47 & 0.41 & 0.51 & 0.86 & 0.87 & 0.89 \\
Lung Opacity & 0.42 & 0.22 & 0.55 & 0.68 & 0.69 & 0.82 \\
Pleural Effusion & 0.67 & 0.46 & 0.77 & 0.80 & 0.81 & 0.93 \\
\cmidrule(lr){1-7}
Macro avg.\ (four selected) & 0.43 & 0.29 & 0.52 & 0.74 & 0.75 & 0.84 \\
Macro avg.\ (all 12) & 0.33 & 0.21 & 0.44 & 0.87 & 0.87 & 0.93 \\
Micro avg.\ (all 12) & 0.45 & 0.29 & 0.55 & 0.87 & 0.87 & 0.93 \\
\bottomrule
\end{tabular}
\caption{XQA-RR inter-annotator agreement for image labels in the four selected categories, including macro-average scores for the four selected categories and macro/micro-averaged results for all 12 pathologies. $\kappa_{\text{w}}$ denotes quadratically weighted Cohen's $\kappa$ computed on the original three-class labels (negative, uncertain, positive). $\kappa_{\text{low}}$ and $\kappa_{\text{high}}$ denote unweighted Cohen's $\kappa$ under pessimistic and optimistic uncertainty mappings, respectively. F1${\text{low}}$ and F1${\text{high}}$ are computed under the same mappings.}
\label{tab:cuh_images_ann1_vs_ann2}
\end{table}

\begin{table}[H]
\centering
\begin{tabular}{lrrrrrr}
\toprule
Pathology & $\kappa_{\text{w}}$ & $\kappa_{\text{low}}$ & $\kappa_{\text{high}}$ & F1 & F1$_{\text{low}}$ & F1$_{\text{high}}$ \\
\midrule
Atelectasis & 0.80 & 0.78 & 0.81 & 0.96 & 0.96 & 0.96 \\
Consolidation & 0.94 & 0.89 & 0.96 & 0.97 & 0.96 & 0.98 \\
Lung Opacity & 0.84 & 0.83 & 0.85 & 0.93 & 0.93 & 0.94 \\
Pleural Effusion & 0.89 & 0.87 & 0.90 & 0.97 & 0.97 & 0.98 \\
\cmidrule(lr){1-7}
Macro avg.\ (four selected) & 0.87 & 0.84 & 0.88 & 0.96 & 0.96 & 0.97 \\
Macro avg.\ (all 12) & 0.82 & 0.76 & 0.84 & 0.98 & 0.98 & 0.98 \\
Micro avg.\ (all 12) & 0.88 & 0.85 & 0.90 & 0.98 & 0.98 & 0.99 \\
\bottomrule
\end{tabular}
\caption{XQA-RR inter-annotator agreement for report labels in the four selected categories, including macro-average scores for the four selected categories and macro/micro-averaged results for all 12 pathologies. $\kappa_{\text{w}}$ denotes quadratically weighted Cohen's $\kappa$ computed on the original three-class labels (negative, uncertain, positive). $\kappa_{\text{low}}$ and $\kappa_{\text{high}}$ denote unweighted Cohen's $\kappa$ under pessimistic and optimistic uncertainty mappings, respectively. F1${\text{low}}$ and F1${\text{high}}$ are computed under the same mappings.}
\label{tab:cuh_reports_ann1_vs_ann2}
\end{table}

\clearpage

\begin{table}[H]
\centering
\begin{tabular}{lrrrrrr}
\toprule
Pathology & $\kappa_{\text{w}}$ & $\kappa_{\text{low}}$ & $\kappa_{\text{high}}$ & F1 & F1$_{\text{low}}$ & F1$_{\text{high}}$ \\
\midrule
\multicolumn{7}{c}{CR vs.\ Rep1} \\
\midrule
Atelectasis          & 0.23 & 0.16 & 0.28 & 0.79 & 0.78 & 0.82 \\
Consolidation        & 0.22 & 0.16 & 0.27 & 0.70 & 0.69 & 0.74 \\
Lung Opacity         & 0.43 & 0.33 & 0.50 & 0.73 & 0.71 & 0.78 \\
Pleural Effusion     & 0.56 & 0.46 & 0.61 & 0.86 & 0.85 & 0.89 \\
\cmidrule(lr){1-7}
Macro avg.\ (four selected) & 0.36 & 0.28 & 0.41 & 0.77 & 0.76 & 0.81 \\
Macro avg.\ (all 12) & 0.34 & 0.24 & 0.46 & 0.90 & 0.89 & 0.92 \\
Micro avg.\ (all 12) & 0.43 & 0.33 & 0.49 & 0.90 & 0.89 & 0.92 \\
\midrule
\multicolumn{7}{c}{CR vs.\ Rep2} \\
\midrule
Atelectasis          & 0.25 & 0.17 & 0.31 & 0.78 & 0.77 & 0.82 \\
Consolidation        & 0.24 & 0.19 & 0.28 & 0.72 & 0.71 & 0.75 \\
Lung Opacity         & 0.40 & 0.30 & 0.47 & 0.74 & 0.72 & 0.79 \\
Pleural Effusion     & 0.54 & 0.45 & 0.60 & 0.85 & 0.84 & 0.89 \\
\cmidrule(lr){1-7}
Macro avg.\ (four selected) & 0.36 & 0.28 & 0.41 & 0.77 & 0.76 & 0.81 \\
Macro avg.\ (all 12) & 0.32 & 0.22 & 0.44 & 0.89 & 0.89 & 0.92 \\
Micro avg.\ (all 12) & 0.42 & 0.32 & 0.48 & 0.90 & 0.89 & 0.92 \\
\bottomrule
\end{tabular}
\caption{XQA-RR agreement between conflict-resolved (CR) image labels and the labels from the two individual report annotators (Rep1 and Rep2) in the four selected categories, including macro-average scores for the four selected categories and macro/micro-averaged results for all 12 pathologies. $\kappa_{\text{w}}$ denotes quadratically weighted Cohen's $\kappa$ computed on the original three-class labels (negative, uncertain, positive). $\kappa_{\text{low}}$ and $\kappa_{\text{high}}$ denote unweighted Cohen's $\kappa$ under pessimistic and optimistic uncertainty mappings, respectively. F1${\text{low}}$ and F1${\text{high}}$ are computed under the same mappings.}
\label{tab:cuh_conflict_resolver_vs_report_ann1_ann2}
\end{table}

\begin{table}[H]
\centering
\begin{tabular}{lrrrrrr}
\toprule
Pathology & $\kappa_{\text{w}}$ & $\kappa_{\text{low}}$ & $\kappa_{\text{high}}$ & F1 & F1$_{\text{low}}$ & F1$_{\text{high}}$ \\
\midrule
Atelectasis          & 0.27 &  0.00 & 0.49 & 0.66 & 0.67 & 0.85 \\
Consolidation        & 0.38 &  0.04 & 0.64 & 0.67 & 0.70 & 0.89 \\
Lung Opacity         & 0.11 & -0.06 & 0.31 & 0.49 & 0.47 & 0.79 \\
Pleural Effusion     & 0.62 &  0.37 & 0.75 & 0.73 & 0.72 & 0.90 \\
\cmidrule(lr){1-7}
Macro avg.\ (four selected) & 0.34 & 0.09 & 0.55 & 0.64 & 0.64 & 0.86 \\
Macro avg.\ (all 12) & 0.38 & 0.16 & 0.64 & 0.82 & 0.83 & 0.94 \\
Micro avg.\ (all 12) & 0.44 & 0.16 & 0.65 & 0.82 & 0.83 & 0.95 \\
\bottomrule
\end{tabular}
\caption{MIMIC-CXR inter-annotator agreement for image labels in the four selected categories, including macro-average scores for the four selected categories and macro/micro-averaged results for all 12 pathologies. $\kappa_{\text{w}}$ denotes quadratically weighted Cohen's $\kappa$ computed on the original three-class labels (negative, uncertain, positive). $\kappa_{\text{low}}$ and $\kappa_{\text{high}}$ denote unweighted Cohen's $\kappa$ under pessimistic and optimistic uncertainty mappings, respectively. F1${\text{low}}$ and F1${\text{high}}$ are computed under the same mappings.}
\label{tab:mimic_images_ann1_vs_ann2}
\end{table}

\begin{table}[t]
\centering
\begin{tabular}{lrrrrrr}
\toprule
Pathology & $\kappa_{\text{w}}$ & $\kappa_{\text{low}}$ & $\kappa_{\text{high}}$ & F1 & F1$_{\text{low}}$ & F1$_{\text{high}}$ \\
\midrule
Atelectasis          & 0.20 & 0.11 & 0.27 & 0.72 & 0.72 & 0.79 \\
Consolidation        & 0.45 & 0.34 & 0.51 & 0.91 & 0.91 & 0.94 \\
Lung Opacity         & 0.38 & 0.38 & 0.38 & 0.79 & 0.79 & 0.79 \\
Pleural Effusion     & 0.33 & 0.26 & 0.38 & 0.72 & 0.72 & 0.78 \\
\cmidrule(lr){1-7}
Macro avg.\ (four selected) & 0.34 & 0.27 & 0.39 & 0.78 & 0.78 & 0.83 \\
Macro avg.\ (all 12) & 0.33 & 0.25 & 0.40 & 0.84 & 0.84 & 0.89 \\
Micro avg.\ (all 12) & 0.36 & 0.27 & 0.42 & 0.84 & 0.84 & 0.89 \\
\bottomrule
\end{tabular}
\caption{MIMIC-CXR agreement between labels from report findings and impressions sections in the four selected categories, including macro-average scores for the four selected categories and macro/micro-averaged results for all 12 pathologies. $\kappa_{\text{w}}$ denotes quadratically weighted Cohen's $\kappa$ computed on the original three-class labels (negative, uncertain, positive). $\kappa_{\text{low}}$ and $\kappa_{\text{high}}$ denote unweighted Cohen's $\kappa$ under pessimistic and optimistic uncertainty mappings, respectively. F1${\text{low}}$ and F1${\text{high}}$ are computed under the same mappings.}
\label{tab:mimic_reports_findings_vs_impressions}
\end{table}

\begin{longtable}{lrrrrrr}
\toprule
Pathology & $\kappa_{\text{w}}$ & $\kappa_{\text{low}}$ & $\kappa_{\text{high}}$ & F1 & F1$_{\text{low}}$ & F1$_{\text{high}}$ \\
\midrule
\endfirsthead
\toprule
Pathology & $\kappa_{\text{w}}$ & $\kappa_{\text{low}}$ & $\kappa_{\text{high}}$ & F1 & F1$_{\text{low}}$ & F1$_{\text{high}}$ \\
\midrule
\endhead
\midrule
\multicolumn{7}{r}{\textit{Continued on next page}}\\
\endfoot
\bottomrule\\
\caption{MIMIC-CXR agreement between image-only aggregation algorithm labels and report findings labels in the four selected categories, including macro-average scores for the four selected categories and macro/micro-averaged results for all 12 pathologies. $\kappa_{\text{w}}$ denotes quadratically weighted Cohen's $\kappa$ computed on the original three-class labels (negative, uncertain, positive). $\kappa_{\text{low}}$ and $\kappa_{\text{high}}$ denote unweighted Cohen's $\kappa$ under pessimistic and optimistic uncertainty mappings, respectively. F1${\text{low}}$ and F1${\text{high}}$ are computed under the same mappings.}
\label{tab:mimic_image_fusions_vs_findings}
\endlastfoot
\multicolumn{7}{c}{MACE} \\
\midrule
Atelectasis          & 0.17 & -0.07 & 0.37 & 0.57 & 0.54 & 0.75 \\
Consolidation        & 0.07 & -0.01 & 0.16 & 0.68 & 0.69 & 0.82 \\
Lung Opacity         & 0.19 & -0.07 & 0.42 & 0.59 & 0.54 & 0.76 \\
Pleural Effusion     & 0.44 &  0.16 & 0.63 & 0.67 & 0.64 & 0.85 \\
\cmidrule(lr){1-7}
Macro avg.\ (four selected) & 0.22 & 0.00 & 0.39 & 0.63 & 0.60 & 0.79 \\
Macro avg.\ (all 12) & 0.23 & 0.03 & 0.43 & 0.75 & 0.73 & 0.88 \\
Micro avg.\ (all 12) & 0.27 & 0.05 & 0.46 & 0.75 & 0.74 & 0.88 \\
\midrule
\multicolumn{7}{c}{FUSE\_OR} \\
\midrule
Atelectasis          & 0.17 & -0.05 & 0.36 & 0.57 & 0.55 & 0.74 \\
Consolidation        & 0.08 & -0.01 & 0.20 & 0.65 & 0.65 & 0.82 \\
Lung Opacity         & 0.21 & -0.10 & 0.47 & 0.58 & 0.52 & 0.78 \\
Pleural Effusion     & 0.45 &  0.25 & 0.57 & 0.69 & 0.67 & 0.82 \\
\cmidrule(lr){1-7}
Macro avg.\ (four selected) & 0.23 & 0.02 & 0.40 & 0.62 & 0.60 & 0.79 \\
Macro avg.\ (all 12) & 0.23 & 0.04 & 0.43 & 0.75 & 0.73 & 0.88 \\
Micro avg.\ (all 12) & 0.29 & 0.06 & 0.47 & 0.75 & 0.74 & 0.88 \\
\midrule
\multicolumn{7}{c}{NN Fusion} \\
\midrule
Atelectasis          & 0.12 & -0.12 & 0.34 & 0.60 & 0.57 & 0.77 \\
Consolidation        & 0.10 & -0.03 & 0.27 & 0.66 & 0.66 & 0.87 \\
Lung Opacity         & 0.20 & -0.12 & 0.47 & 0.59 & 0.53 & 0.79 \\
Pleural Effusion     & 0.44 &  0.24 & 0.57 & 0.70 & 0.68 & 0.83 \\
\cmidrule(lr){1-7}
Macro avg.\ (four selected) & 0.21 & -0.01 & 0.41 & 0.64 & 0.61 & 0.81 \\
Macro avg.\ (all 12) & 0.18 & 0.03 & 0.31 & 0.75 & 0.74 & 0.87 \\
Micro avg.\ (all 12) & 0.23 & 0.03 & 0.41 & 0.76 & 0.74 & 0.87 \\
\midrule
\multicolumn{7}{c}{NN Fusion (PS)} \\
\midrule
Atelectasis          & 0.14 & -0.09 & 0.35 & 0.61 & 0.58 & 0.77 \\
Consolidation        & 0.09 & -0.03 & 0.25 & 0.66 & 0.66 & 0.87 \\
Lung Opacity         & 0.21 & -0.09 & 0.47 & 0.59 & 0.53 & 0.79 \\
Pleural Effusion     & 0.44 &  0.22 & 0.57 & 0.69 & 0.66 & 0.83 \\
\cmidrule(lr){1-7}
Macro avg.\ (four selected) & 0.22 & 0.00 & 0.41 & 0.64 & 0.61 & 0.81 \\
Macro avg.\ (all 12) & 0.17 & 0.02 & 0.33 & 0.74 & 0.73 & 0.87 \\
Micro avg.\ (all 12) & 0.23 & 0.03 & 0.42 & 0.75 & 0.74 & 0.87 \\
\end{longtable}

\begin{longtable}{lrrrrrr}
\toprule
Pathology & $\kappa_{\text{w}}$ & $\kappa_{\text{low}}$ & $\kappa_{\text{high}}$ & F1 & F1$_{\text{low}}$ & F1$_{\text{high}}$ \\
\midrule
\endfirsthead
\toprule
Pathology & $\kappa_{\text{w}}$ & $\kappa_{\text{low}}$ & $\kappa_{\text{high}}$ & F1 & F1$_{\text{low}}$ & F1$_{\text{high}}$ \\
\midrule
\endhead
\midrule
\multicolumn{7}{r}{\textit{Continued on next page}}\\
\endfoot
\bottomrule\\
\caption{MIMIC-CXR agreement between image-only aggregation algorithm labels and report impressions labels in the four selected categories, including macro-average scores for the four selected categories and macro/micro-averaged results for all 12 pathologies. $\kappa_{\text{w}}$ denotes quadratically weighted Cohen's $\kappa$ computed on the original three-class labels (negative, uncertain, positive). $\kappa_{\text{low}}$ and $\kappa_{\text{high}}$ denote unweighted Cohen's $\kappa$ under pessimistic and optimistic uncertainty mappings, respectively. F1${\text{low}}$ and F1${\text{high}}$ are computed under the same mappings.}
\label{tab:mimic_image_fusions_vs_impressions}
\endlastfoot
\multicolumn{7}{c}{MACE} \\
\midrule
Atelectasis          & 0.14 & -0.04 & 0.31 & 0.59 & 0.59 & 0.78 \\
Consolidation        & 0.06 & -0.02 & 0.14 & 0.68 & 0.68 & 0.81 \\
Lung Opacity         & 0.14 & -0.08 & 0.35 & 0.62 & 0.58 & 0.78 \\
Pleural Effusion     & 0.36 &  0.12 & 0.54 & 0.66 & 0.63 & 0.82 \\
\cmidrule(lr){1-7}
Macro avg.\ (four selected) & 0.17 & -0.00 & 0.34 & 0.64 & 0.62 & 0.80 \\
Macro avg.\ (all 12) & 0.20 & 0.02 & 0.38 & 0.78 & 0.77 & 0.89 \\
Micro avg.\ (all 12) & 0.24 & 0.05 & 0.42 & 0.79 & 0.78 & 0.89 \\
\midrule
\multicolumn{7}{c}{FUSE\_OR} \\
\midrule
Atelectasis          & 0.13 & -0.04 & 0.31 & 0.59 & 0.58 & 0.78 \\
Consolidation        & 0.06 & -0.02 & 0.17 & 0.65 & 0.65 & 0.81 \\
Lung Opacity         & 0.13 & -0.10 & 0.37 & 0.59 & 0.55 & 0.79 \\
Pleural Effusion     & 0.35 &  0.18 & 0.47 & 0.66 & 0.64 & 0.79 \\
\cmidrule(lr){1-7}
Macro avg.\ (four selected) & 0.17 & 0.00 & 0.33 & 0.62 & 0.60 & 0.79 \\
Macro avg.\ (all 12) & 0.20 & 0.02 & 0.38 & 0.78 & 0.77 & 0.89 \\
Micro avg.\ (all 12) & 0.25 & 0.05 & 0.42 & 0.79 & 0.77 & 0.89 \\
\midrule
\multicolumn{7}{c}{NN Fusion} \\
\midrule
Atelectasis          & 0.10 & -0.12 & 0.36 & 0.68 & 0.66 & 0.86 \\
Consolidation        & 0.08 & -0.04 & 0.24 & 0.66 & 0.65 & 0.86 \\
Lung Opacity         & 0.13 & -0.13 & 0.41 & 0.62 & 0.58 & 0.81 \\
Pleural Effusion     & 0.35 &  0.18 & 0.46 & 0.68 & 0.67 & 0.80 \\
\cmidrule(lr){1-7}
Macro avg.\ (four selected) & 0.17 & -0.03 & 0.37 & 0.66 & 0.64 & 0.83 \\
Macro avg.\ (all 12) & 0.15 & 0.00 & 0.29 & 0.79 & 0.78 & 0.89 \\
Micro avg.\ (all 12) & 0.22 & 0.03 & 0.39 & 0.80 & 0.79 & 0.90 \\
\midrule
\multicolumn{7}{c}{NN Fusion (PS)} \\
\midrule
Atelectasis          & 0.12 & -0.10 & 0.37 & 0.68 & 0.67 & 0.85 \\
Consolidation        & 0.07 & -0.04 & 0.23 & 0.65 & 0.65 & 0.85 \\
Lung Opacity         & 0.13 & -0.11 & 0.39 & 0.62 & 0.57 & 0.80 \\
Pleural Effusion     & 0.35 &  0.16 & 0.48 & 0.67 & 0.65 & 0.80 \\
\cmidrule(lr){1-7}
Macro avg.\ (four selected) & 0.17 & -0.02 & 0.37 & 0.66 & 0.64 & 0.83 \\
Macro avg.\ (all 12) & 0.15 & -0.00 & 0.32 & 0.79 & 0.78 & 0.90 \\
Micro avg.\ (all 12) & 0.23 & 0.03 & 0.41 & 0.80 & 0.79 & 0.90 \\
\end{longtable}

\clearpage

\section{Supplementary Agreement and Rank Correlation Figures}

\begin{figure}[!htbp]
    \centering
    \includegraphics[width=0.82\textwidth]{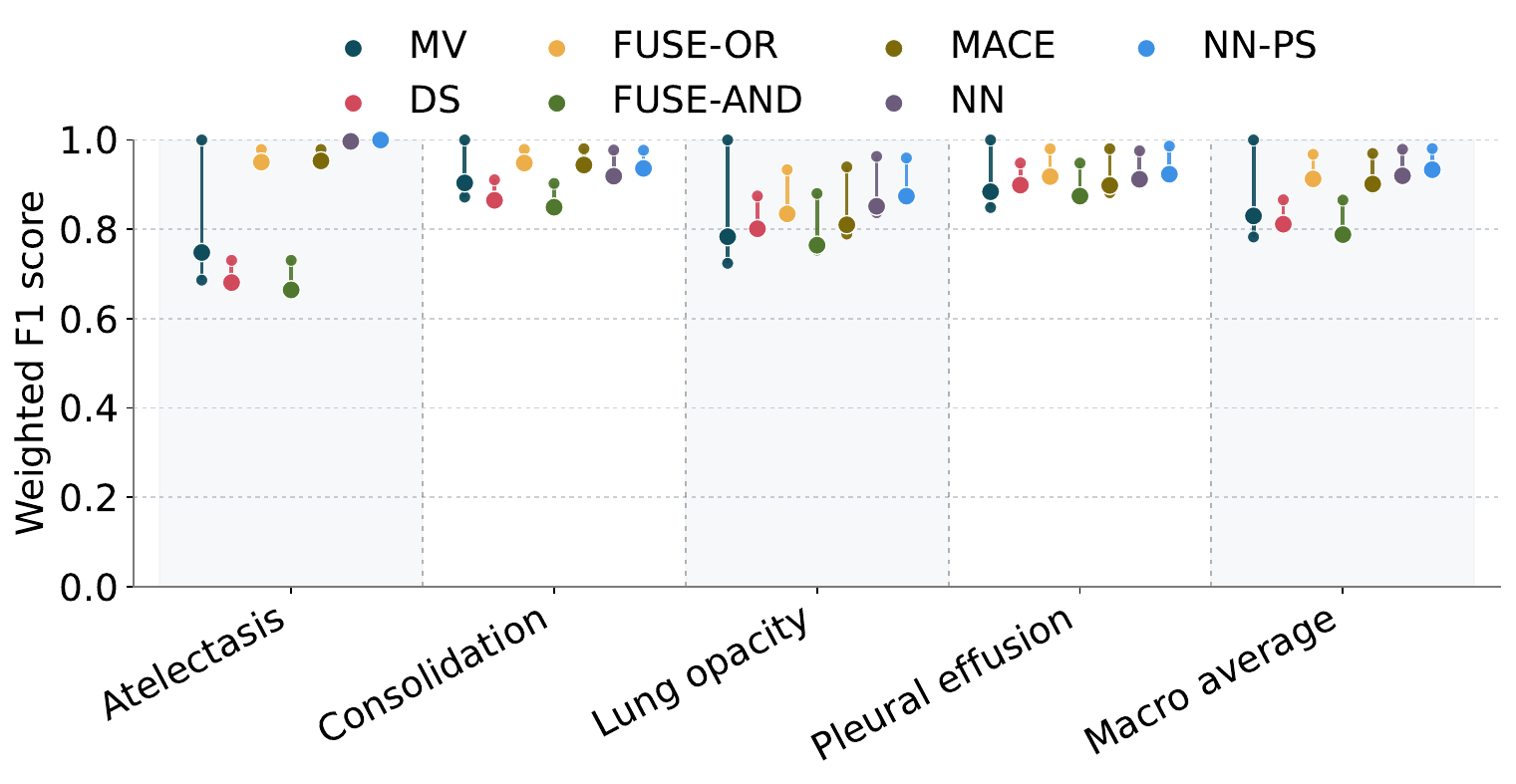}
    \caption{Agreement between each image-derived label fusion method and the XQA-RR conflict-resolved (CR) image-derived labels, measured by weighted F1. Intervals indicate the range induced by the pessimistic and optimistic uncertainty mappings. Macro-average scores are calculated across the four selected pathologies.}
    \label{fig:cuh_fusion_agreement_f1_app}
\end{figure}

\begin{figure}[!htbp]
    \centering
    \includegraphics[width=0.82\textwidth]{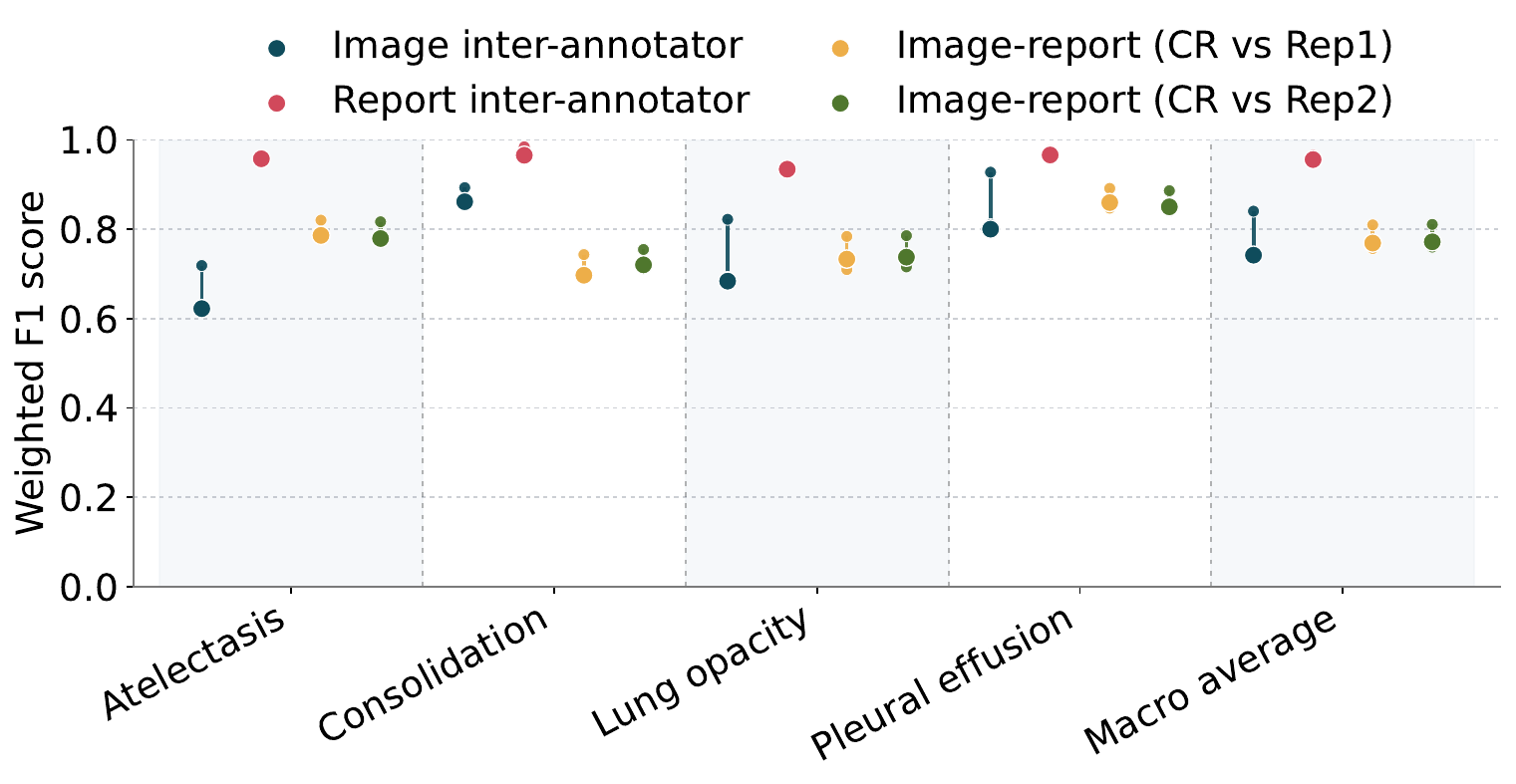}
    \caption{XQA-RR inter-annotator and cross-source label agreement measured by weighted F1. Intervals indicate the range induced by the pessimistic and optimistic uncertainty mappings. Macro-average scores are calculated across the four selected pathologies.}
    \label{fig:cuh_agreement_f1_app}
\end{figure}

\begin{figure}[!htbp]
    \centering
    \includegraphics[width=0.82\textwidth]{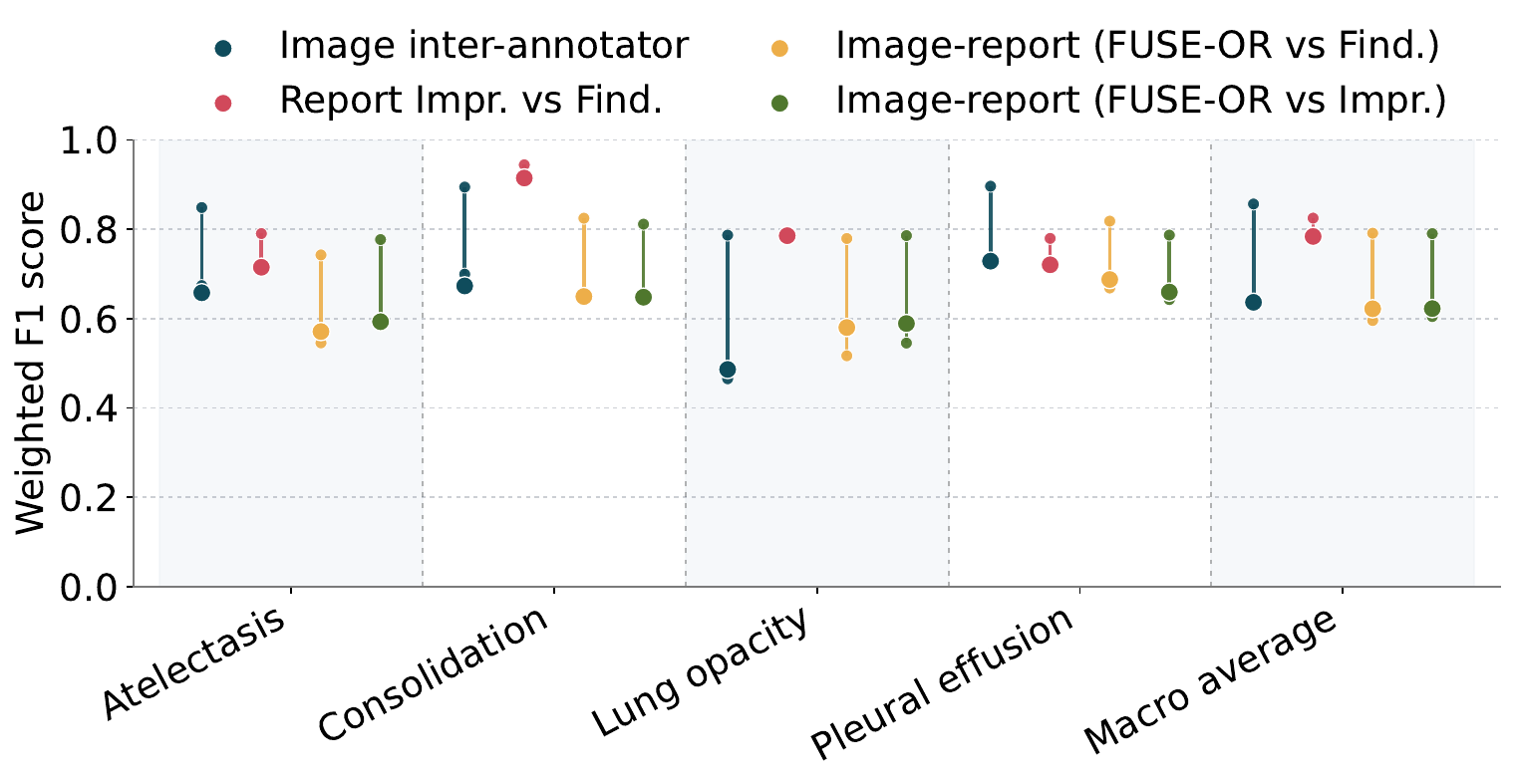}
    \caption{MIMIC-CXR inter-annotator and cross-source agreement measured by weighted F1. Intervals indicate the range induced by the pessimistic and optimistic uncertainty mappings. Macro-average scores are calculated across the four selected pathologies.}
    \label{fig:mimic_agreement_f1_app}
\end{figure}

\begin{figure}[!htbp]
    \centering
    \includegraphics[width=0.84\textwidth]{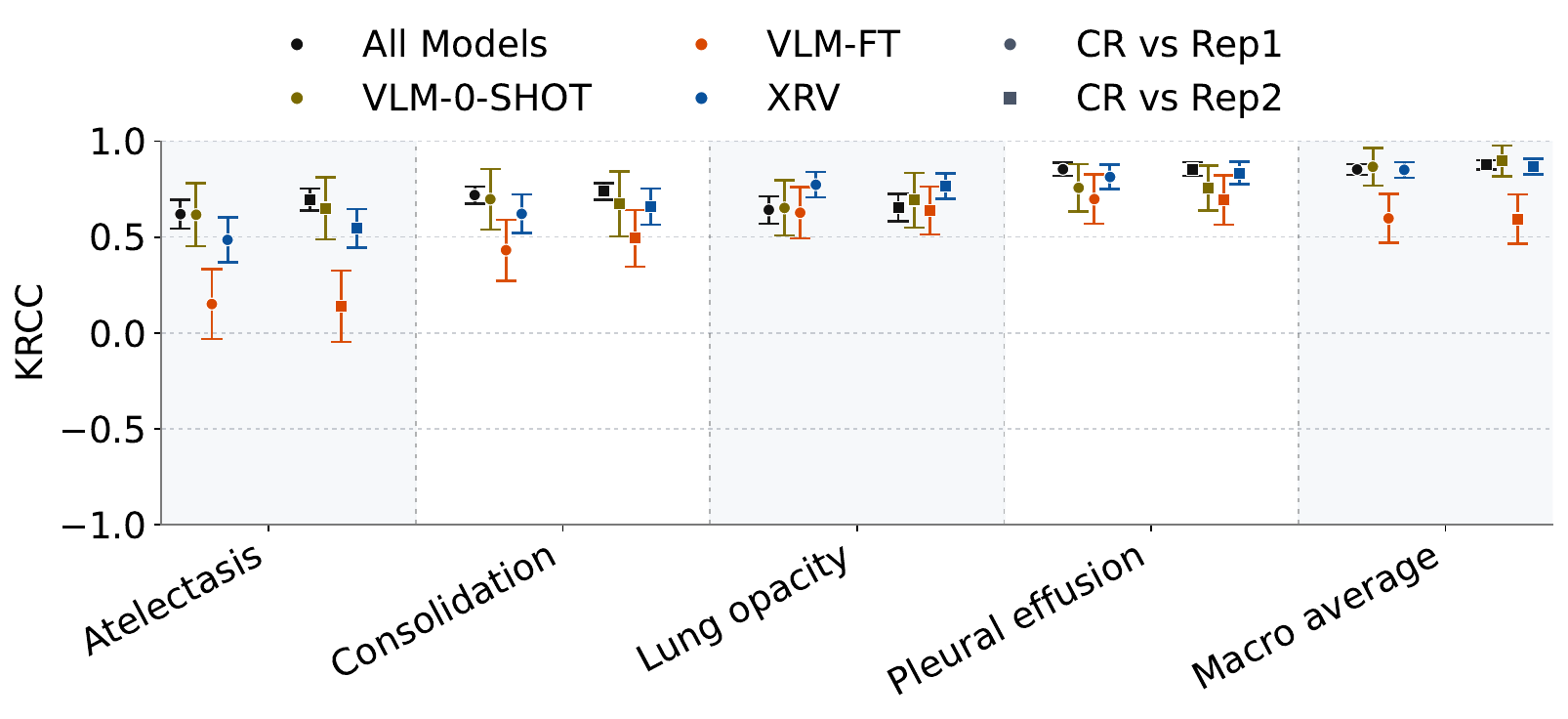}
    \caption{XQA-RR rank correlation coefficients between model rankings induced by each pair of label sources, measured by Kendall rank correlation coefficient (KRCC). CR denotes conflict-resolved image-derived labels; Rep1/Rep2 denote report-derived labels from the two report annotators. Results are broken down by pathology and model group. Macro average refers to rankings induced by ROC-AUC macro-averaged across the four selected pathologies.}
    \label{fig:cuh_rcc_krcc_app}
\end{figure}

\begin{figure}[!htbp]
    \centering
    \includegraphics[width=0.84\textwidth]{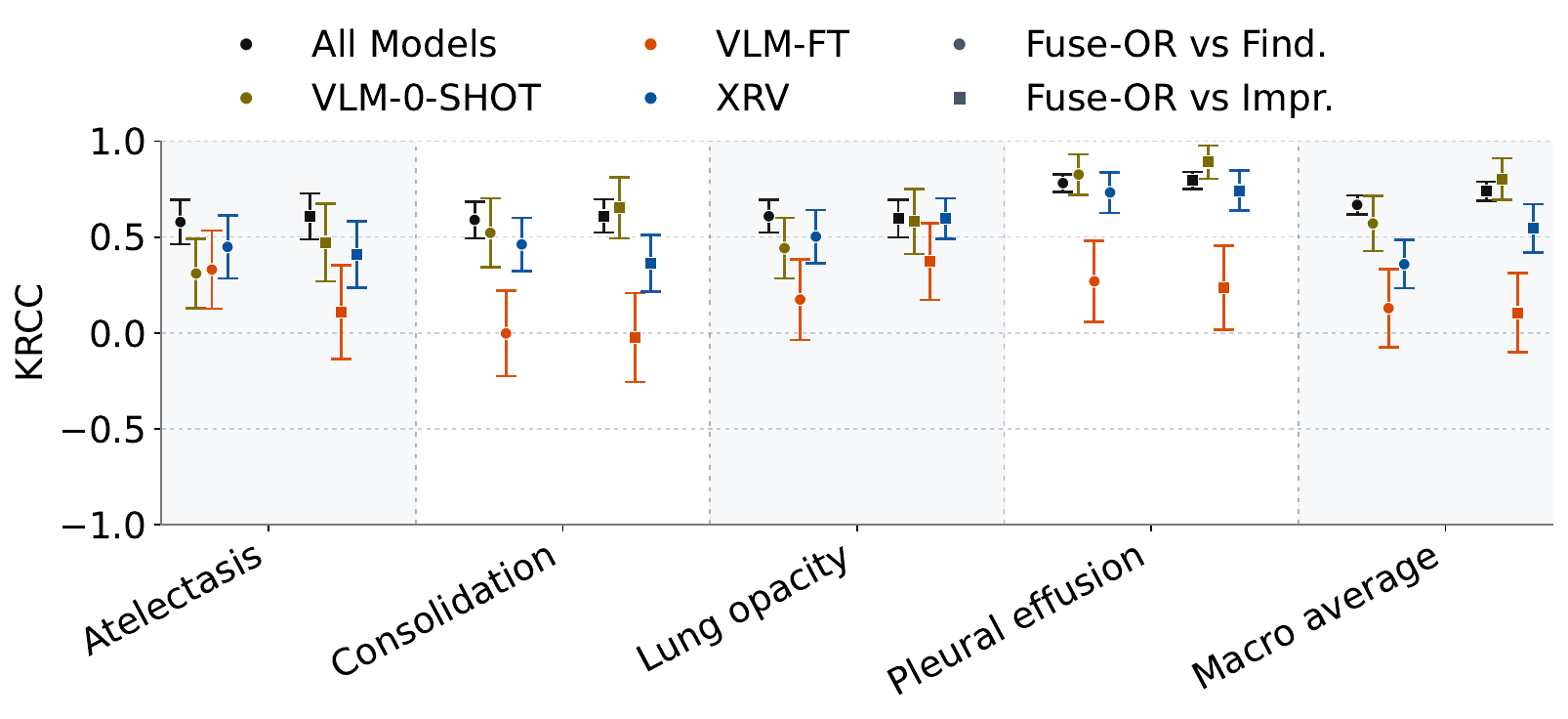}
    \caption{MIMIC-CXR rank correlation coefficients between model rankings induced by each pair of label sources, measured by Kendall rank correlation coefficient (KRCC). FUSE\_OR denotes automatically fused image-derived labels; Find.\ and Impr.\ denote labels derived from the Findings and Impressions sections of the radiology report. Results are broken down by pathology and model group. Macro average refers to rankings induced by ROC-AUC macro-averaged across the four selected pathologies.}
    \label{fig:mimic_rcc_krcc_app}
\end{figure}

\clearpage

\section{ROC-AUC Tables}

\begin{table}[H]
\centering
\begin{tabular}{lccc}
\hline
Model & CR & Rep1 & Rep2 \\
\hline
D-POOL & 0.70 (0.66, 0.73) & 0.67 (0.61, 0.73) & 0.71 (0.66, 0.76) \\
D-CheX & 0.65 (0.61, 0.69) & 0.60 (0.54, 0.66) & 0.61 (0.56, 0.67) \\
D-MC\_ch & 0.57 (0.53, 0.61) & 0.57 (0.51, 0.63) & 0.59 (0.53, 0.64) \\
D-MC\_nb & 0.58 (0.54, 0.62) & 0.59 (0.53, 0.66) & 0.61 (0.55, 0.67) \\
D-NIH & 0.62 (0.58, 0.66) & 0.64 (0.59, 0.70) & 0.67 (0.61, 0.72) \\
D-PC & 0.64 (0.60, 0.68) & 0.68 (0.63, 0.74) & 0.71 (0.66, 0.76) \\
D-RSNA & 0.50 (0.50, 0.50) & 0.50 (0.50, 0.50) & 0.50 (0.50, 0.50) \\
R-POOL & 0.67 (0.63, 0.71) & 0.71 (0.65, 0.76) & 0.75 (0.70, 0.80) \\
JFH & 0.73 (0.69, 0.76) & 0.68 (0.62, 0.74) & 0.73 (0.67, 0.78) \\
D-POOL-CUH & 0.72 (0.69, 0.76) & 0.70 (0.65, 0.75) & 0.74 (0.69, 0.79) \\
D-CheX-CUH & 0.68 (0.64, 0.72) & 0.60 (0.54, 0.66) & 0.65 (0.59, 0.70) \\
D-MC\_ch-CUH & 0.69 (0.65, 0.72) & 0.63 (0.57, 0.70) & 0.65 (0.59, 0.71) \\
D-MC\_nb-CUH & 0.68 (0.64, 0.72) & 0.64 (0.58, 0.70) & 0.67 (0.61, 0.73) \\
D-NIH-CUH & 0.70 (0.66, 0.74) & 0.66 (0.60, 0.73) & 0.70 (0.65, 0.76) \\
D-PC-CUH & 0.69 (0.65, 0.73) & 0.67 (0.61, 0.73) & 0.70 (0.64, 0.75) \\
D-RSNA-CUH & 0.64 (0.60, 0.68) & 0.59 (0.52, 0.65) & 0.65 (0.59, 0.71) \\
D-NORMAL & 0.67 (0.63, 0.71) & 0.66 (0.60, 0.72) & 0.66 (0.61, 0.72) \\
CONVIRT-CUH & 0.73 (0.70, 0.77) & 0.74 (0.68, 0.79) & 0.77 (0.72, 0.82) \\
GLORIA-G-CUH & 0.73 (0.69, 0.77) & 0.77 (0.71, 0.82) & 0.81 (0.76, 0.85) \\
R-MedCLIP-CUH & 0.74 (0.70, 0.77) & 0.74 (0.68, 0.79) & 0.76 (0.71, 0.81) \\
V-MedCLIP-CUH & 0.75 (0.71, 0.78) & 0.71 (0.66, 0.77) & 0.76 (0.71, 0.80) \\
MedKLIP-CUH & 0.73 (0.70, 0.76) & 0.72 (0.66, 0.77) & 0.76 (0.71, 0.81) \\
MFLAG-CUH & 0.61 (0.57, 0.65) & 0.56 (0.50, 0.62) & 0.63 (0.58, 0.69) \\
R-MGCA-CUH & 0.75 (0.71, 0.78) & 0.72 (0.67, 0.78) & 0.76 (0.70, 0.81) \\
MRM-CUH & 0.77 (0.73, 0.80) & 0.74 (0.69, 0.79) & 0.77 (0.72, 0.81) \\
V-MGCA-CUH & 0.73 (0.69, 0.77) & 0.76 (0.71, 0.81) & 0.79 (0.74, 0.83) \\
REFERS-CUH & 0.75 (0.72, 0.79) & 0.75 (0.69, 0.80) & 0.78 (0.73, 0.82) \\
CONVIRT & 0.41 (0.36, 0.45) & 0.50 (0.44, 0.57) & 0.49 (0.42, 0.55) \\
GLORIA-G & 0.61 (0.57, 0.66) & 0.66 (0.60, 0.71) & 0.67 (0.62, 0.72) \\
GLORIA-B & 0.63 (0.59, 0.67) & 0.68 (0.62, 0.74) & 0.66 (0.60, 0.71) \\
R-MGCA & 0.45 (0.41, 0.50) & 0.45 (0.39, 0.52) & 0.49 (0.43, 0.54) \\
V-MGCA & 0.48 (0.44, 0.53) & 0.48 (0.42, 0.54) & 0.51 (0.44, 0.58) \\
PTUNIFIER & 0.61 (0.56, 0.65) & 0.67 (0.61, 0.73) & 0.64 (0.59, 0.69) \\
REFERS & 0.61 (0.57, 0.65) & 0.66 (0.60, 0.72) & 0.63 (0.57, 0.68) \\
\hline
Mean & 0.65 (0.62, 0.69) & 0.65 (0.59, 0.71) & 0.68 (0.62, 0.73) \\
\hline
\end{tabular}
\caption{XQA-RR ROC-AUC for Atelectasis when evaluated against conflict-resolved (CR) image labels and labels from the two report annotators (Rep1 and Rep2). The mean scores represent the average ROC-AUC performance across all tested models.}
\label{tab:cuh_atelectasis_combined_auc}
\end{table}

\begin{table}[tbp]
\centering
\begin{tabular}{lccc}
\hline
Model & CR & Rep1 & Rep2 \\
\hline
D-POOL & 0.88 (0.85, 0.90) & 0.75 (0.72, 0.79) & 0.76 (0.72, 0.79) \\
D-CheX & 0.78 (0.75, 0.82) & 0.71 (0.67, 0.74) & 0.70 (0.67, 0.74) \\
D-MC\_ch & 0.66 (0.61, 0.71) & 0.68 (0.64, 0.72) & 0.66 (0.62, 0.71) \\
D-MC\_nb & 0.81 (0.78, 0.85) & 0.70 (0.66, 0.73) & 0.69 (0.65, 0.72) \\
D-NIH & 0.76 (0.72, 0.80) & 0.68 (0.64, 0.72) & 0.67 (0.63, 0.71) \\
D-PC & 0.82 (0.79, 0.85) & 0.71 (0.68, 0.75) & 0.71 (0.68, 0.75) \\
D-RSNA & 0.50 (0.50, 0.50) & 0.50 (0.50, 0.50) & 0.50 (0.50, 0.50) \\
R-POOL & 0.81 (0.78, 0.85) & 0.68 (0.64, 0.72) & 0.69 (0.65, 0.73) \\
JFH & 0.90 (0.88, 0.92) & 0.77 (0.73, 0.80) & 0.77 (0.74, 0.81) \\
D-POOL-CUH & 0.85 (0.82, 0.88) & 0.78 (0.75, 0.81) & 0.79 (0.76, 0.82) \\
D-CheX-CUH & 0.85 (0.82, 0.88) & 0.76 (0.73, 0.79) & 0.77 (0.74, 0.80) \\
D-MC\_ch-CUH & 0.82 (0.79, 0.85) & 0.75 (0.72, 0.79) & 0.75 (0.72, 0.78) \\
D-MC\_nb-CUH & 0.83 (0.80, 0.86) & 0.76 (0.72, 0.79) & 0.76 (0.72, 0.79) \\
D-NIH-CUH & 0.83 (0.79, 0.86) & 0.75 (0.71, 0.79) & 0.75 (0.71, 0.79) \\
D-PC-CUH & 0.85 (0.82, 0.87) & 0.77 (0.74, 0.80) & 0.77 (0.74, 0.81) \\
D-RSNA-CUH & 0.83 (0.80, 0.86) & 0.75 (0.71, 0.78) & 0.75 (0.71, 0.78) \\
D-NORMAL & 0.80 (0.76, 0.84) & 0.74 (0.70, 0.78) & 0.74 (0.71, 0.78) \\
CONVIRT-CUH & 0.87 (0.84, 0.89) & 0.79 (0.75, 0.82) & 0.79 (0.76, 0.82) \\
GLORIA-G-CUH & 0.87 (0.84, 0.89) & 0.78 (0.75, 0.82) & 0.79 (0.75, 0.82) \\
R-MedCLIP-CUH & 0.89 (0.86, 0.91) & 0.80 (0.77, 0.83) & 0.81 (0.77, 0.83) \\
V-MedCLIP-CUH & 0.86 (0.83, 0.89) & 0.80 (0.77, 0.83) & 0.80 (0.77, 0.83) \\
MedKLIP-CUH & 0.84 (0.81, 0.87) & 0.78 (0.74, 0.81) & 0.78 (0.75, 0.81) \\
MFLAG-CUH & 0.77 (0.73, 0.81) & 0.75 (0.72, 0.79) & 0.75 (0.71, 0.78) \\
R-MGCA-CUH & 0.85 (0.83, 0.88) & 0.78 (0.74, 0.81) & 0.78 (0.75, 0.81) \\
MRM-CUH & 0.85 (0.83, 0.88) & 0.80 (0.77, 0.83) & 0.80 (0.76, 0.83) \\
V-MGCA-CUH & 0.88 (0.85, 0.90) & 0.81 (0.77, 0.84) & 0.81 (0.78, 0.84) \\
REFERS-CUH & 0.86 (0.83, 0.88) & 0.78 (0.74, 0.81) & 0.78 (0.74, 0.81) \\
CONVIRT & 0.58 (0.54, 0.63) & 0.58 (0.54, 0.62) & 0.59 (0.55, 0.63) \\
GLORIA-G & 0.69 (0.65, 0.73) & 0.62 (0.58, 0.66) & 0.62 (0.58, 0.66) \\
GLORIA-B & 0.69 (0.64, 0.73) & 0.62 (0.58, 0.66) & 0.61 (0.57, 0.66) \\
R-MGCA & 0.40 (0.35, 0.45) & 0.45 (0.40, 0.49) & 0.44 (0.40, 0.49) \\
V-MGCA & 0.43 (0.38, 0.47) & 0.44 (0.40, 0.48) & 0.44 (0.40, 0.48) \\
PTUNIFIER & 0.71 (0.66, 0.75) & 0.64 (0.60, 0.67) & 0.64 (0.60, 0.68) \\
REFERS & 0.69 (0.64, 0.74) & 0.61 (0.56, 0.65) & 0.61 (0.57, 0.65) \\
\hline
Mean & 0.77 (0.74, 0.81) & 0.71 (0.67, 0.74) & 0.71 (0.67, 0.74) \\
\hline
\end{tabular}
\caption{XQA-RR ROC-AUC for Consolidation when evaluated against conflict-resolved (CR) image labels and labels from the two report annotators (Rep1 and Rep2). The mean scores represent the average ROC-AUC performance across all tested models.}
\label{tab:cuh_consolidation_combined_auc}
\end{table}

\begin{table}[tbp]
\centering
\begin{tabular}{lccc}
\hline
Model & CR & Rep1 & Rep2 \\
\hline
D-POOL & 0.85 (0.82, 0.87) & 0.75 (0.71, 0.78) & 0.75 (0.72, 0.79) \\
D-CheX & 0.71 (0.68, 0.75) & 0.61 (0.57, 0.65) & 0.60 (0.56, 0.64) \\
D-MC\_ch & 0.71 (0.67, 0.74) & 0.62 (0.58, 0.65) & 0.63 (0.59, 0.66) \\
D-MC\_nb & 0.81 (0.78, 0.84) & 0.67 (0.64, 0.71) & 0.69 (0.66, 0.73) \\
D-NIH & 0.50 (0.50, 0.50) & 0.50 (0.50, 0.50) & 0.50 (0.50, 0.50) \\
D-PC & 0.50 (0.50, 0.50) & 0.50 (0.50, 0.50) & 0.50 (0.50, 0.50) \\
D-RSNA & 0.82 (0.79, 0.85) & 0.75 (0.72, 0.78) & 0.75 (0.72, 0.79) \\
R-POOL & 0.67 (0.63, 0.71) & 0.61 (0.57, 0.65) & 0.60 (0.56, 0.65) \\
JFH & 0.50 (0.50, 0.50) & 0.50 (0.50, 0.50) & 0.50 (0.50, 0.50) \\
D-POOL-CUH & 0.86 (0.83, 0.88) & 0.80 (0.77, 0.83) & 0.82 (0.79, 0.85) \\
D-CheX-CUH & 0.82 (0.79, 0.85) & 0.77 (0.73, 0.80) & 0.78 (0.75, 0.81) \\
D-MC\_ch-CUH & 0.84 (0.81, 0.87) & 0.76 (0.73, 0.79) & 0.78 (0.75, 0.81) \\
D-MC\_nb-CUH & 0.84 (0.81, 0.87) & 0.77 (0.74, 0.81) & 0.79 (0.75, 0.82) \\
D-NIH-CUH & 0.83 (0.80, 0.86) & 0.77 (0.74, 0.80) & 0.79 (0.76, 0.82) \\
D-PC-CUH & 0.85 (0.83, 0.88) & 0.79 (0.76, 0.82) & 0.80 (0.77, 0.84) \\
D-RSNA-CUH & 0.86 (0.84, 0.89) & 0.78 (0.75, 0.81) & 0.79 (0.76, 0.82) \\
D-NORMAL & 0.81 (0.78, 0.84) & 0.75 (0.72, 0.78) & 0.76 (0.72, 0.79) \\
CONVIRT-CUH & 0.85 (0.82, 0.87) & 0.82 (0.79, 0.84) & 0.83 (0.80, 0.86) \\
GLORIA-G-CUH & 0.86 (0.84, 0.89) & 0.82 (0.79, 0.85) & 0.82 (0.80, 0.85) \\
R-MedCLIP-CUH & 0.86 (0.83, 0.88) & 0.83 (0.80, 0.86) & 0.83 (0.80, 0.86) \\
V-MedCLIP-CUH & 0.86 (0.84, 0.89) & 0.83 (0.80, 0.86) & 0.83 (0.81, 0.86) \\
MedKLIP-CUH & 0.84 (0.81, 0.87) & 0.80 (0.77, 0.83) & 0.80 (0.77, 0.83) \\
MFLAG-CUH & 0.79 (0.76, 0.82) & 0.71 (0.68, 0.75) & 0.71 (0.67, 0.75) \\
R-MGCA-CUH & 0.80 (0.77, 0.83) & 0.80 (0.77, 0.83) & 0.79 (0.76, 0.82) \\
MRM-CUH & 0.87 (0.85, 0.90) & 0.83 (0.80, 0.86) & 0.85 (0.82, 0.87) \\
V-MGCA-CUH & 0.84 (0.81, 0.87) & 0.82 (0.79, 0.84) & 0.82 (0.79, 0.84) \\
REFERS-CUH & 0.85 (0.83, 0.88) & 0.82 (0.79, 0.85) & 0.83 (0.80, 0.86) \\
CONVIRT & 0.82 (0.79, 0.85) & 0.77 (0.74, 0.80) & 0.79 (0.76, 0.82) \\
GLORIA-G & 0.84 (0.82, 0.87) & 0.81 (0.78, 0.84) & 0.82 (0.79, 0.85) \\
GLORIA-B & 0.83 (0.80, 0.86) & 0.82 (0.79, 0.84) & 0.83 (0.80, 0.86) \\
R-MGCA & 0.49 (0.45, 0.53) & 0.53 (0.49, 0.57) & 0.54 (0.50, 0.58) \\
V-MGCA & 0.47 (0.43, 0.51) & 0.47 (0.43, 0.51) & 0.48 (0.44, 0.53) \\
PTUNIFIER & 0.82 (0.80, 0.85) & 0.82 (0.79, 0.85) & 0.83 (0.80, 0.86) \\
REFERS & 0.79 (0.76, 0.82) & 0.78 (0.75, 0.81) & 0.78 (0.74, 0.81) \\
\hline
Mean & 0.77 (0.75, 0.80) & 0.73 (0.70, 0.76) & 0.74 (0.71, 0.77) \\
\hline
\end{tabular}
\caption{XQA-RR ROC-AUC for Lung opacity when evaluated against conflict-resolved (CR) image labels and labels from the two report annotators (Rep1 and Rep2). The mean scores represent the average ROC-AUC performance across all tested models.}
\label{tab:cuh_lung_opacity_combined_auc}
\end{table}

\begin{table}[tbp]
\centering
\begin{tabular}{lccc}
\hline
Model & CR & Rep1 & Rep2 \\
\hline
D-POOL & 0.86 (0.82, 0.89) & 0.83 (0.79, 0.87) & 0.82 (0.78, 0.86) \\
D-CheX & 0.75 (0.71, 0.78) & 0.74 (0.69, 0.78) & 0.73 (0.69, 0.77) \\
D-MC\_ch & 0.70 (0.66, 0.74) & 0.73 (0.68, 0.77) & 0.70 (0.66, 0.75) \\
D-MC\_nb & 0.70 (0.66, 0.74) & 0.69 (0.64, 0.74) & 0.67 (0.62, 0.72) \\
D-NIH & 0.80 (0.76, 0.83) & 0.77 (0.73, 0.81) & 0.74 (0.70, 0.78) \\
D-PC & 0.86 (0.83, 0.88) & 0.82 (0.78, 0.86) & 0.82 (0.78, 0.86) \\
D-RSNA & 0.50 (0.50, 0.50) & 0.50 (0.50, 0.50) & 0.50 (0.50, 0.50) \\
R-POOL & 0.88 (0.85, 0.91) & 0.83 (0.79, 0.87) & 0.82 (0.78, 0.86) \\
JFH & 0.91 (0.89, 0.93) & 0.86 (0.83, 0.89) & 0.85 (0.82, 0.89) \\
D-POOL-CUH & 0.89 (0.86, 0.91) & 0.86 (0.83, 0.89) & 0.85 (0.81, 0.88) \\
D-CheX-CUH & 0.85 (0.82, 0.88) & 0.82 (0.78, 0.86) & 0.81 (0.76, 0.85) \\
D-MC\_ch-CUH & 0.84 (0.80, 0.86) & 0.82 (0.78, 0.85) & 0.79 (0.75, 0.83) \\
D-MC\_nb-CUH & 0.83 (0.80, 0.86) & 0.82 (0.78, 0.86) & 0.80 (0.76, 0.84) \\
D-NIH-CUH & 0.85 (0.82, 0.88) & 0.82 (0.78, 0.86) & 0.80 (0.76, 0.84) \\
D-PC-CUH & 0.87 (0.85, 0.90) & 0.85 (0.82, 0.89) & 0.84 (0.81, 0.88) \\
D-RSNA-CUH & 0.80 (0.77, 0.84) & 0.79 (0.74, 0.83) & 0.78 (0.73, 0.83) \\
D-NORMAL & 0.81 (0.78, 0.84) & 0.79 (0.75, 0.83) & 0.78 (0.74, 0.82) \\
CONVIRT-CUH & 0.93 (0.91, 0.94) & 0.89 (0.86, 0.92) & 0.88 (0.85, 0.91) \\
GLORIA-G-CUH & 0.91 (0.89, 0.94) & 0.87 (0.84, 0.91) & 0.87 (0.83, 0.90) \\
R-MedCLIP-CUH & 0.91 (0.89, 0.93) & 0.88 (0.85, 0.91) & 0.87 (0.83, 0.90) \\
V-MedCLIP-CUH & 0.93 (0.91, 0.95) & 0.91 (0.88, 0.93) & 0.89 (0.86, 0.92) \\
MedKLIP-CUH & 0.89 (0.87, 0.92) & 0.87 (0.84, 0.90) & 0.86 (0.82, 0.89) \\
MFLAG-CUH & 0.81 (0.77, 0.84) & 0.80 (0.76, 0.84) & 0.77 (0.73, 0.82) \\
R-MGCA-CUH & 0.88 (0.86, 0.91) & 0.86 (0.82, 0.89) & 0.85 (0.82, 0.89) \\
MRM-CUH & 0.93 (0.90, 0.94) & 0.91 (0.88, 0.93) & 0.89 (0.86, 0.92) \\
V-MGCA-CUH & 0.92 (0.90, 0.94) & 0.89 (0.86, 0.92) & 0.88 (0.85, 0.91) \\
REFERS-CUH & 0.92 (0.89, 0.94) & 0.88 (0.85, 0.91) & 0.87 (0.83, 0.90) \\
CONVIRT & 0.71 (0.66, 0.75) & 0.76 (0.71, 0.81) & 0.75 (0.70, 0.80) \\
GLORIA-G & 0.79 (0.76, 0.82) & 0.77 (0.73, 0.81) & 0.77 (0.72, 0.80) \\
GLORIA-B & 0.72 (0.68, 0.76) & 0.71 (0.66, 0.76) & 0.70 (0.65, 0.74) \\
R-MGCA & 0.57 (0.52, 0.61) & 0.56 (0.50, 0.61) & 0.55 (0.50, 0.60) \\
V-MGCA & 0.58 (0.53, 0.62) & 0.58 (0.53, 0.63) & 0.57 (0.51, 0.62) \\
PTUNIFIER & 0.83 (0.80, 0.87) & 0.83 (0.79, 0.87) & 0.82 (0.78, 0.86) \\
REFERS & 0.68 (0.63, 0.73) & 0.74 (0.69, 0.79) & 0.74 (0.69, 0.79) \\
\hline
Mean & 0.81 (0.78, 0.84) & 0.80 (0.76, 0.83) & 0.78 (0.74, 0.82) \\
\hline
\end{tabular}
\caption{XQA-RR ROC-AUC for Pleural effusion when evaluated against conflict-resolved (CR) image labels and labels from the two report annotators (Rep1 and Rep2). The mean scores represent the average ROC-AUC performance across all tested models.}
\label{tab:cuh_pleural_effusion_combined_auc}
\end{table}

\begin{table}[tbp]
\centering
\begin{tabular}{lccc}
\hline
Model & CR & Rep1 & Rep2 \\
\hline
D-POOL & 0.82 (0.79, 0.85) & 0.75 (0.71, 0.79) & 0.76 (0.72, 0.80) \\
D-CheX & 0.72 (0.69, 0.76) & 0.66 (0.62, 0.71) & 0.66 (0.62, 0.70) \\
D-MC\_ch & 0.66 (0.62, 0.70) & 0.65 (0.60, 0.70) & 0.65 (0.60, 0.69) \\
D-MC\_nb & 0.73 (0.69, 0.76) & 0.66 (0.62, 0.71) & 0.66 (0.62, 0.71) \\
D-NIH & 0.67 (0.64, 0.70) & 0.65 (0.61, 0.68) & 0.64 (0.61, 0.68) \\
D-PC & 0.70 (0.68, 0.73) & 0.68 (0.64, 0.71) & 0.69 (0.65, 0.72) \\
D-RSNA & 0.58 (0.57, 0.59) & 0.56 (0.55, 0.57) & 0.56 (0.55, 0.57) \\
R-POOL & 0.76 (0.72, 0.79) & 0.71 (0.66, 0.75) & 0.72 (0.67, 0.76) \\
JFH & 0.76 (0.74, 0.78) & 0.70 (0.67, 0.73) & 0.71 (0.68, 0.74) \\
D-POOL-CUH & 0.83 (0.80, 0.86) & 0.79 (0.75, 0.82) & 0.80 (0.76, 0.84) \\
D-CheX-CUH & 0.80 (0.77, 0.83) & 0.74 (0.69, 0.78) & 0.75 (0.71, 0.79) \\
D-MC\_ch-CUH & 0.80 (0.76, 0.83) & 0.74 (0.70, 0.78) & 0.74 (0.70, 0.78) \\
D-MC\_nb-CUH & 0.80 (0.76, 0.83) & 0.75 (0.71, 0.79) & 0.75 (0.71, 0.79) \\
D-NIH-CUH & 0.80 (0.77, 0.83) & 0.75 (0.71, 0.79) & 0.76 (0.72, 0.80) \\
D-PC-CUH & 0.82 (0.79, 0.85) & 0.77 (0.73, 0.81) & 0.78 (0.74, 0.82) \\
D-RSNA-CUH & 0.78 (0.75, 0.82) & 0.73 (0.68, 0.77) & 0.74 (0.70, 0.79) \\
D-NORMAL & 0.77 (0.74, 0.81) & 0.74 (0.69, 0.78) & 0.74 (0.69, 0.78) \\
CONVIRT-CUH & 0.84 (0.82, 0.87) & 0.81 (0.77, 0.84) & 0.82 (0.78, 0.85) \\
GLORIA-G-CUH & 0.84 (0.82, 0.87) & 0.81 (0.77, 0.85) & 0.82 (0.79, 0.86) \\
R-MedCLIP-CUH & 0.85 (0.82, 0.87) & 0.81 (0.78, 0.85) & 0.82 (0.78, 0.85) \\
V-MedCLIP-CUH & 0.85 (0.82, 0.88) & 0.81 (0.78, 0.85) & 0.82 (0.79, 0.86) \\
MedKLIP-CUH & 0.83 (0.80, 0.86) & 0.79 (0.75, 0.83) & 0.80 (0.76, 0.83) \\
MFLAG-CUH & 0.74 (0.71, 0.78) & 0.71 (0.66, 0.75) & 0.71 (0.67, 0.76) \\
R-MGCA-CUH & 0.82 (0.79, 0.85) & 0.79 (0.75, 0.83) & 0.79 (0.76, 0.83) \\
MRM-CUH & 0.86 (0.83, 0.88) & 0.82 (0.79, 0.85) & 0.83 (0.79, 0.86) \\
V-MGCA-CUH & 0.84 (0.81, 0.87) & 0.82 (0.78, 0.85) & 0.83 (0.79, 0.86) \\
REFERS-CUH & 0.84 (0.82, 0.87) & 0.81 (0.77, 0.84) & 0.81 (0.78, 0.85) \\
CONVIRT & 0.63 (0.59, 0.67) & 0.65 (0.61, 0.70) & 0.65 (0.61, 0.70) \\
GLORIA-G & 0.73 (0.70, 0.77) & 0.71 (0.67, 0.75) & 0.72 (0.68, 0.76) \\
GLORIA-B & 0.72 (0.68, 0.76) & 0.71 (0.66, 0.75) & 0.70 (0.66, 0.74) \\
R-MGCA & 0.48 (0.43, 0.52) & 0.50 (0.45, 0.55) & 0.50 (0.46, 0.55) \\
V-MGCA & 0.49 (0.44, 0.53) & 0.49 (0.44, 0.54) & 0.50 (0.45, 0.55) \\
PTUNIFIER & 0.74 (0.70, 0.78) & 0.74 (0.70, 0.78) & 0.73 (0.69, 0.77) \\
REFERS & 0.69 (0.65, 0.73) & 0.70 (0.65, 0.74) & 0.69 (0.64, 0.73) \\
\hline
Mean & 0.75 (0.72, 0.78) & 0.72 (0.68, 0.76) & 0.72 (0.69, 0.76) \\
\hline
\end{tabular}
\caption{XQA-RR ROC-AUC macro averages across the four selected pathologies (Atelectasis, Consolidation, Lung opacity, and Pleural effusion) when evaluated against conflict-resolved (CR) image labels and labels from the two report annotators (Rep1 and Rep2). The mean scores represent the average ROC-AUC performance across all tested models.}
\label{tab:cuh_s_overall_1_combined_auc}
\end{table}

\begin{table}[tbp]
\centering
\begin{tabular}{lccc}
\hline
Model & FUSE-OR & Impression & Findings \\
\hline
D-POOL & 0.67 (0.64, 0.70) & 0.64 (0.59, 0.69) & 0.61 (0.58, 0.65) \\
D-CheX & 0.63 (0.60, 0.67) & 0.62 (0.57, 0.67) & 0.58 (0.54, 0.61) \\
D-MC\_ch & 0.63 (0.60, 0.67) & 0.62 (0.57, 0.67) & 0.60 (0.57, 0.64) \\
D-MC\_nb & 0.61 (0.58, 0.65) & 0.60 (0.55, 0.65) & 0.59 (0.55, 0.62) \\
D-NIH & 0.69 (0.66, 0.72) & 0.59 (0.54, 0.65) & 0.62 (0.58, 0.65) \\
D-PC & 0.64 (0.61, 0.67) & 0.62 (0.57, 0.67) & 0.65 (0.61, 0.69) \\
D-RSNA & 0.50 (0.50, 0.50) & 0.50 (0.50, 0.50) & 0.50 (0.50, 0.50) \\
R-POOL & 0.64 (0.61, 0.67) & 0.62 (0.57, 0.67) & 0.65 (0.62, 0.69) \\
JFH & 0.69 (0.66, 0.72) & 0.65 (0.60, 0.70) & 0.66 (0.62, 0.70) \\
CONVIRT-MIMIC & 0.71 (0.68, 0.75) & 0.64 (0.60, 0.69) & 0.62 (0.58, 0.66) \\
GLORIA-G-MIMIC & 0.73 (0.70, 0.77) & 0.65 (0.60, 0.69) & 0.64 (0.61, 0.68) \\
R-MedCLIP-MIMIC & 0.72 (0.68, 0.75) & 0.63 (0.58, 0.68) & 0.63 (0.59, 0.67) \\
V-MedCLIP-MIMIC & 0.74 (0.71, 0.78) & 0.65 (0.60, 0.70) & 0.65 (0.61, 0.69) \\
MedKLIP-MIMIC & 0.72 (0.69, 0.76) & 0.63 (0.58, 0.69) & 0.63 (0.59, 0.67) \\
MFLAG-MIMIC & 0.72 (0.68, 0.75) & 0.65 (0.60, 0.69) & 0.65 (0.61, 0.69) \\
R-MGCA-MIMIC & 0.72 (0.69, 0.75) & 0.64 (0.59, 0.69) & 0.64 (0.60, 0.68) \\
MRM-MIMIC & 0.74 (0.70, 0.77) & 0.64 (0.59, 0.69) & 0.66 (0.62, 0.70) \\
V-MGCA-MIMIC & 0.73 (0.69, 0.76) & 0.64 (0.59, 0.69) & 0.66 (0.62, 0.69) \\
REFERS-MIMIC & 0.72 (0.69, 0.76) & 0.65 (0.60, 0.70) & 0.65 (0.61, 0.69) \\
CONVIRT & 0.41 (0.37, 0.45) & 0.49 (0.44, 0.55) & 0.50 (0.46, 0.54) \\
GLORIA-G & 0.55 (0.52, 0.59) & 0.52 (0.48, 0.57) & 0.61 (0.57, 0.65) \\
GLORIA-B & 0.55 (0.51, 0.59) & 0.56 (0.51, 0.61) & 0.60 (0.55, 0.64) \\
R-MGCA & 0.47 (0.43, 0.51) & 0.51 (0.46, 0.57) & 0.50 (0.45, 0.54) \\
V-MGCA & 0.52 (0.48, 0.56) & 0.53 (0.48, 0.58) & 0.49 (0.45, 0.53) \\
PTUNIFIER & 0.43 (0.39, 0.47) & 0.52 (0.47, 0.57) & 0.56 (0.52, 0.60) \\
REFERS & 0.41 (0.37, 0.46) & 0.43 (0.37, 0.48) & 0.53 (0.48, 0.57) \\
\hline
Mean & 0.63 (0.59, 0.66) & 0.59 (0.55, 0.64) & 0.60 (0.56, 0.64) \\
\hline
\end{tabular}
\caption{MIMIC-CXR ROC-AUC results for Atelectasis when evaluated against FUSE-OR image labels and labels from the findings and impressions report sections. The mean scores represent the average ROC-AUC performance across all tested models.}
\label{tab:mimic_atelectasis_combined_auc}
\end{table}

\begin{table}[tbp]
\centering
\begin{tabular}{lccc}
\hline
Model & FUSE-OR & Impression & Findings \\
\hline
D-POOL & 0.74 (0.71, 0.77) & 0.66 (0.59, 0.72) & 0.59 (0.52, 0.66) \\
D-CheX & 0.68 (0.65, 0.72) & 0.55 (0.49, 0.62) & 0.50 (0.43, 0.57) \\
D-MC\_ch & 0.71 (0.68, 0.74) & 0.66 (0.59, 0.72) & 0.62 (0.54, 0.70) \\
D-MC\_nb & 0.72 (0.69, 0.75) & 0.65 (0.59, 0.70) & 0.63 (0.56, 0.70) \\
D-NIH & 0.68 (0.65, 0.71) & 0.57 (0.49, 0.64) & 0.52 (0.43, 0.59) \\
D-PC & 0.72 (0.69, 0.75) & 0.68 (0.61, 0.74) & 0.67 (0.60, 0.74) \\
D-RSNA & 0.50 (0.50, 0.50) & 0.50 (0.50, 0.50) & 0.50 (0.50, 0.50) \\
R-POOL & 0.66 (0.63, 0.69) & 0.69 (0.62, 0.75) & 0.59 (0.51, 0.66) \\
JFH & 0.78 (0.75, 0.81) & 0.66 (0.59, 0.72) & 0.64 (0.57, 0.72) \\
CONVIRT-MIMIC & 0.79 (0.76, 0.82) & 0.72 (0.66, 0.78) & 0.71 (0.64, 0.78) \\
GLORIA-G-MIMIC & 0.81 (0.78, 0.84) & 0.71 (0.65, 0.77) & 0.70 (0.62, 0.77) \\
R-MedCLIP-MIMIC & 0.81 (0.78, 0.84) & 0.74 (0.67, 0.80) & 0.72 (0.64, 0.79) \\
V-MedCLIP-MIMIC & 0.80 (0.77, 0.83) & 0.75 (0.69, 0.81) & 0.74 (0.66, 0.81) \\
MedKLIP-MIMIC & 0.82 (0.79, 0.85) & 0.71 (0.65, 0.77) & 0.70 (0.63, 0.77) \\
MFLAG-MIMIC & 0.80 (0.77, 0.83) & 0.73 (0.66, 0.79) & 0.70 (0.62, 0.77) \\
R-MGCA-MIMIC & 0.82 (0.79, 0.84) & 0.75 (0.69, 0.81) & 0.71 (0.63, 0.78) \\
MRM-MIMIC & 0.81 (0.78, 0.84) & 0.74 (0.68, 0.80) & 0.74 (0.67, 0.81) \\
V-MGCA-MIMIC & 0.80 (0.77, 0.83) & 0.73 (0.66, 0.79) & 0.69 (0.61, 0.76) \\
REFERS-MIMIC & 0.81 (0.78, 0.83) & 0.76 (0.70, 0.82) & 0.73 (0.66, 0.80) \\
CONVIRT & 0.58 (0.54, 0.62) & 0.56 (0.49, 0.63) & 0.61 (0.53, 0.70) \\
GLORIA-G & 0.62 (0.58, 0.66) & 0.65 (0.57, 0.71) & 0.61 (0.53, 0.67) \\
GLORIA-B & 0.60 (0.56, 0.64) & 0.66 (0.59, 0.73) & 0.58 (0.50, 0.66) \\
R-MGCA & 0.44 (0.40, 0.49) & 0.45 (0.38, 0.53) & 0.47 (0.39, 0.55) \\
V-MGCA & 0.45 (0.41, 0.49) & 0.41 (0.35, 0.48) & 0.36 (0.29, 0.44) \\
PTUNIFIER & 0.73 (0.70, 0.77) & 0.69 (0.62, 0.75) & 0.65 (0.58, 0.72) \\
REFERS & 0.58 (0.54, 0.63) & 0.65 (0.57, 0.72) & 0.66 (0.59, 0.74) \\
\hline
Mean & 0.70 (0.67, 0.73) & 0.66 (0.59, 0.72) & 0.63 (0.56, 0.70) \\
\hline
\end{tabular}
\caption{MIMIC-CXR ROC-AUC results for Consolidation when evaluated against FUSE-OR image labels and labels from the findings and impressions report sections. The mean scores represent the average ROC-AUC performance across all tested models.}
\label{tab:mimic_consolidation_combined_auc}
\end{table}

\begin{table}[tbp]
\centering
\begin{tabular}{lccc}
\hline
Model & FUSE-OR & Impression & Findings \\
\hline
D-POOL & 0.74 (0.71, 0.77) & 0.62 (0.57, 0.66) & 0.56 (0.52, 0.60) \\
D-CheX & 0.69 (0.65, 0.72) & 0.53 (0.49, 0.57) & 0.54 (0.50, 0.58) \\
D-MC\_ch & 0.70 (0.67, 0.73) & 0.57 (0.53, 0.62) & 0.56 (0.52, 0.60) \\
D-MC\_nb & 0.72 (0.69, 0.75) & 0.58 (0.54, 0.63) & 0.57 (0.52, 0.60) \\
D-NIH & 0.50 (0.50, 0.50) & 0.50 (0.50, 0.50) & 0.50 (0.50, 0.50) \\
D-PC & 0.50 (0.50, 0.50) & 0.50 (0.50, 0.50) & 0.50 (0.50, 0.50) \\
D-RSNA & 0.70 (0.67, 0.73) & 0.58 (0.54, 0.63) & 0.56 (0.52, 0.60) \\
R-POOL & 0.66 (0.63, 0.69) & 0.65 (0.61, 0.70) & 0.59 (0.56, 0.63) \\
JFH & 0.50 (0.50, 0.50) & 0.50 (0.50, 0.50) & 0.50 (0.50, 0.50) \\
CONVIRT-MIMIC & 0.79 (0.75, 0.82) & 0.64 (0.60, 0.68) & 0.63 (0.60, 0.67) \\
GLORIA-G-MIMIC & 0.80 (0.76, 0.83) & 0.65 (0.61, 0.70) & 0.62 (0.58, 0.66) \\
R-MedCLIP-MIMIC & 0.79 (0.75, 0.82) & 0.65 (0.61, 0.69) & 0.64 (0.60, 0.67) \\
V-MedCLIP-MIMIC & 0.79 (0.76, 0.82) & 0.65 (0.60, 0.69) & 0.64 (0.60, 0.67) \\
MedKLIP-MIMIC & 0.79 (0.76, 0.83) & 0.64 (0.59, 0.68) & 0.62 (0.58, 0.65) \\
MFLAG-MIMIC & 0.78 (0.74, 0.81) & 0.64 (0.60, 0.68) & 0.61 (0.57, 0.64) \\
R-MGCA-MIMIC & 0.80 (0.77, 0.84) & 0.67 (0.63, 0.71) & 0.63 (0.59, 0.66) \\
MRM-MIMIC & 0.80 (0.77, 0.83) & 0.67 (0.63, 0.72) & 0.65 (0.61, 0.69) \\
V-MGCA-MIMIC & 0.77 (0.74, 0.81) & 0.63 (0.59, 0.67) & 0.62 (0.59, 0.66) \\
REFERS-MIMIC & 0.78 (0.74, 0.81) & 0.65 (0.60, 0.69) & 0.63 (0.59, 0.66) \\
CONVIRT & 0.73 (0.69, 0.77) & 0.65 (0.60, 0.69) & 0.63 (0.59, 0.66) \\
GLORIA-G & 0.74 (0.70, 0.78) & 0.64 (0.60, 0.69) & 0.58 (0.54, 0.61) \\
GLORIA-B & 0.72 (0.68, 0.76) & 0.65 (0.61, 0.70) & 0.61 (0.57, 0.65) \\
R-MGCA & 0.50 (0.46, 0.55) & 0.48 (0.44, 0.52) & 0.48 (0.44, 0.52) \\
V-MGCA & 0.44 (0.39, 0.48) & 0.50 (0.46, 0.55) & 0.45 (0.41, 0.48) \\
PTUNIFIER & 0.72 (0.68, 0.76) & 0.62 (0.58, 0.66) & 0.61 (0.57, 0.65) \\
REFERS & 0.67 (0.62, 0.71) & 0.61 (0.57, 0.65) & 0.61 (0.58, 0.65) \\
\hline
Mean & 0.70 (0.66, 0.73) & 0.60 (0.57, 0.64) & 0.58 (0.55, 0.61) \\
\hline
\end{tabular}
\caption{MIMIC-CXR ROC-AUC results for Lung Opacity when evaluated against FUSE-OR image labels and labels from the findings and impressions report sections. The mean scores represent the average ROC-AUC performance across all tested models.}
\label{tab:mimic_lung_opacity_combined_auc}
\end{table}

\begin{table}[tbp]
\centering
\begin{tabular}{lccc}
\hline
Model & FUSE-OR & Impression & Findings \\
\hline
D-POOL & 0.84 (0.82, 0.87) & 0.76 (0.73, 0.80) & 0.75 (0.71, 0.78) \\
D-CheX & 0.80 (0.77, 0.82) & 0.70 (0.66, 0.74) & 0.69 (0.65, 0.72) \\
D-MC\_ch & 0.80 (0.77, 0.82) & 0.71 (0.67, 0.75) & 0.70 (0.66, 0.73) \\
D-MC\_nb & 0.80 (0.77, 0.82) & 0.71 (0.67, 0.75) & 0.70 (0.66, 0.73) \\
D-NIH & 0.78 (0.75, 0.80) & 0.70 (0.66, 0.74) & 0.69 (0.65, 0.73) \\
D-PC & 0.81 (0.78, 0.83) & 0.77 (0.73, 0.80) & 0.76 (0.73, 0.80) \\
D-RSNA & 0.50 (0.50, 0.50) & 0.50 (0.50, 0.50) & 0.50 (0.50, 0.50) \\
R-POOL & 0.83 (0.81, 0.86) & 0.75 (0.71, 0.79) & 0.76 (0.73, 0.80) \\
JFH & 0.90 (0.88, 0.92) & 0.82 (0.79, 0.85) & 0.81 (0.78, 0.84) \\
CONVIRT-MIMIC & 0.93 (0.92, 0.95) & 0.84 (0.81, 0.86) & 0.83 (0.80, 0.85) \\
GLORIA-G-MIMIC & 0.93 (0.91, 0.95) & 0.83 (0.80, 0.86) & 0.82 (0.79, 0.85) \\
R-MedCLIP-MIMIC & 0.94 (0.92, 0.95) & 0.85 (0.83, 0.88) & 0.84 (0.81, 0.86) \\
V-MedCLIP-MIMIC & 0.94 (0.92, 0.95) & 0.86 (0.83, 0.88) & 0.85 (0.82, 0.87) \\
MedKLIP-MIMIC & 0.93 (0.92, 0.95) & 0.83 (0.80, 0.86) & 0.82 (0.79, 0.85) \\
MFLAG-MIMIC & 0.93 (0.91, 0.95) & 0.83 (0.80, 0.86) & 0.83 (0.80, 0.86) \\
R-MGCA-MIMIC & 0.93 (0.91, 0.94) & 0.83 (0.80, 0.86) & 0.83 (0.80, 0.86) \\
MRM-MIMIC & 0.93 (0.92, 0.95) & 0.84 (0.82, 0.87) & 0.83 (0.80, 0.86) \\
V-MGCA-MIMIC & 0.93 (0.91, 0.95) & 0.83 (0.81, 0.86) & 0.83 (0.80, 0.86) \\
REFERS-MIMIC & 0.92 (0.90, 0.94) & 0.84 (0.81, 0.86) & 0.82 (0.79, 0.85) \\
CONVIRT & 0.71 (0.68, 0.75) & 0.68 (0.63, 0.72) & 0.71 (0.68, 0.75) \\
GLORIA-G & 0.78 (0.75, 0.81) & 0.72 (0.69, 0.76) & 0.71 (0.68, 0.75) \\
GLORIA-B & 0.65 (0.61, 0.68) & 0.64 (0.60, 0.68) & 0.64 (0.61, 0.68) \\
R-MGCA & 0.51 (0.48, 0.55) & 0.52 (0.48, 0.57) & 0.49 (0.45, 0.53) \\
V-MGCA & 0.60 (0.56, 0.64) & 0.55 (0.51, 0.59) & 0.54 (0.50, 0.58) \\
PTUNIFIER & 0.80 (0.78, 0.83) & 0.77 (0.74, 0.80) & 0.76 (0.72, 0.79) \\
REFERS & 0.74 (0.70, 0.77) & 0.72 (0.67, 0.75) & 0.71 (0.67, 0.75) \\
\hline
Macro Avg. & 0.81 (0.79, 0.84) & 0.75 (0.71, 0.78) & 0.74 (0.71, 0.77) \\
\hline
\end{tabular}
\caption{MIMIC-CXR ROC-AUC results for Pleural effusion when evaluated against FUSE-OR image labels and labels from the findings and impressions report sections. The mean scores refer to the average ROC-AUC performance of all tested models.}
\label{tab:mimic_pleural_effusion_combined_auc}
\end{table}

\begin{table}[tbp]
\centering
\begin{tabular}{lccc}
\hline
Model & FUSE-OR & Impression & Findings \\
\hline
D-POOL & 0.75 (0.72, 0.78) & 0.67 (0.62, 0.72) & 0.63 (0.58, 0.67) \\
D-CheX & 0.70 (0.67, 0.73) & 0.60 (0.55, 0.65) & 0.58 (0.53, 0.62) \\
D-MC\_ch & 0.71 (0.68, 0.74) & 0.64 (0.59, 0.69) & 0.62 (0.57, 0.67) \\
D-MC\_nb & 0.71 (0.68, 0.74) & 0.64 (0.59, 0.68) & 0.62 (0.57, 0.66) \\
D-NIH & 0.66 (0.64, 0.68) & 0.59 (0.55, 0.63) & 0.58 (0.54, 0.62) \\
D-PC & 0.67 (0.64, 0.69) & 0.64 (0.60, 0.68) & 0.65 (0.61, 0.68) \\
D-RSNA & 0.55 (0.54, 0.56) & 0.52 (0.51, 0.53) & 0.52 (0.51, 0.52) \\
R-POOL & 0.70 (0.67, 0.73) & 0.68 (0.63, 0.73) & 0.65 (0.60, 0.69) \\
JFH & 0.72 (0.70, 0.74) & 0.66 (0.62, 0.69) & 0.65 (0.62, 0.69) \\
CONVIRT-MIMIC & 0.81 (0.78, 0.83) & 0.71 (0.66, 0.75) & 0.70 (0.65, 0.74) \\
GLORIA-G-MIMIC & 0.82 (0.79, 0.85) & 0.71 (0.66, 0.75) & 0.69 (0.65, 0.74) \\
R-MedCLIP-MIMIC & 0.81 (0.78, 0.84) & 0.72 (0.67, 0.76) & 0.71 (0.66, 0.75) \\
V-MedCLIP-MIMIC & 0.82 (0.79, 0.84) & 0.73 (0.68, 0.77) & 0.72 (0.67, 0.76) \\
MedKLIP-MIMIC & 0.82 (0.79, 0.84) & 0.70 (0.66, 0.75) & 0.69 (0.65, 0.74) \\
MFLAG-MIMIC & 0.81 (0.78, 0.84) & 0.71 (0.67, 0.76) & 0.70 (0.65, 0.74) \\
R-MGCA-MIMIC & 0.82 (0.79, 0.84) & 0.72 (0.68, 0.77) & 0.70 (0.65, 0.74) \\
MRM-MIMIC & 0.82 (0.79, 0.85) & 0.72 (0.68, 0.77) & 0.72 (0.68, 0.77) \\
V-MGCA-MIMIC & 0.81 (0.78, 0.84) & 0.71 (0.66, 0.75) & 0.70 (0.65, 0.74) \\
REFERS-MIMIC & 0.81 (0.78, 0.84) & 0.72 (0.68, 0.77) & 0.71 (0.66, 0.75) \\
CONVIRT & 0.61 (0.57, 0.65) & 0.60 (0.54, 0.65) & 0.61 (0.56, 0.67) \\
GLORIA-G & 0.67 (0.64, 0.71) & 0.63 (0.58, 0.68) & 0.62 (0.58, 0.67) \\
GLORIA-B & 0.63 (0.59, 0.67) & 0.63 (0.58, 0.68) & 0.61 (0.56, 0.65) \\
R-MGCA & 0.48 (0.44, 0.52) & 0.49 (0.44, 0.54) & 0.48 (0.43, 0.53) \\
V-MGCA & 0.50 (0.46, 0.54) & 0.50 (0.45, 0.55) & 0.46 (0.41, 0.51) \\
PTUNIFIER & 0.67 (0.64, 0.71) & 0.65 (0.60, 0.70) & 0.64 (0.60, 0.69) \\
REFERS & 0.60 (0.56, 0.64) & 0.60 (0.55, 0.65) & 0.63 (0.58, 0.68) \\
\hline
Mean & 0.71 (0.68, 0.74) & 0.65 (0.60, 0.69) & 0.64 (0.59, 0.68) \\
\hline
\end{tabular}
\caption{MIMIC-CXR ROC-AUC macro averages across the four selected pathologies (Atelectasis, Consolidation, Lung opacity, and Pleural effusion) when evaluated against FUSE-OR image labels and labels from the findings and impressions report sections. The mean scores represent the average ROC-AUC performance across all tested models.}
\label{tab:mimic_s_overall_1_combined_auc}
\end{table}

\clearpage

\section{Rank Correlation Tables}
\begin{table}[H]
\centering
\small
\setlength{\tabcolsep}{4pt}
\begin{tabular}{llcccc}
\hline
\makecell[l]{Finding /\\summary target} & Pair & All Models & VLM-0-SHOT & VLM-FT & XRV \\
\hline
\multirow{2}{*}{Atelectasis} & vs Rep1 & $0.80 \pm 0.07$ & $0.79 \pm 0.11$ & $0.21 \pm 0.22$ & $0.62 \pm 0.14$ \\
                             & vs Rep2 & $0.87 \pm 0.05$ & $0.80 \pm 0.11$ & $0.20 \pm 0.22$ & $0.69 \pm 0.11$ \\
\hline
\multirow{2}{*}{Consolidation} & vs Rep1 & $0.88 \pm 0.04$ & $0.82 \pm 0.12$ & $0.55 \pm 0.18$ & $0.78 \pm 0.10$ \\
                               & vs Rep2 & $0.89 \pm 0.03$ & $0.80 \pm 0.14$ & $0.63 \pm 0.16$ & $0.82 \pm 0.09$ \\
\hline
\multirow{2}{*}{Lung opacity} & vs Rep1 & $0.80 \pm 0.07$ & $0.79 \pm 0.12$ & $0.77 \pm 0.13$ & $0.90 \pm 0.05$ \\
                              & vs Rep2 & $0.81 \pm 0.06$ & $0.82 \pm 0.11$ & $0.78 \pm 0.11$ & $0.89 \pm 0.05$ \\
\hline
\multirow{2}{*}{Pleural effusion} & vs Rep1 & $0.96 \pm 0.02$ & $0.86 \pm 0.09$ & $0.83 \pm 0.11$ & $0.93 \pm 0.04$ \\
                                  & vs Rep2 & $0.96 \pm 0.02$ & $0.87 \pm 0.08$ & $0.82 \pm 0.11$ & $0.94 \pm 0.04$ \\
\hline
\multirow{2}{*}{Macro Avg.} & vs Rep1 & $0.96 \pm 0.01$ & $0.94 \pm 0.05$ & $0.73 \pm 0.12$ & $0.95 \pm 0.02$ \\
                            & vs Rep2 & $0.97 \pm 0.01$ & $0.96 \pm 0.03$ & $0.73 \pm 0.12$ & $0.96 \pm 0.02$ \\
\hline
\end{tabular}
\caption{XQA-RR SRCC across pathologies and model groups. Pairs compare conflict-resolved (CR) image labels across the two report annotators' labels (Rep1 and Rep2). Macro average refers to the SRCC of rankings derived from the macro-averaged ROC-AUCs of the four selected pathologies. SRCC is calculated across model groups that include TorchXRayVision supervised models (XRV), BenchX VLM 0-shot models (VLM-0-shot), BenchX VLM fine-tuned models (VLM-FT), and all models combined.}
\label{tab:cuh_all_pairs_spearman_all_groups}
\end{table}

\begin{table}[H]
\centering
\small
\setlength{\tabcolsep}{4pt}
\begin{tabular}{llcccc}
\hline
\makecell[l]{Finding /\\summary target} & Pair & All Models & VLM-0-SHOT & VLM-FT & XRV \\
\hline
\multirow{2}{*}{Atelectasis} & vs Rep1 & $0.62 \pm 0.08$ & $0.62 \pm 0.16$ & $0.15 \pm 0.18$ & $0.49 \pm 0.12$ \\
                             & vs Rep2 & $0.69 \pm 0.06$ & $0.65 \pm 0.16$ & $0.14 \pm 0.19$ & $0.55 \pm 0.10$ \\
\hline
\multirow{2}{*}{Consolidation} & vs Rep1 & $0.72 \pm 0.05$ & $0.70 \pm 0.16$ & $0.43 \pm 0.16$ & $0.62 \pm 0.10$ \\
                               & vs Rep2 & $0.74 \pm 0.04$ & $0.67 \pm 0.17$ & $0.49 \pm 0.15$ & $0.66 \pm 0.09$ \\
\hline
\multirow{2}{*}{Lung opacity} & vs Rep1 & $0.64 \pm 0.07$ & $0.65 \pm 0.14$ & $0.63 \pm 0.13$ & $0.77 \pm 0.07$ \\
                              & vs Rep2 & $0.65 \pm 0.07$ & $0.69 \pm 0.14$ & $0.64 \pm 0.13$ & $0.77 \pm 0.07$ \\
\hline
\multirow{2}{*}{Pleural effusion} & vs Rep1 & $0.85 \pm 0.03$ & $0.76 \pm 0.12$ & $0.70 \pm 0.13$ & $0.81 \pm 0.06$ \\
                                  & vs Rep2 & $0.85 \pm 0.04$ & $0.76 \pm 0.12$ & $0.69 \pm 0.13$ & $0.83 \pm 0.06$ \\
\hline
\multirow{2}{*}{Macro Avg.} & vs Rep1 & $0.85 \pm 0.03$ & $0.87 \pm 0.10$ & $0.60 \pm 0.13$ & $0.85 \pm 0.04$ \\
                            & vs Rep2 & $0.88 \pm 0.02$ & $0.90 \pm 0.08$ & $0.59 \pm 0.13$ & $0.87 \pm 0.04$ \\
\hline
\end{tabular}
\caption{XQA-RR KRCC across pathologies and model groups. Pairs compare conflict-resolved (CR) image labels across the two report annotators' labels (Rep1 and Rep2). Macro average refers to the KRCC of rankings derived from the macro-averaged ROC-AUC of the four selected pathologies. KRCC is calculated across model groups that include TorchXRayVision supervised models (XRV), BenchX VLM 0-shot models (VLM-0-shot), BenchX VLM fine-tuned models (VLM-FT), and all models combined.}
\label{tab:cuh_all_pairs_kendalltau_all_groups}
\end{table}

\begin{table}[H]
\centering
\small
\setlength{\tabcolsep}{4pt}
\begin{tabular}{llcccc}
\hline
\makecell[l]{Finding /\\summary target} & Pair & All Models & VLM-0-SHOT & VLM-FT & XRV \\
\hline
\multirow{2}{*}{Atelectasis} & vs Find. & $0.76 \pm 0.12$ & $0.42 \pm 0.21$ & $0.44 \pm 0.25$ & $0.57 \pm 0.19$ \\
                             & vs Impr. & $0.78 \pm 0.12$ & $0.61 \pm 0.23$ & $0.15 \pm 0.33$ & $0.51 \pm 0.19$ \\
\hline
\multirow{2}{*}{Consolidation} & vs Find. & $0.77 \pm 0.11$ & $0.66 \pm 0.17$ & $0.00 \pm 0.30$ & $0.63 \pm 0.15$ \\
                               & vs Impr. & $0.79 \pm 0.09$ & $0.79 \pm 0.13$ & $-0.04 \pm 0.31$ & $0.46 \pm 0.16$ \\
\hline
\multirow{2}{*}{Lung opacity} & vs Find. & $0.78 \pm 0.09$ & $0.58 \pm 0.15$ & $0.24 \pm 0.28$ & $0.66 \pm 0.12$ \\
                              & vs Impr. & $0.75 \pm 0.10$ & $0.74 \pm 0.15$ & $0.49 \pm 0.24$ & $0.71 \pm 0.09$ \\
\hline
\multirow{2}{*}{Pleural effusion} & vs Find. & $0.92 \pm 0.02$ & $0.91 \pm 0.06$ & $0.36 \pm 0.27$ & $0.86 \pm 0.08$ \\
                                  & vs Impr. & $0.93 \pm 0.02$ & $0.95 \pm 0.04$ & $0.32 \pm 0.29$ & $0.86 \pm 0.08$ \\
\hline
\multirow{2}{*}{Macro Avg.} & vs Find. & $0.86 \pm 0.06$ & $0.89 \pm 0.07$ & $0.73 \pm 0.12$ & $0.82 \pm 0.08$ \\
                            & vs Impr. & $0.96 \pm 0.01$ & $0.84 \pm 0.08$ & $0.59 \pm 0.17$ & $0.91 \pm 0.06$ \\
\hline
\multirow{2}{*}{Micro Avg.} & vs Find. & $0.93 \pm 0.01$ & $0.94 \pm 0.05$ & $0.62 \pm 0.15$ & $0.74 \pm 0.07$ \\
                            & vs Impr. & $0.96 \pm 0.01$ & $0.90 \pm 0.06$ & $0.53 \pm 0.18$ & $0.89 \pm 0.04$ \\
\hline
\end{tabular}
\caption{MIMIC SRCC across pathologies and model groups. Pairs compare FUSE-OR image labels against labels from radiology report findings (Find.) or impressions (Impr.) section annotations. Macro average refers to the SRCC of rankings derived from the macro-averaged ROC-AUCs of the four selected pathologies.  SRCC is calculated across model groups that include TorchXRayVision supervised models (XRV), BenchX VLM 0-shot models (VLM-0-shot), BenchX VLM fine-tuned models (VLM-FT), and all models combined.}
\label{tab:mimic_all_pairs_spearman_all_groups}
\end{table}

\begin{table}[H]
\centering
\small
\setlength{\tabcolsep}{4pt}
\begin{tabular}{llcccc}
\hline
\makecell[l]{Finding /\\summary target} & Pair & All Models & VLM-0-SHOT & VLM-FT & XRV \\
\hline
\multirow{2}{*}{Atelectasis} & vs Find. & $0.58 \pm 0.11$ & $0.31 \pm 0.18$ & $0.33 \pm 0.20$ & $0.45 \pm 0.16$ \\
                             & vs Impr. & $0.61 \pm 0.12$ & $0.47 \pm 0.20$ & $0.11 \pm 0.24$ & $0.41 \pm 0.17$ \\
\hline
\multirow{2}{*}{Consolidation} & vs Find. & $0.59 \pm 0.10$ & $0.52 \pm 0.18$ & $-0.00 \pm 0.22$ & $0.46 \pm 0.14$ \\
                               & vs Impr. & $0.61 \pm 0.09$ & $0.65 \pm 0.16$ & $-0.02 \pm 0.23$ & $0.36 \pm 0.15$ \\
\hline
\multirow{2}{*}{Lung opacity} & vs Find. & $0.61 \pm 0.08$ & $0.44 \pm 0.16$ & $0.17 \pm 0.21$ & $0.50 \pm 0.14$ \\
                              & vs Impr. & $0.60 \pm 0.10$ & $0.58 \pm 0.17$ & $0.37 \pm 0.20$ & $0.60 \pm 0.11$ \\
\hline
\multirow{2}{*}{Pleural effusion} & vs Find. & $0.78 \pm 0.05$ & $0.83 \pm 0.11$ & $0.27 \pm 0.21$ & $0.73 \pm 0.11$ \\
                                  & vs Impr. & $0.80 \pm 0.05$ & $0.89 \pm 0.09$ & $0.24 \pm 0.22$ & $0.74 \pm 0.10$ \\
\hline
\multirow{2}{*}{Macro Avg.} & vs Find. & $0.67 \pm 0.05$ & $0.57 \pm 0.15$ & $0.13 \pm 0.20$ & $0.36 \pm 0.13$ \\
                            & vs Impr. & $0.74 \pm 0.05$ & $0.80 \pm 0.11$ & $0.11 \pm 0.21$ & $0.55 \pm 0.13$ \\
\hline
\end{tabular}
\caption{MIMIC KRCC across pathologies and model groups. Pairs compare FUSE-OR image labels against labels from radiology report findings (Find.) or impressions (Impr.) section annotations. Macro average refers to the KRCC of rankings derived from the macro-averaged ROC-AUC of the four selected pathologies.  KRCC is calculated across model groups that include TorchXRayVision supervised models (XRV), BenchX VLM 0-shot models (VLM-0-shot), BenchX VLM fine-tuned models (VLM-FT), and all models combined.}
\label{tab:mimic_all_pairs_kendalltau_all_groups}
\end{table}

\clearpage

\section{Label Fusion Methods: Agreement with Conflict-Resolved Image-derived Labels}

Figure~\ref{fig:cuh_fusion_agreement_k} and Appendix Figure~\ref{fig:cuh_fusion_agreement_f1_app} show the agreement between each image-derived label fusion method and the XQA-RR conflict-resolved (CR) labels, supporting the use of FUSE\_OR as a proxy for CR in MIMIC-CXR, where no conflict resolver is available (Table~\ref{tab:cuh_image_algorithms_vs_conflict_resolver}).

\begin{longtable}{lrrrrrr}
\toprule
Pathology & $\kappa_{\text{w}}$ & $\kappa_{\text{low}}$ & $\kappa_{\text{high}}$ & F1 & F1$_{\text{low}}$ & F1$_{\text{high}}$ \\
\midrule
\endfirsthead
\toprule
Pathology & $\kappa_{\text{w}}$ & $\kappa_{\text{low}}$ & $\kappa_{\text{high}}$ & F1 & F1$_{\text{low}}$ & F1$_{\text{high}}$ \\
\midrule
\endhead
\midrule
\multicolumn{7}{r}{\textit{Continued on next page}}\\
\endfoot
\bottomrule\\
\caption{XQA-RR agreement between image-only algorithm labels and the conflict-resolved image labels. $\kappa_{\text{w}}$ denotes quadratically weighted Cohen's $\kappa$ computed on the original three-class labels (negative, uncertain, positive). $\kappa_{\text{low}}$ and $\kappa_{\text{high}}$ denote unweighted Cohen's $\kappa$ under pessimistic and optimistic uncertainty mappings, respectively. F1${\text{low}}$ and F1${\text{high}}$ are computed under the same mappings.}
\label{tab:cuh_image_algorithms_vs_conflict_resolver}
\endlastfoot
\multicolumn{7}{c}{MV} \\
\midrule
Atelectasis          & 0.76 & 0.14 & 1.00 & 0.75 & 0.69 & 1.00 \\
Consolidation        & 0.87 & 0.51 & 1.00 & 0.90 & 0.87 & 1.00 \\
Lung Opacity         & 0.79 & 0.30 & 1.00 & 0.78 & 0.72 & 1.00 \\
Pleural Effusion     & 0.87 & 0.54 & 1.00 & 0.88 & 0.85 & 1.00 \\
\cmidrule(lr){1-7}
Macro avg.\ (four selected) & 0.83 & 0.37 & 1.00 & 0.83 & 0.78 & 1.00 \\
Macro avg.\ (all 12) & 0.76 & 0.35 & 1.00 & 0.92 & 0.90 & 1.00 \\
Micro avg.\ (all 12) & 0.82 & 0.40 & 1.00 & 0.92 & 0.90 & 1.00 \\
\midrule
\multicolumn{7}{c}{DS-EM} \\
\midrule
Atelectasis          & 0.23 & 0.20 & 0.25 & 0.68 & 0.67 & 0.73 \\
Consolidation        & 0.60 & 0.51 & 0.65 & 0.86 & 0.86 & 0.91 \\
Lung Opacity         & 0.65 & 0.56 & 0.69 & 0.80 & 0.81 & 0.87 \\
Pleural Effusion     & 0.81 & 0.73 & 0.84 & 0.90 & 0.91 & 0.95 \\
\cmidrule(lr){1-7}
Macro avg.\ (four selected) & 0.57 & 0.50 & 0.61 & 0.81 & 0.81 & 0.87 \\
Macro avg.\ (all 12) & 0.54 & 0.42 & 0.60 & 0.92 & 0.92 & 0.95 \\
Micro avg.\ (all 12) & 0.64 & 0.53 & 0.69 & 0.92 & 0.92 & 0.95 \\
\midrule
\multicolumn{7}{c}{FUSE\_OR} \\
\midrule
Atelectasis          & 0.93 & 0.87 & 0.95 & 0.95 & 0.95 & 0.98 \\
Consolidation        & 0.90 & 0.83 & 0.93 & 0.95 & 0.95 & 0.98 \\
Lung Opacity         & 0.78 & 0.61 & 0.85 & 0.83 & 0.83 & 0.93 \\
Pleural Effusion     & 0.89 & 0.75 & 0.94 & 0.92 & 0.91 & 0.98 \\
\cmidrule(lr){1-7}
Macro avg.\ (four selected) & 0.87 & 0.77 & 0.92 & 0.91 & 0.91 & 0.97 \\
Macro avg.\ (all 12) & 0.78 & 0.59 & 0.89 & 0.95 & 0.95 & 0.98 \\
Micro avg.\ (all 12) & 0.82 & 0.69 & 0.87 & 0.94 & 0.94 & 0.98 \\
\midrule
\multicolumn{7}{c}{FUSE\_AND} \\
\midrule
Atelectasis          & 0.21 & 0.17 & 0.25 & 0.66 & 0.66 & 0.73 \\
Consolidation        & 0.57 & 0.48 & 0.61 & 0.85 & 0.85 & 0.90 \\
Lung Opacity         & 0.60 & 0.44 & 0.70 & 0.76 & 0.75 & 0.88 \\
Pleural Effusion     & 0.77 & 0.62 & 0.84 & 0.87 & 0.87 & 0.95 \\
\cmidrule(lr){1-7}
Macro avg.\ (four selected) & 0.54 & 0.43 & 0.60 & 0.79 & 0.78 & 0.87 \\
Macro avg.\ (all 12) & 0.48 & 0.37 & 0.54 & 0.91 & 0.91 & 0.95 \\
Micro avg.\ (all 12) & 0.60 & 0.47 & 0.66 & 0.91 & 0.91 & 0.95 \\
\midrule
\multicolumn{7}{c}{MACE} \\
\midrule
Atelectasis          & 0.93 & 0.87 & 0.94 & 0.95 & 0.95 & 0.98 \\
Consolidation        & 0.90 & 0.81 & 0.94 & 0.94 & 0.95 & 0.98 \\
Lung Opacity         & 0.75 & 0.50 & 0.86 & 0.81 & 0.79 & 0.94 \\
Pleural Effusion     & 0.86 & 0.66 & 0.94 & 0.90 & 0.88 & 0.98 \\
\cmidrule(lr){1-7}
Macro avg.\ (four selected) & 0.86 & 0.71 & 0.92 & 0.90 & 0.89 & 0.97 \\
Macro avg.\ (all 12) & 0.77 & 0.56 & 0.88 & 0.94 & 0.94 & 0.98 \\
Micro avg.\ (all 12) & 0.81 & 0.66 & 0.87 & 0.94 & 0.94 & 0.98 \\
\midrule
\multicolumn{7}{c}{NN Fusion} \\
\midrule
Atelectasis          & 1.00 & 0.99 & 1.00 & 1.00 & 1.00 & 1.00 \\
Consolidation        & 0.86 & 0.69 & 0.92 & 0.92 & 0.91 & 0.98 \\
Lung Opacity         & 0.82 & 0.61 & 0.91 & 0.85 & 0.84 & 0.96 \\
Pleural Effusion     & 0.87 & 0.73 & 0.93 & 0.91 & 0.90 & 0.98 \\
\cmidrule(lr){1-7}
Macro avg.\ (four selected) & 0.89 & 0.76 & 0.94 & 0.92 & 0.91 & 0.98 \\
Macro avg.\ (all 12) & 0.71 & 0.53 & 0.82 & 0.96 & 0.96 & 0.99 \\
Micro avg.\ (all 12) & 0.89 & 0.75 & 0.94 & 0.96 & 0.96 & 0.99 \\
\midrule
\multicolumn{7}{c}{NN Fusion (PS)} \\
\midrule
Atelectasis          & 1.00 & 1.00 & 1.00 & 1.00 & 1.00 & 1.00 \\
Consolidation        & 0.87 & 0.76 & 0.92 & 0.94 & 0.93 & 0.98 \\
Lung Opacity         & 0.84 & 0.68 & 0.90 & 0.87 & 0.87 & 0.96 \\
Pleural Effusion     & 0.91 & 0.76 & 0.96 & 0.92 & 0.92 & 0.99 \\
\cmidrule(lr){1-7}
Macro avg.\ (four selected) & 0.90 & 0.80 & 0.95 & 0.93 & 0.93 & 0.98 \\
Macro avg.\ (all 12) & 0.69 & 0.56 & 0.79 & 0.96 & 0.96 & 0.99 \\
Micro avg.\ (all 12) & 0.89 & 0.77 & 0.94 & 0.96 & 0.96 & 0.99 \\
\end{longtable}

\begin{figure}[!htbp]
    \centering
    \includegraphics[width=0.88\textwidth]{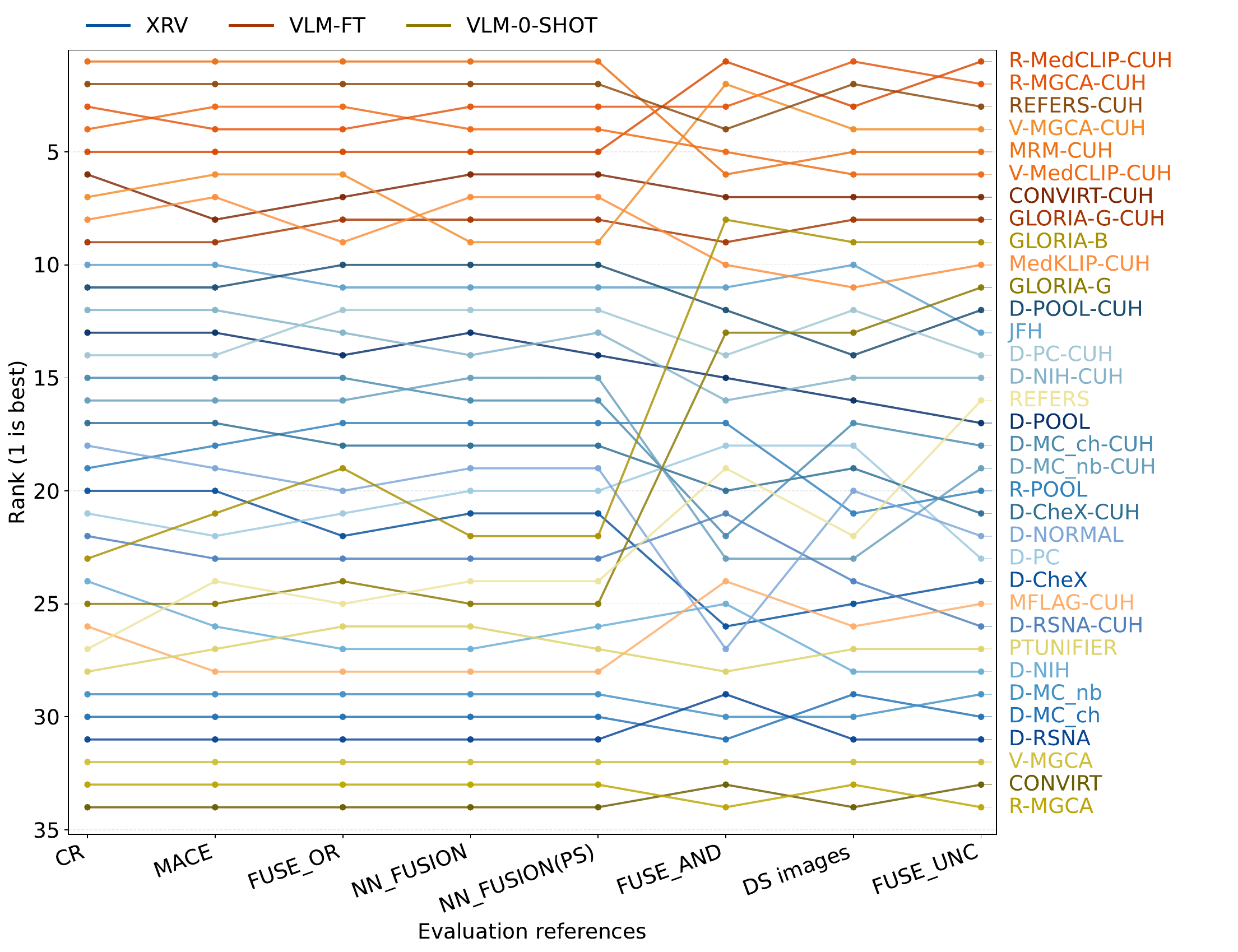}
    \caption{XQA-RR model rankings using fusion labels as reference for \textbf{Atelectasis}.}
    \label{fig:cuh_fusion_atelectasis}
\end{figure}

\begin{figure}[!htbp]
    \centering
    \includegraphics[width=0.88\textwidth]{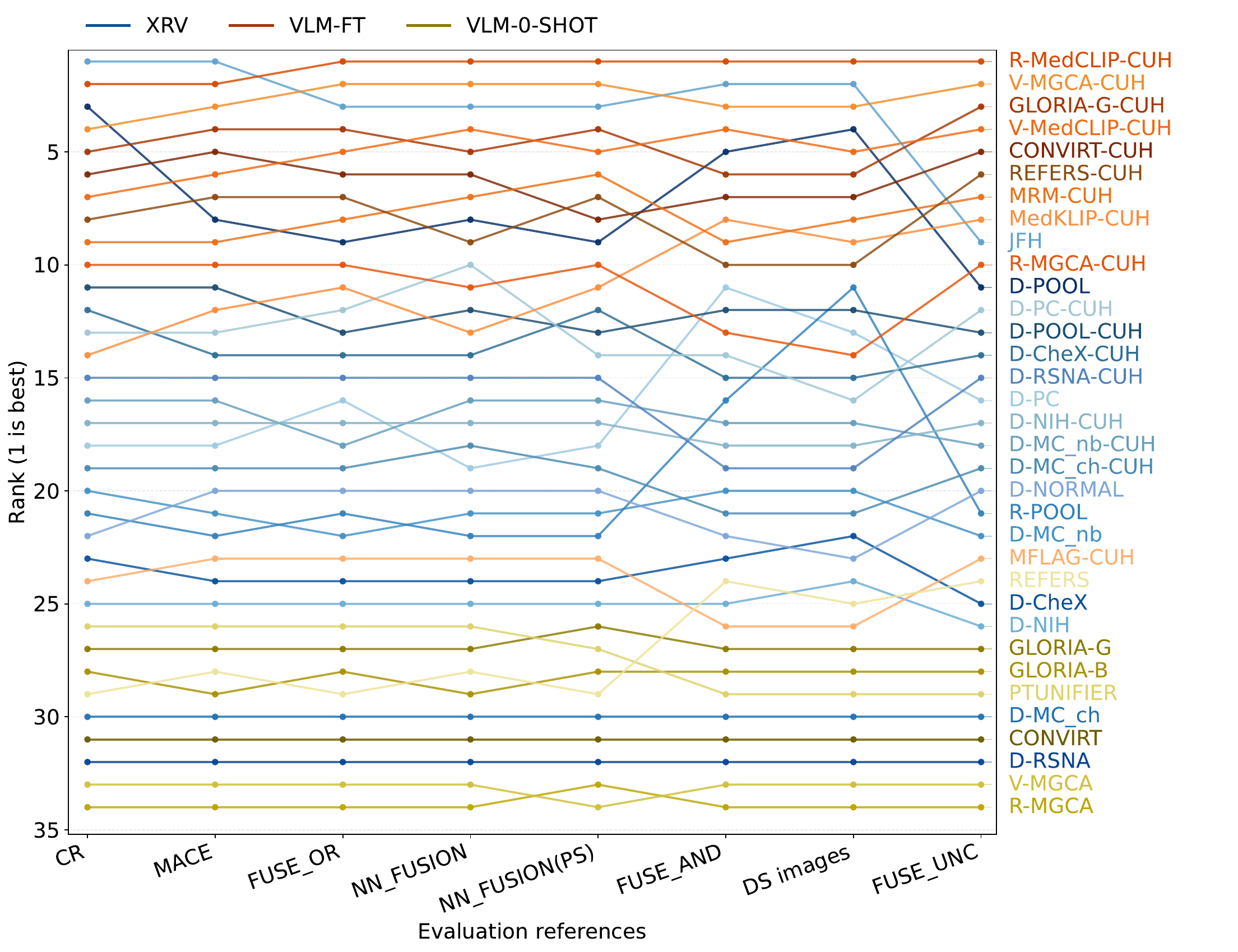}
    \caption{XQA-RR model rankings using fusion labels as reference for \textbf{Consolidation}.}
    \label{fig:cuh_fusion_consolidation}
\end{figure}

\begin{figure}[!htbp]
    \centering
    \includegraphics[width=0.88\textwidth]{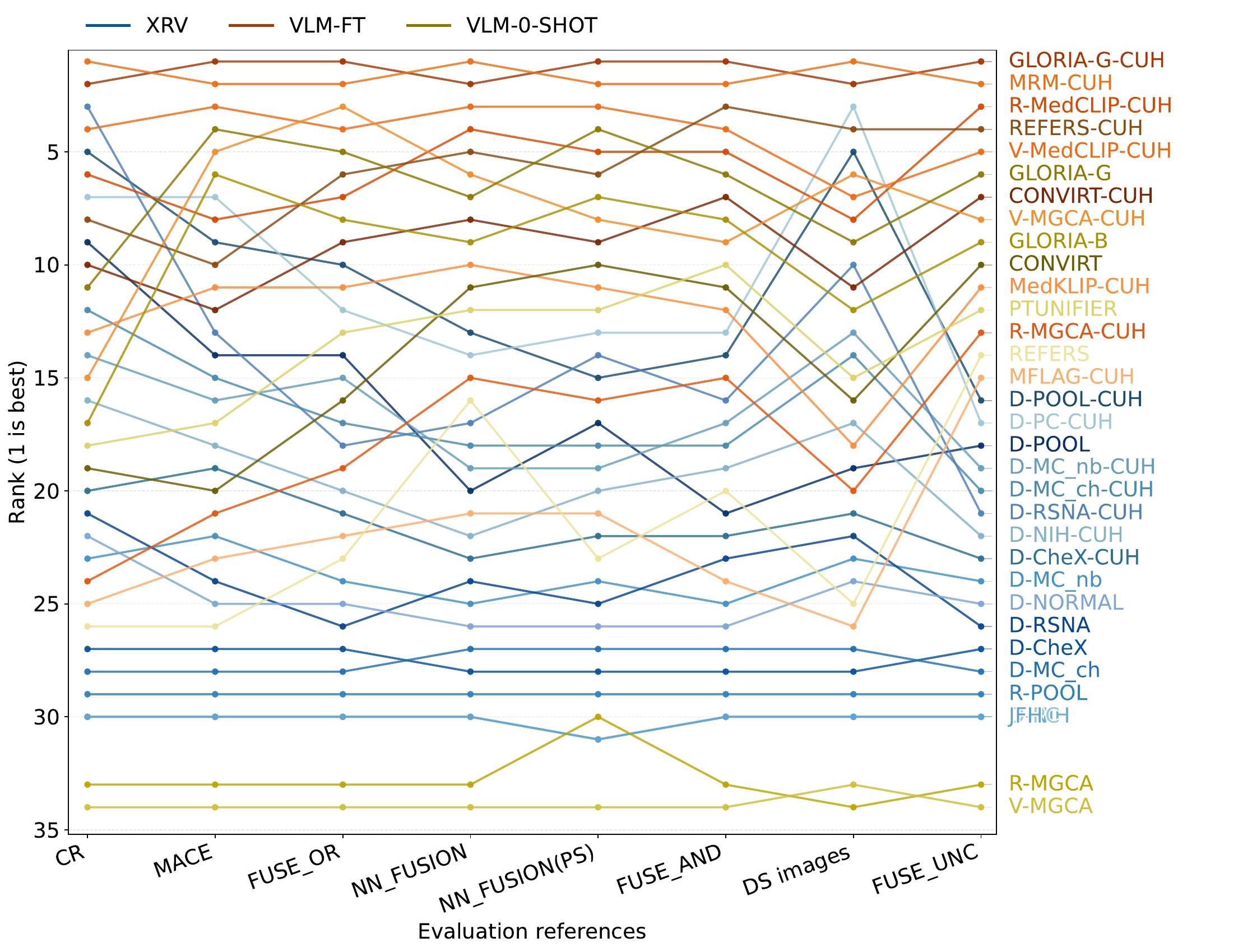}
    \caption{XQA-RR model rankings using fusion labels as reference for \textbf{Lung opacity}.}
    \label{fig:cuh_fusion_lung_opacity}
\end{figure}

\begin{figure}[!htbp]
    \centering
    \includegraphics[width=0.88\textwidth]{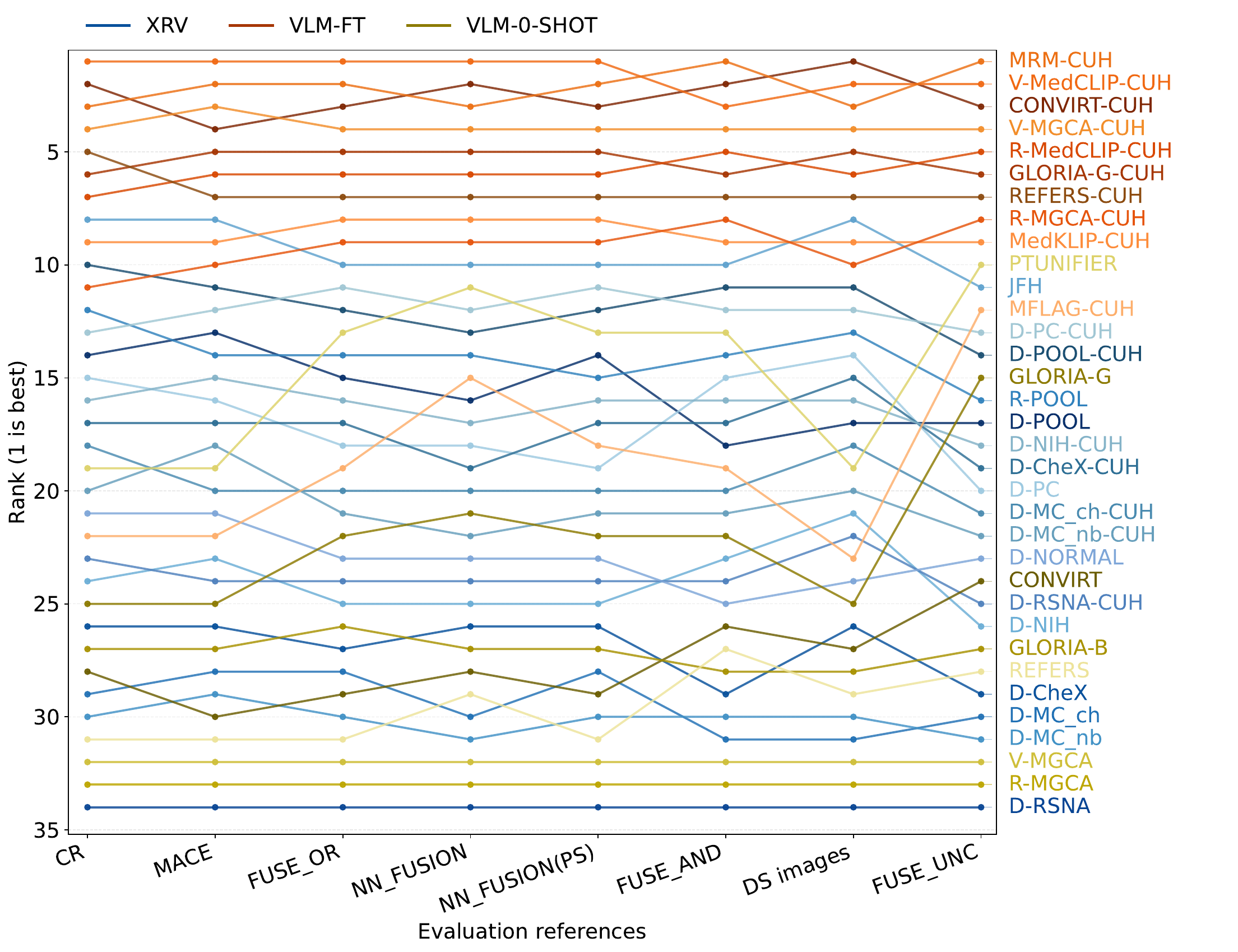}
    \caption{XQA-RR model rankings using fusion labels as reference for \textbf{Pleural effusion}.}
    \label{fig:cuh_fusion_pleural_effusion}
\end{figure}

\begin{figure}[!htbp]
    \centering
    \includegraphics[width=0.88\textwidth]{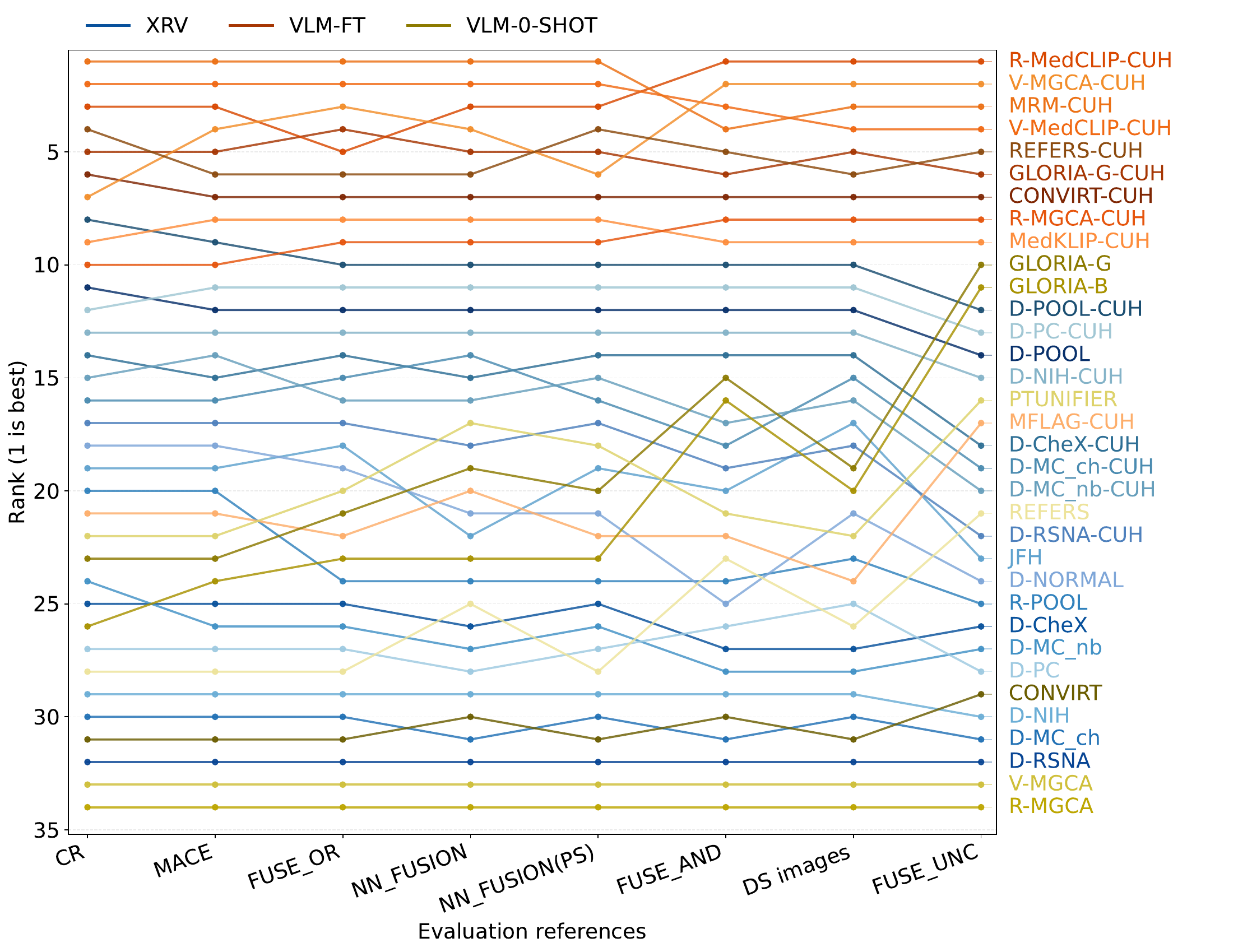}
    \caption{XQA-RR model rankings using fusion labels as reference on ROC-AUC macro-averaged across the four selected pathologies.}
    \label{fig:cuh_fusion_global}
\end{figure}

Figures~\ref{fig:cuh_fusion_atelectasis}--\ref{fig:cuh_fusion_global} show model rankings on XQA-RR using fusion-derived labels as the reference.

\clearpage

\section{MIMIC: Findings vs.\ Impressions Label Source}

\begin{figure}[!htbp]
    \centering

    \begin{subfigure}[b]{0.31\textwidth}
        \centering
        \includegraphics[width=\textwidth]{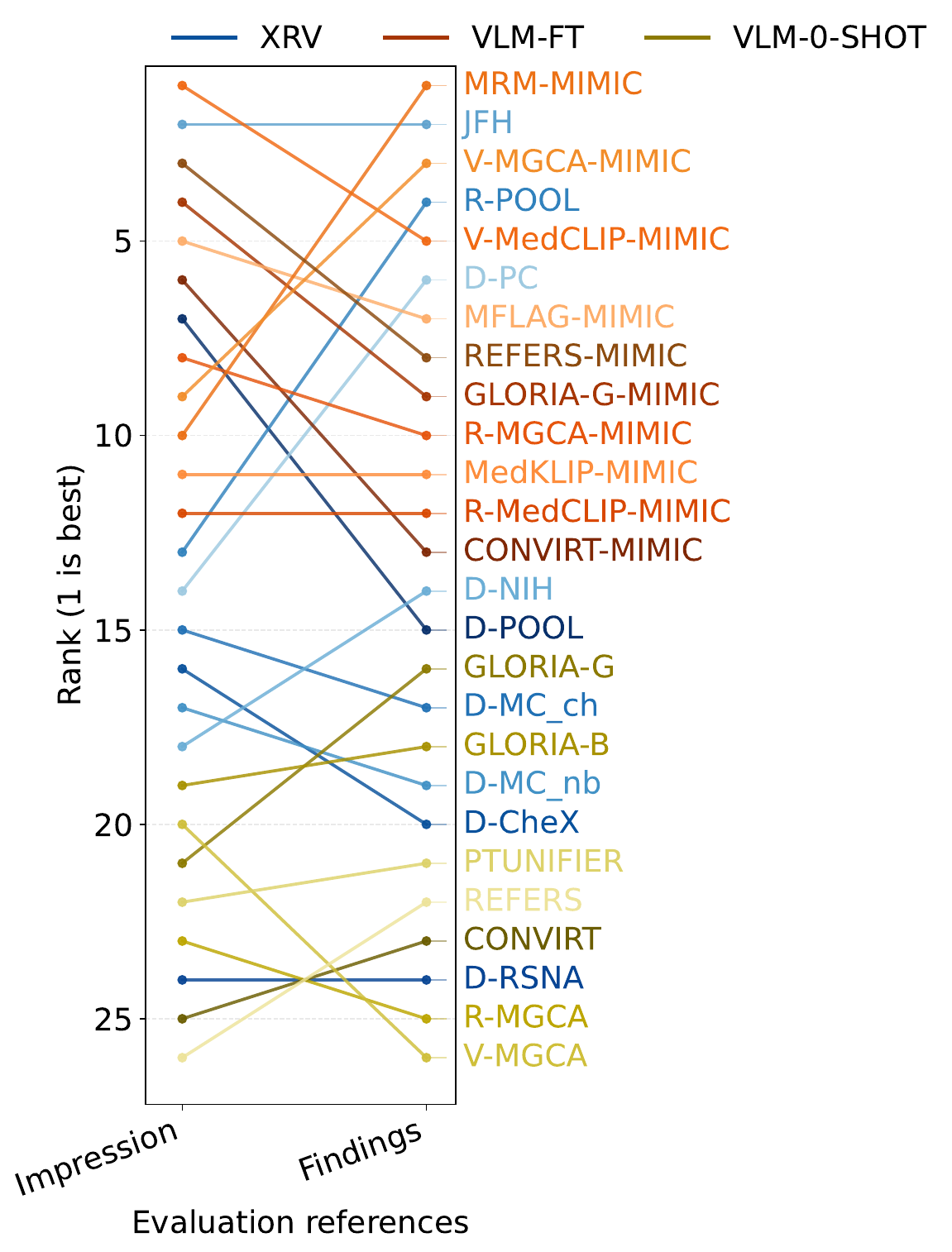}
        \caption{Atelectasis}
        \label{fig:mimic_impr_atelectasis}
    \end{subfigure}
    \hfill
    \begin{subfigure}[b]{0.31\textwidth}
        \centering
        \includegraphics[width=\textwidth]{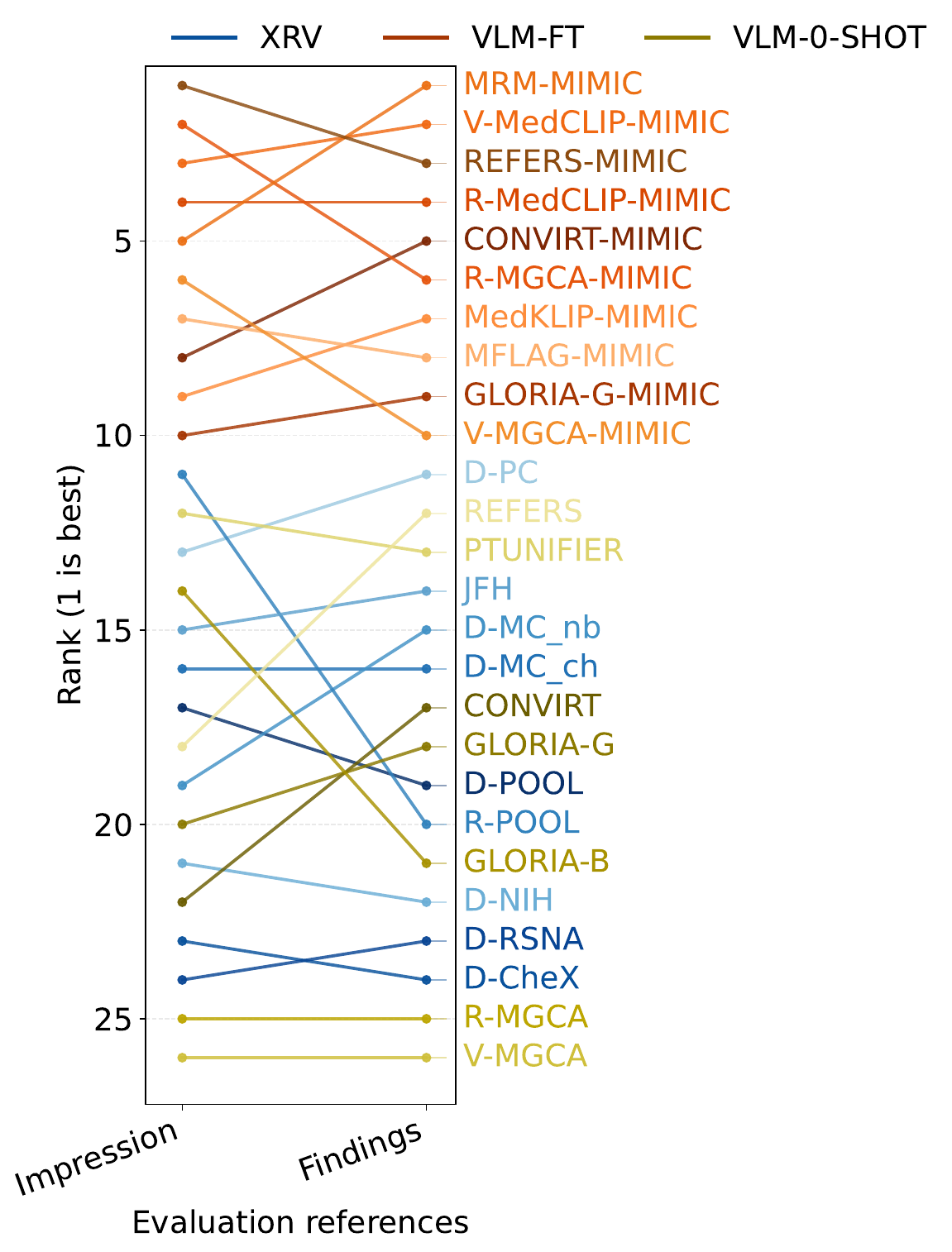}
        \caption{Consolidation}
        \label{fig:mimic_impr_consolidation}
    \end{subfigure}
    \hfill
    \begin{subfigure}[b]{0.31\textwidth}
        \centering
        \includegraphics[width=\textwidth]{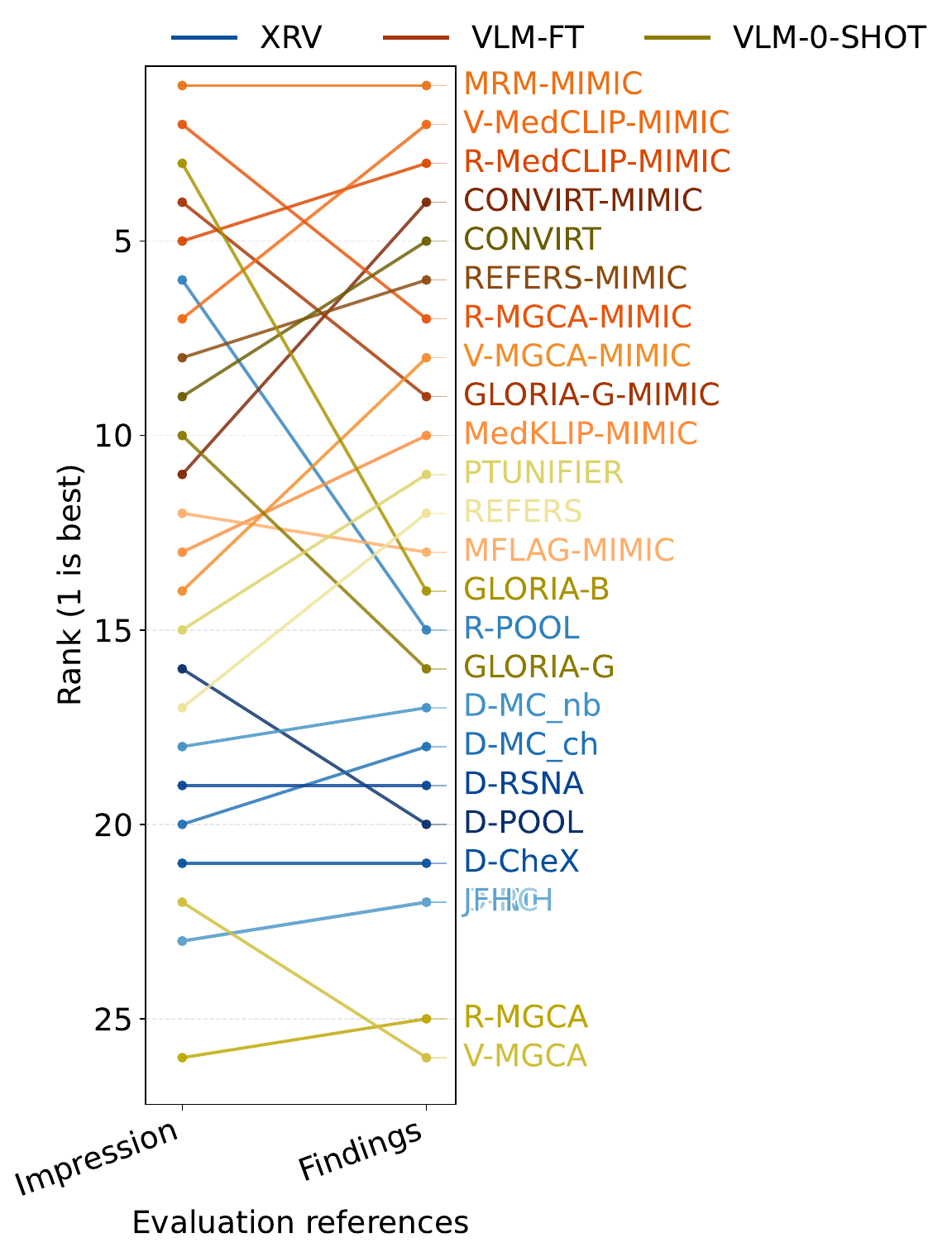}
        \caption{Lung opacity}
        \label{fig:mimic_impr_lung_opacity}
    \end{subfigure}

    \vspace{0.5em}

    \begin{subfigure}[b]{0.31\textwidth}
        \centering
        \includegraphics[width=\textwidth]{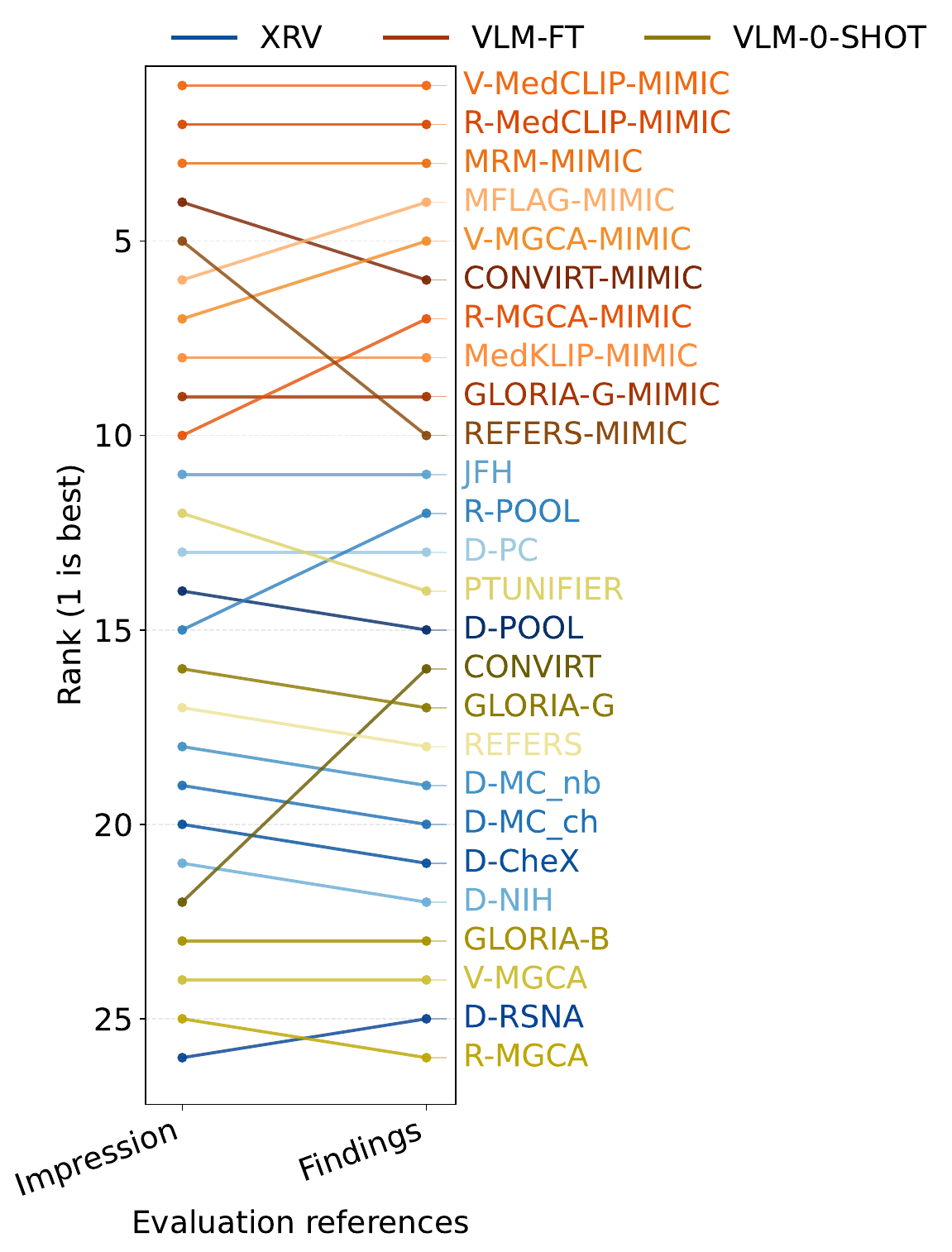}
        \caption{Pleural effusion}
        \label{fig:mimic_impr_pleural_effusion}
    \end{subfigure}
    \hspace{0.06\textwidth}
    \begin{subfigure}[b]{0.31\textwidth}
        \centering
        \includegraphics[width=\textwidth]{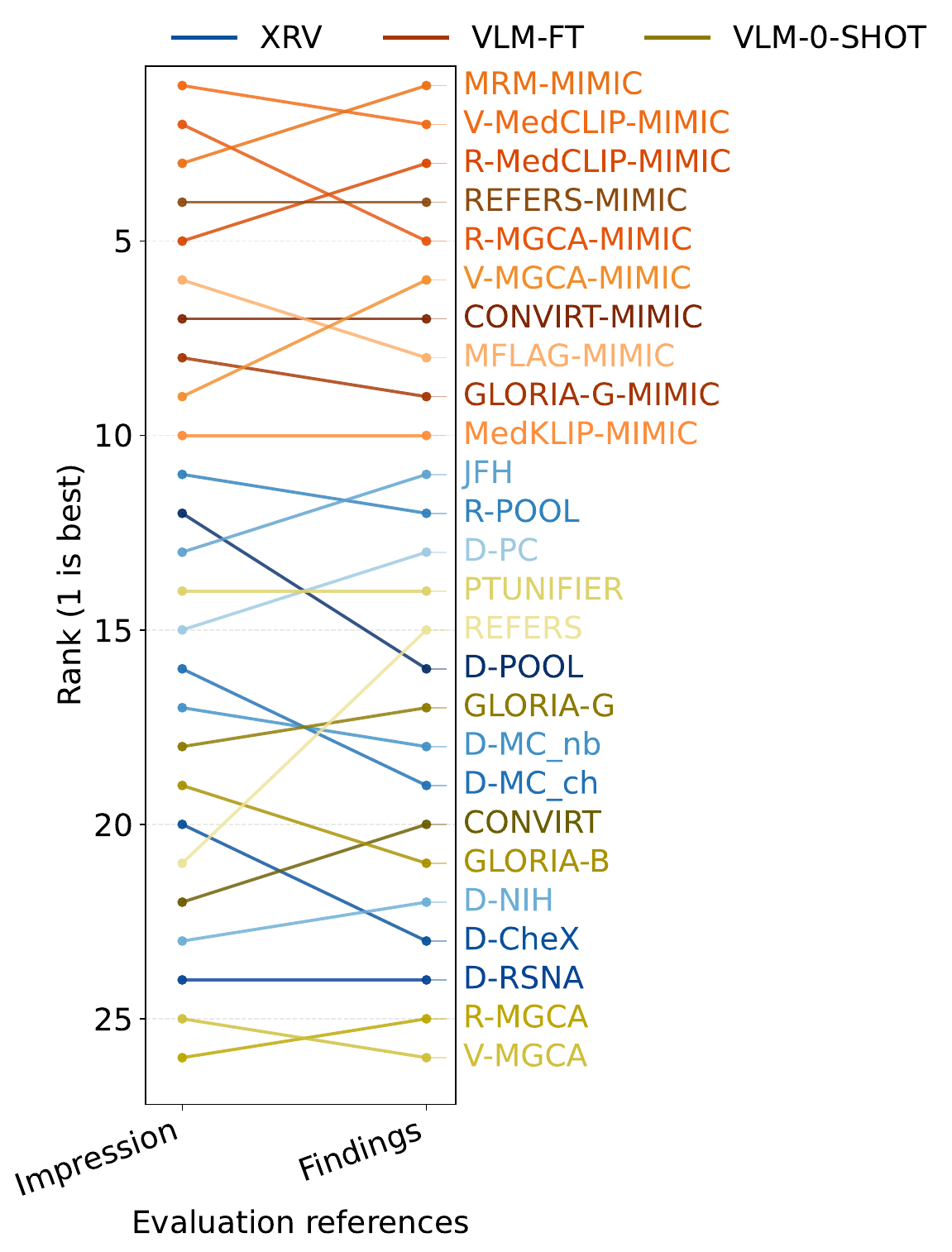}
        \caption{Macro-average}
        \label{fig:mimic_impr_global}
    \end{subfigure}

    \caption{MIMIC-CXR model rankings induced by labels derived from the Findings and Impressions sections of the radiology report.}
    \label{fig:mimic_impr_rankings}
\end{figure}

Figures~\ref{fig:mimic_impr_atelectasis}--\ref{fig:mimic_impr_global} show how model rankings on MIMIC-CXR change depending on whether labels are derived from the Findings or Impressions section of the report.

\end{document}